\documentclass[fleqn,usenatbib]{mnras}
\usepackage{newtxtext,newtxmath}
\usepackage[T1]{fontenc}
\DeclareRobustCommand{\VAN}[3]{#2}
\let\VANthebibliography\thebibliography
\def\thebibliography{\DeclareRobustCommand{\VAN}[3]{##3}\VANthebibliography}

\usepackage{float}
\usepackage{graphicx}	
\usepackage{amsmath}	
\usepackage{subcaption}
\usepackage{xcolor}
\usepackage{comment}
\usepackage{wrapfig,lipsum,booktabs}
\usepackage{anyfontsize}

\newcommand{\di}[1]{{\textcolor{black}{#1}}}
\newcommand{\dii}[1]{{\textcolor{black}{#1}}}
\newcommand{\diii}[1]{{\textcolor{black}{#1}}}

\usepackage[percent]{overpic}
\def\gtsim {>\kern-1.2em\lower1.1ex\hbox{$\sim$}~}   
\def\ltsim {<\kern-1.2em\lower1.1ex\hbox{$\sim$}~}   

\title[Metallicity gradients]{\dii{The impact of supernova feedback on metallicity-gradient evolution in cosmological simulations}}

\author[Ibrahim \& Kobayashi]{
Dyna Ibrahim$^{1}$\thanks{E-mail: d.ibrahim3@herts.ac.uk}
and Chiaki Kobayashi$^{1}$
\\
$^{1}$Centre for Astrophysics Research, Department of Physics, Astronomy and Mathematics University of Hertfordshire, College Lane, Hatfield AL10 9AB, UK}

\date{Received 19th January 2025}

\pubyear{2025}

\begin{document}
\label{firstpage}
\pagerange{\pageref{firstpage}--\pageref{lastpage}}
\maketitle

\begin{abstract}
Tracing the cosmic path of galaxies requires an understanding of
their chemical enrichment and merging histories. One of the most important constraints is the internal structure of galaxies, notably the internal distribution of elements. 
Using our cosmological chemodynamical simulations, including all relevant physical processes and the latest nucleosynthesis yields, we investigate the evolution of radial metallicity gradients of stellar populations and the interstellar medium within each galaxy. This work explores the role of supernova feedback on the metallicity gradients by comparing three feedback models, ejecting energy in thermal, stochastic and mechanical forms. At $z=0$, the mechanical feedback model produces the gradient--mass relations of stars and gas both in {excellent agreement} with observations; gradients are the steepest at intermediate-mass ($M_*\sim10^{10}M_\odot$) and flatten in massive galaxies, probably by major mergers. For each model, we predict 
similar gradient--mass relations up to $z=4$ and find that the mechanical feedback model gives flatter gradients of both stars and gas for lower-mass galaxies ($M_*<10^{10}M_\odot$) possibly due to suppressed star formation and metal ejection by stellar feedback. With all feedback models, most galaxies have negative gas-phase metallicity gradients up to $z=5$, suggesting an inside-out growth, which is consistent with other cosmological simulations but not with recent observations at $z\sim1$--2.5. We find a mild redshift evolution of gradients up to $z=4$, 
with a transition at $z=5$, where gradients steepen for both stars and gas. 
\diii{These should be investigated with higher-resolution simulations and observations.}

\end{abstract}



\section{Introduction}
The chemical composition of galaxies provides key insights into galaxy formation and evolution and has been intensively studied with large-scale galaxy surveys. 
For example, the correlation between galaxy stellar mass and metallicity (MZR), which we focused on in our previous paper \citep{Ibrahim_Kobayashi_2024}, is important to explain global galaxy properties over cosmic time. 
Moreover, studying the spatial distribution of metals inside galaxies gives more constraints on 
the chemical enrichment and star formation histories in the galaxies \di{(\cite{Maiolino_Mannucci2019} for a review)}. 
Namely, radial metallicity gradients can constrain the growth of disc galaxies (e.g. \citealt{Larson1976, Kobayashi_Nakasato_2011, Pilkington2012, Vincenzo2020}), and the merging histories of the early-type galaxies (e.g. \citealt{White1978, Kobayashi2004, Hopkins2009}).

Present-day metallicity gradients in galaxies seem to depend on various properties 
such as galaxy stellar mass, which was studied with slit observations of a limited sample of early-type galaxies (ETGs) (e.g. \citealt{Davies1993}; \citealt{Kobayashi1999}; \citealt{Spolaor_Kobayashi_2010}),
the merging histories of galaxies \citep{Rich2012},  
galaxy type \citep{Kuntschner_2010,Goddard_2017}, and environment (\citealt{Zheng2017}).

For further studies of the internal structure of galaxies, large-scale observational surveys have been performed with integral field units (IFU) such as SAURON (e.g. \citealt{Kuntschner_2010}), CALIFA (e.g. \di{\citealt{Sanchez_2012, sanchez2014, sanchez_Menguiano2016}}), MaNGA (e.g. \citealt{Bundy_2015}),  SAMI (e.g. \citealt{Ho_2014}) and its successor HECTOR (e.g. \citealt{Bland_Hawthorn2015}),  and MUSE with higher resolution \di{(\citealt{Bacon2010,sanchez_Menguiano2018})}, where metallicity gradients are obtained for thousands of nearby galaxies.


In the nearby universe, the inner regions of most spiral galaxies are more chemically enriched than the outskirts, which produces {\it negative} metallicity gradients (e.g. \citealt{Searle1971, Belfiore2017}). 
The negative gradient is observed for individual H II regions{, planetary nebulae, Cephids, and open clusters} in the Milky Way (e.g., \citealt{Magrini_Stanghellini_2010,Stanghellini_Haywood_2010}), and in nearby galaxies (e.g. \citealt{Zaritsky_Kennicutt_1994, Kewley_2010,Bresolin_2011, Berg_2015}).
One way to explain negative gradients of gas-phase (and also stellar) metallicity is the `inside-out' growth scenario of galaxy formation (\citealt{Larson1974, Matteucci_Francois_1989, Samland_Hensler_1997, Portinari_Chiosi_1999, Prantzos_Boissier_2000};  \citealt{Kobayashi_Nakasato_2011}).
Several physical processes can `invert' this negative gas gradient, making the outskirt equally or more metal-enriched than the centre of the galaxy. The gas-phase gradient can be flattened due to the accretion of metal-enriched gas into the outskirt \citep{Bresolin_2012}, supernova-driven winds \citep{Gibson_Pilkington_2013, Ma_2017}, 
or galactic mergers and interactions \citep{Rich2012}.
Finally, the gas-phase gradients can become positive (inverted) due to the accretion of pristine gas to the galactic centre (e.g., \citealt{Jones2013, Almeida2018}) or strong metal-rich galactic outflows (e.g. \citealt{Tissera_2022}).
Merging histories of galaxies, notably major mergers, can also impact the flattening of the {\it stellar} metallicity gradients. However, gas-rich galaxy mergers can cause central star formation from metal-rich gas, recreating negative gradients of stars. \citet{Kobayashi2004} studied these effects in a cosmological context. 

Therefore, tracing the evolution of the metallicity gradients across cosmic time is crucial to understanding the role of star formation, gas flows, and feedback processes during galaxy evolution. Despite being well observed in the nearby universe, metallicity gradients remain poorly constrained at higher redshifts, giving diverse conclusions (e.g. \citealt{Cresci2010, Queyrel_2012, Yuan2011, Jones2013, Wang_2017, Wang2022, Venturi2024, Ju2024}). Direct measurements of metallicity gradients at high redshifts require the near-infrared IFU (e.g., KMOS on the Very Large Telescope (VLT); \citealt{Sharples_2013, Curti2020_klever}), with surveys targeting hundreds of galaxies at $z \sim$ 1–2 (e.g. \citealt{Stott_2014, Wuyts_2016}). Furthermore, NIRSpec IFU on the James Webb Space Telescope (JWST) can expand this study including higher redshifts $z>6$ (e.g., \citealt{Venturi2024, Ju2024}).

\di{
Metallicity gradients have been extensively studied using hydrodynamical simulations, including 
{\citet{Taylor_2017},} IllustrisTNG \citep{Hemler2021,Garcia_2023}, EAGLE \citep{Tissera_2022}, CIELO \citep{Tapia_Contreras_2025}, FIRE \citep{Ma_2017}, and FOGGIE \citep{Acharyya2025}, in addition to isolated galaxy simulations \dii{\citep{Kobayashi2004,Kobayashi_Nakasato_2011}}, MUGS and MaGICC \citep{Gibson_Pilkington_2013}. 
\dii{These studies consistently find that gradients are typically negative at $z=0$, flatten with increasing stellar mass and are shaped by inside-out growth, stellar feedback, radial mixing, and gas accretion. }
At higher redshifts, however, the picture becomes more varied: some simulations predict flatter gradients such as EAGLE \citep{Tissera_2022} and enhanced MAGICC \citep{Gibson_Pilkington_2013}, while others find \dii{steeper} gradients, e.g., TNG50 \citep{Hemler2021} and MUGS \citep{Gibson_Pilkington_2013}. 
Some studies also report breaks between inner and outer metallicity profiles, linked to structural and dynamical transitions in galaxies \citep{Garcia_2023}. While quantitative differences exist depending on the feedback models and resolution, the qualitative trends across simulations remain consistent. 
Semi-analytical models have also explored the origin and evolution of metallicity gradients in galaxies \citep[e.g.,][]{Lian_Thomas2018, Belfiore2019,Yates2021, Sharda2021, Sharda2024,Stevens_2024}, offering complementary insights on the roles of star formation efficiency, gas inflows, and outflows.
These models are particularly useful for isolating the effects of feedback strength, enrichment timescales, and metal mixing processes such as turbulent diffusion, which can be parameterized to explore a wider range of physical scenarios.
}

To predict and constrain chemical enrichment within galaxies, hydrodynamical simulations are necessary. In \citet{Ibrahim_Kobayashi_2024}, we implemented and compared four models of supernova feedback (thermal, stochastic, kinetic, and mechanical) with our cosmological hydrodynamical simulations, and concluded that the mechanical feedback model gives the best match with the local stellar and gas-phase MZRs, as well as the observed cosmic star formation rates (SFRs). 
\di{Similar comparative studies of supernova feedback have also been carried out in other cosmological simulations \citep[e.g.][]{Gentry2017, Chaikin2022}, highlighting the importance of feedback modeling in reproducing observed galaxy properties.}

\dii{Most previous cosmological simulations have focused on the global mass–metallicity relation, while only a few have examined metallicity gradients in detail, and typically with a single feedback prescription (e.g. \citealt{Taylor_2017, Tissera_2022}). Previous work systematically comparing different supernova feedback schemes and their impact on metallicity gradients remains  limited (e.g. \citealt{Valentini_2017}), despite the central role of feedback in redistributing metals.
Furthermore, uncertainties in nucleosynthesis yields directly affect the predicted enrichment and therefore the gradients themselves. 
In this paper, we address these gaps by employing updated nucleosynthesis yields and by contrasting thermal, stochastic, and mechanical feedback models within otherwise identical chemodynamical simulations. This allows us to isolate the impact of feedback physics on gradients and provide robust predictions for observations at both low and high redshift.} 
We present both stellar and gas-phase metallicity gradients of the galaxies in our simulations, where various types and masses of galaxies are included based on $\Lambda$ cold dark matter ($\Lambda$CDM) cosmology. Therefore, our prediction can be statistically compared with the ongoing and future observational surveys.

This paper is arranged as follows: In Section \ref{sect_method}, we describe our galaxy samples and define the stellar and gas-phase metallicity gradients used for our simulated galaxies. In Section \ref{sect_results}, we present an example of two galaxies at $z=0.7$ and discuss their stellar and gas-phase gradient and kinematics depending on supernova feedback in detail. Then, we analyze the stellar and gas-phase gradients for all galaxies in our simulations at $z=0.7$, depending on galaxy stellar mass. We also show the stellar mass dependence of metallicity gradients at $z=0$ with our best feedback model. In Section \ref{sect_redshift_evo}, we present the evolution of the metallicity gradients up to $z=5$. In Section \ref{gal_type}, we study the evolution of the gas-phase metallicity gradient depending on galaxy type. Finally, our conclusions are given in Section \ref{sect_conclusions}.

\section{Methods}\label{sect_method} 
\subsection{Our Model}
We perform chemodynamical simulations with our own code based on the GAlaxies with Dark matter and Gas intEracT 3 (\texttt{GADGET-3}) code \citep{Springel2005} including various baryon physics as in our previous work \citep{Ibrahim_Kobayashi_2024}.

In this paper, the simulations are run with the same initial conditions and resolution but in a larger volume, commoving $25 h^{-1}$Mpc cubic box, with periodic boundary conditions. The number of gas and dark matter particles is $N_\mathrm{gas}$= $N_{\mathrm{DM}}$= $320^3$, with mass $M_\mathrm{gas}$=$9.34 \times 10^{6} h^{-1}$ M$_\odot$ and $M_{\mathrm{DM}}$= $5.1 \times 10^{7} h^{-1}$ M$_\odot$. 
We use the same cosmological parameters as in \citet{Ibrahim_Kobayashi_2024}: $\Lambda$CDM cosmology with $h$ = $0.68$, $\Omega_m$ = $0.31$, $\Omega_\Lambda$ = $0.69$ and $\Omega_b$ = $0.048$ \citep{Plank2018}.
The gravitational softening lengths are $\epsilon_{\rm gas}$ = $0.84375$ $h^{-1}$ kpc and $\epsilon_{\rm DM}$ = $1.6875$ $h^{-1}$ kpc for gas and dark matter/stars, respectively.

\di{In this study, we use the same cosmological simulation framework as described in \cite{Ibrahim_Kobayashi_2024}, with key physical prescriptions summarized here for completeness. \dii{We calculate the SFR} based on local gas conditions, following a dynamical timescale as described in \cite{Kobayashi2004}. We adopt a \cite{Kroupa2008} initial mass function (IMF) that is fixed throughout the simulation and does not vary with galaxy mass or redshift. The initial gas composition assumes primordial abundances, {with 75.3\% hydrogen and 24.7\% helium by mass, as in \cite{Kobayashi_Karakas_Lugaro2020}.}}

Our cosmological simulations include various physical processes relevant to galaxy formation and evolution: metallicity-dependent radiative cooling \citep{Kobayashi2004}, star formation \citep{Kobayashi_2007}, black hole physics \citep{Taylor2014}, and element and energy production from asymptotic giant branch stars, Type Ia, Type II supernovae \citep{Kobayashi2004}. 
\di{Our simulations also include hypernovae (HNe), which are energetic core-collapse supernovae from massive progenitors {(20--50$M_\odot$) often associated with gamma-ray bursts \citep{Kobayashi_2006}}. Each HN event releases more energy and metals compared to a typical SN, 
\dii{(each HN releases: $1, 1, 2, 3 \times 10^{52}$ erg for 20, 25, 30, 40 M$_\odot$, respectively)
}, thereby enhancing the effects of stellar feedback and chemical enrichment, particularly in low-metallicity environments. We adopt the same {metallicity-dependent} HN fraction and explosion energy as in \cite{Kobayashi_Nakasato_2011}, calibrated to reproduce elemental abundance patterns in the Milky Way. These values are fixed in our model and not treated as free parameters in this study. Including HNe improves the realism of the simulations and contributes to the regulation of star formation.}

Supernovae feedback modelling was described in detail in \citet{Ibrahim_Kobayashi_2024}; we summarize the key points below.
To study the impact of supernova feedback on the metallicity gradients, we use three out of the four feedback models studied in our previous work \citep{Ibrahim_Kobayashi_2024} since we found that the kinetic feedback model cannot reproduce the global observations of present-day galaxies. In this paper, we use (1) \dii{a model where pure thermal energy is distributed} to the neighbour gas particles, (2) the stochastic feedback (similar to \citealt{Dalla2012}) distributing thermal energy in a stochastic way to a random number of particles with the probability parameter \footnote{ \dii{
 \(f_{\mathrm{stochastic}}\) is the probability parameter introduced in the stochastic feedback model \citep{Dalla2012}. 
It controls the fraction of neighbouring gas particles that are heated such that each receives an energy increase of \(\Delta e = f_{\mathrm{stochastic}} E_{\mathrm{SN}}/N_{\mathrm{ngb}}\) (with $f_{\rm stochastic}$ > 1). 
Larger values of  (e.g. $f_{\rm stochastic}=50$) correspond to fewer particles heated with a proportionally larger energy jump, ensuring efficient feedback even when gas resolution elements are massive. 
By contrast, in the mechanical feedback model, we introduce a separate parameter, \diii{denoted here as $f_{\rm mechanical}$} \citep{Hopkins2018}, to represent the fraction of supernova energy injected in kinetic form (analogous to the kinetic feedback model where \(0 \leq f \leq 1\)). 
Unlike$f_{\rm stochastic}$, which governs the number of particles heated, $f_{\rm mechanical}$ governs the fraction of SN energy between thermal and kinetic models.
}} \dii{$f_{\rm stochastic}$=50}, and (3) the mechanical feedback \citep{Hopkins2018}, which accounts the physics during the Sedov-Taylor phase of supernova expansion, using the fraction parameter \dii{$f_{\rm mechanical}$=1$\%$}. \di{Note that the Sedov–Taylor phase and superbubble evolution are not explicitly resolved due to the resolution limits inherent in cosmological simulations. Similarly, we do not include a sub-grid turbulent metal diffusion model. These physical limitations are discussed further in Section 5 in the context of their potential impact on metallicity gradients.}
It is important to note that the parameter of each feedback model was chosen to match the observed cosmic SFRs in \citet{Ibrahim_Kobayashi_2024}.

\subsection{Galaxy sample}\label{sect_GalaxySample}  
As in \citet{Ibrahim_Kobayashi_2024}, galaxies are identified using the friends-of-friends (FoF) algorithm based on the code used in \cite{Springel2001}. The total stellar mass, $M_*$, is defined within 20 kpc. From the FoF centres,
the centres of galaxies are re-defined as the centre of mass of star particles. We use these galactic centres also for gas-phase gradients at high redshifts, although it is not always possible in observations \citep[e.g.][]{Venturi2024}. We only use galaxies with more than 100 star particles in the 20 kpc radius.

In the case of ongoing mergers or galaxies with satellite companions, our analysis can confound the system with a single galaxy, which can bias our estimation of the metallicity profiles and gradients. To eliminate these contaminations, we apply the following criteria: 
It is considered a single galaxy only if the total stellar mass within $2R_{\rm e}$ from the galactic centre is larger than 75\% of the total stellar mass of the galaxy, otherwise we ignore this object:
\begin{equation}
    M_{*,2 R_{\rm e}}  > 0.75 M_* .
\end{equation}
$M_{*,2R_{\rm e}}$ is the stellar mass within the {projected} radius $r<2 R_{\rm e}$, where $R_{\rm e}$ is the effective radius, i.e. the radius containing half the total stellar mass. 

As a result, 838, 1119, and 591 galaxies with $M_* \sim 10^{8.5-11}$M$_\odot$ are obtained for our thermal, stochastic, and mechanical feedback models, respectively (more details in Section  \ref{sect_Z_profiles}).
In \S \ref{gal_type}, we discuss the dependence of gradients on the galaxy types defined from the star formation main sequence. 

\di{For all gradient calculations and projected maps, we adopt a fixed vertical extent of $\pm$20kpc along the $z$-axis. This corresponds to selecting a cylindrical region centered on the galaxy, with a 20kpc radius in the x–y plane and height of 40kpc. While this choice is held constant across all redshifts and galaxy masses for consistency, we emphasize that all gas quantities are weighted by SFR, which minimizes the influence of non-star-forming gas at large vertical distances (more detail in Appendix \ref{sect_SFMS}).}

\subsection{Metallicity gradients}\label{sect_metallicity_grad}
In this paper, we investigate the radial metallicity profiles although the metallicity distributions can show more detailed structures depending on galaxy types. For each galaxy, we compute the projected radius $r$ with respect to the galactic centre in the $(x,y,z)$ space. Along $z$-axis, all particles in $\pm20$ kpc are projected on the $x$--$y$ plane.

To compare with observations, the stellar metallicity at a given $r$ is weighted by the rest-frame V-band luminosity $L_{\rm V}$ of star particles, where are located most of the absorption lines used in the observations, such as 
\begin{equation}
    Z_{*_{\rm w}} = \frac{\sum (Z_* \times L_\mathrm{V})}{  \sum L_\mathrm{V}} .
\end{equation}
The gas-phase metallicity at a given $r$ is weighted by the SFR of gas particles, as observed with emission lines, such as
\begin{equation}
    Z_{\rm g_w} = \frac{\sum (Z_{\rm g} \times \mathrm{SFR})}{  \sum \mathrm{SFR}} .
\end{equation}
We then produce the metallicity profiles of the gas-phase metallicities against linear $r$ and the stellar metallicity profile against $\log (r/R_{\rm e})$. 
When we estimate the overall metallicity profile for all the galaxies in our simulations at a given redshift $z$ (Sections \ref{sect_Z_profiles} and \ref{sect_gas_profiles}), we first produce the profile for each individual galaxy, bin along $\log (r/R_{\rm e})$ for stars and $r$ for gas, and then calculate the median value of the metallicity in each bin.

Using the median radial profiles obtained, we measure the slope of the profile, which equals the metallicity gradient. 
Before the weighting, the stellar ($\alpha_{*,{\rm nw}}$) and gas-phase ($\alpha_{\rm g,nw}$) metallicity gradient are given by: 
\begin{equation}
    \alpha_{*,{\rm nw}} = \frac{\Delta \log Z_*}{\Delta \log (r/R_{\rm e})}
    ~{\rm [dex~dex^{-1}]}
    \label{grad_str_nonW}
\end{equation}
and
\begin{equation}
    \alpha_{\rm g,nw} = \frac{\Delta \log Z_{\rm g}}{\Delta r} ~~[{\rm dex~kpc}^{-1}] .
    \label{grad_gas_W}
\end{equation}
Replacing the (mass-weighted) metallicities by the weighted metallicities in equations \ref{grad_str_nonW} and \ref{grad_gas_W}, we obtain:
\begin{equation}
    \alpha_{*} = \frac{ \Delta \log (\sum(Z_* \times L_\mathrm{V}) / \sum L_\mathrm{V}) }{\Delta \log (r/R_{\rm e})} 
    ~{\rm [dex~dex^{-1}]}
\label{grad_str_Weighted}
\end{equation}
and
\begin{equation}
    \alpha_{\rm g} = \frac{\Delta \log (\sum(Z_{\rm g} \times \mathrm{SFR_g}) / \sum \mathrm{SFR_g}) }{\Delta r} 
    ~{\rm [dex~kpc^{-1}]} .
    \label{eq_grad_g_w}
\end{equation}

Finally, to calculate the gradient values, the inner and outer boundaries are applied for stars and gas separately (Section \ref{sect_met_maps} for more details).
When we compare to observations \di{at $z=0$} (Section \ref{sect_z0}) with a different definition of gradients in [dex $R_{\rm e}^{-1}$], we divide our gradients by $R_{\rm e}$ [kpc] of individual galaxies to make them in [dex $R_{\rm e}^{-1}$].  
\dii{For stars, this comparison is less straightforward. However, because the stellar metallicity profile flattens at $r<0.1R_{\rm e}$, the metallicity difference from $r\sim0$ to $r=R_{\rm e}$ is well approximated by the stellar gradient $\alpha_*$ [dex $R_{\rm e}^{-1}$]}. 

\section{Results}\label{sect_results} 

\subsection{Metallicity Maps}\label{sect_met_maps}
\begin{figure*}
    \includegraphics[width=0.33\textwidth]{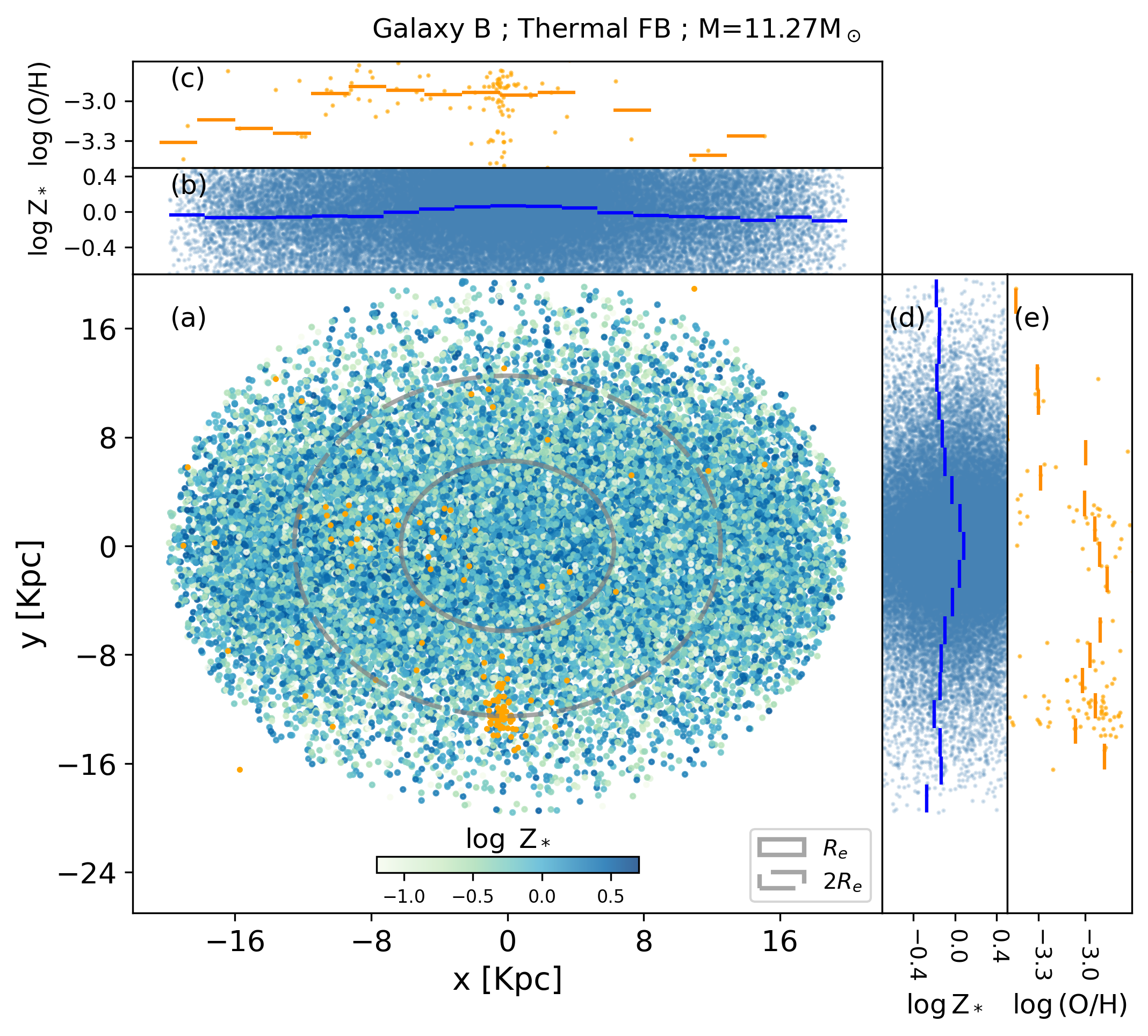}
    \hfill
    \includegraphics[width=0.33\textwidth]{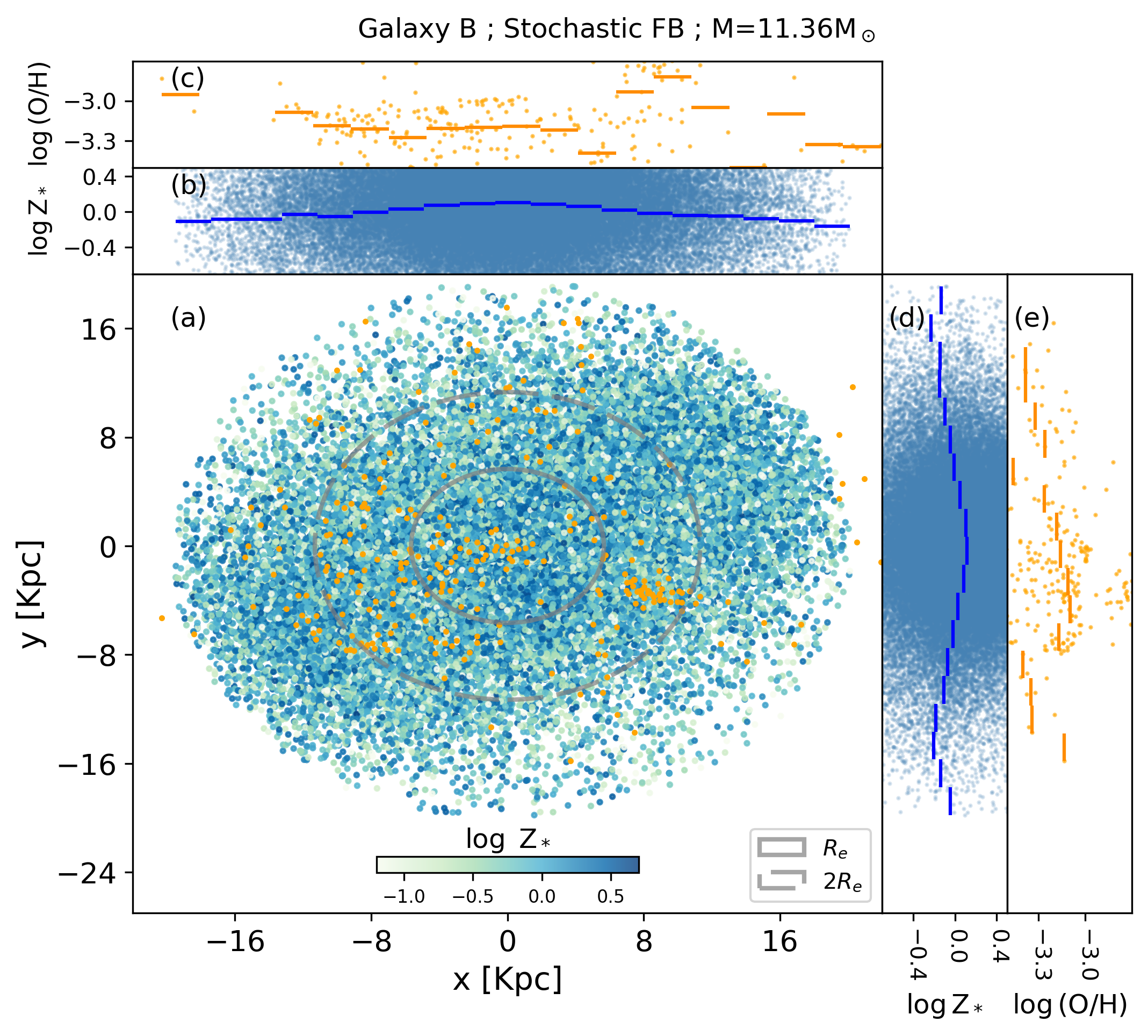}
    \hfill
    \includegraphics[width=0.33\textwidth]{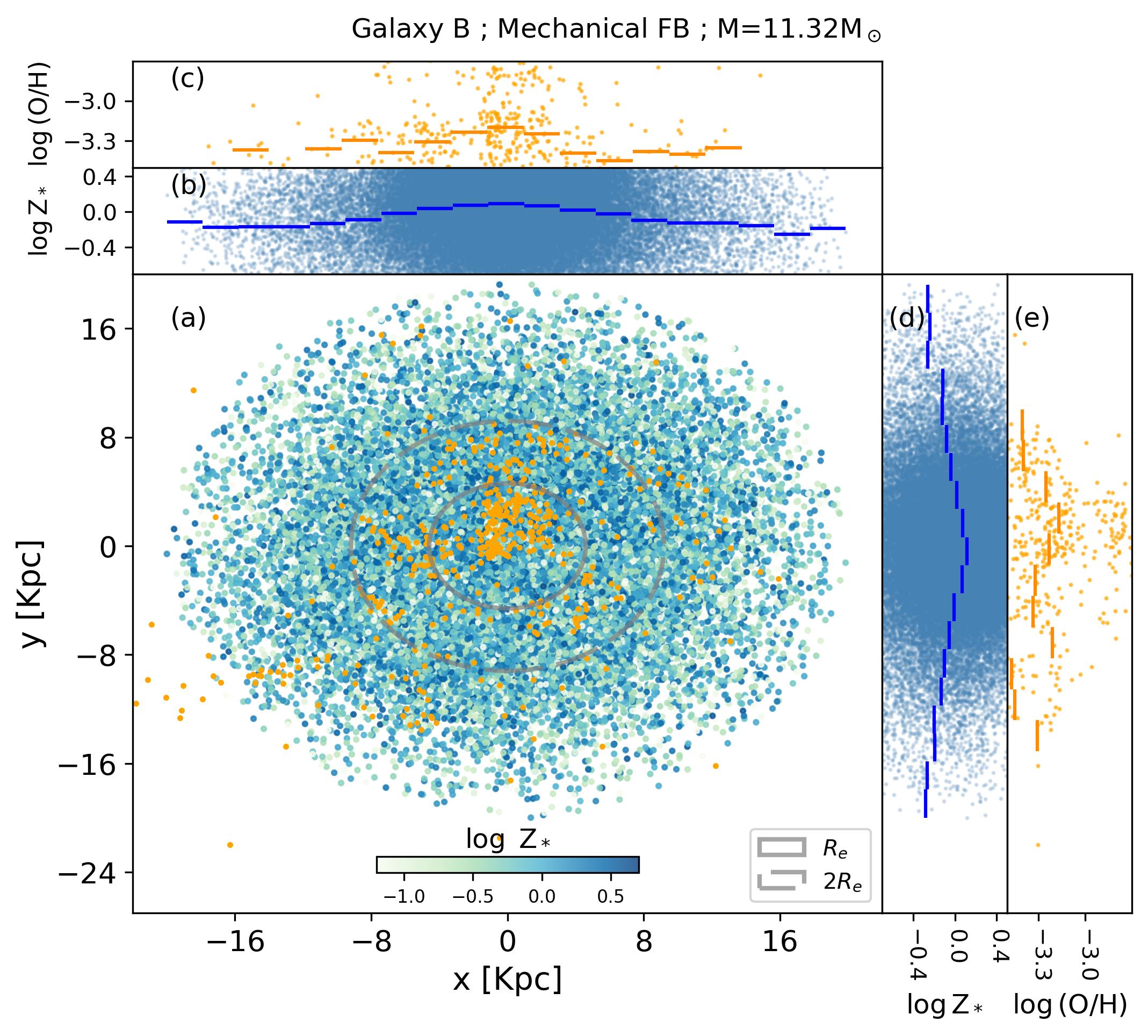}
    \caption{Stellar (blue) and gas (orange) distribution for the same massive galaxy A ($M_*\sim 10^{11}M_\odot$) with the thermal, stochastic and mechanical feedback models (1st, 2nd and 3rd panels, respectively). The grey solid and dashed circles represent $1R_{\rm e}$ and $2R_{\rm e}$, respectively. Panels (b) and (c) show the stellar and gas-phase metallicity distributions, respectively, along the $x$ axis. The solid blue and orange lines are the median metallicity in each bin of $x$ (see main text for details) for stellar and gas-phase metallicity, respectively. Panels (d) and (e) are the same as panels (b) and (c) but along the $y$ axis.}\label{map_highMass}
\end{figure*}
\begin{figure*}
    \includegraphics[width=0.33\textwidth]{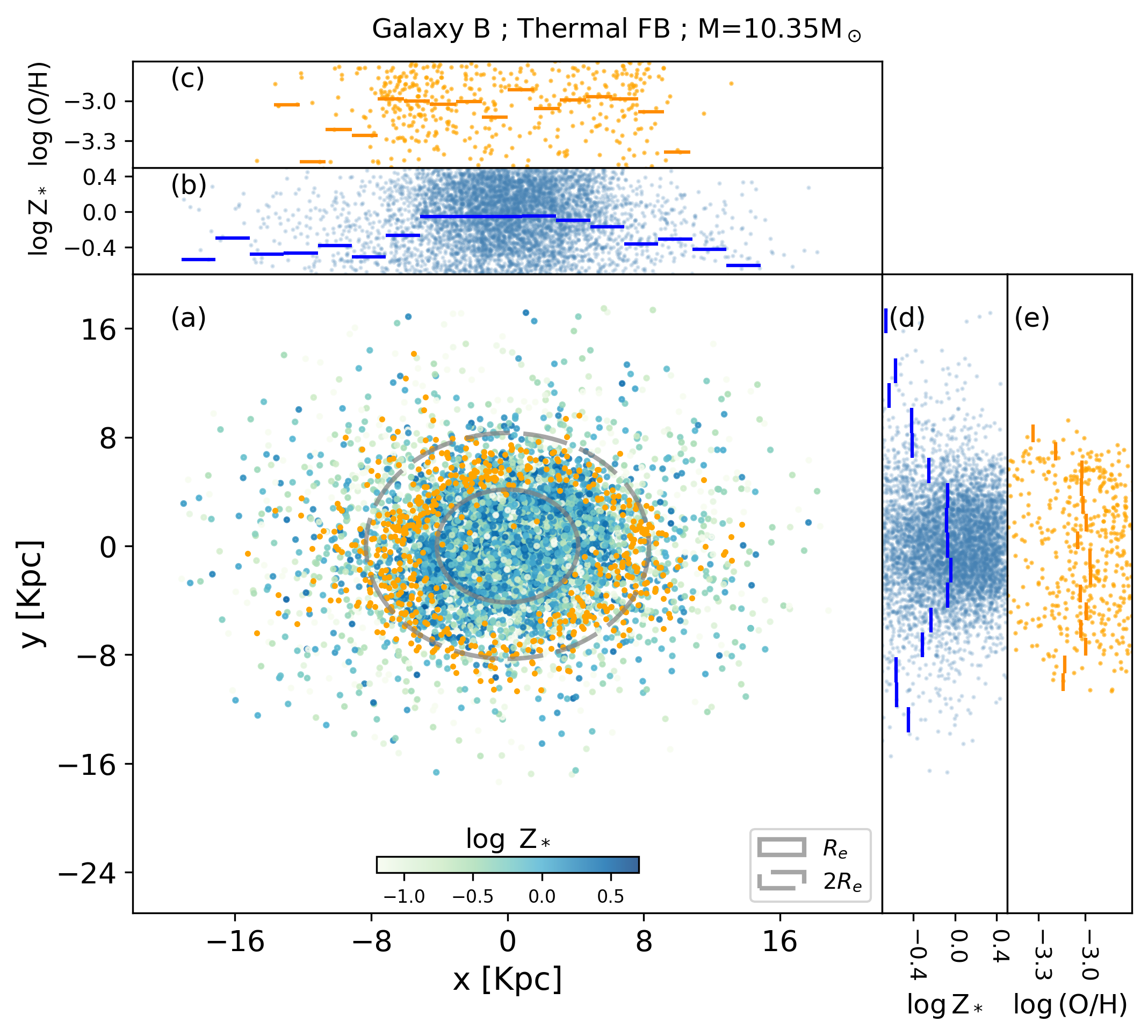}
    \hfill
    \includegraphics[width=0.33\textwidth]{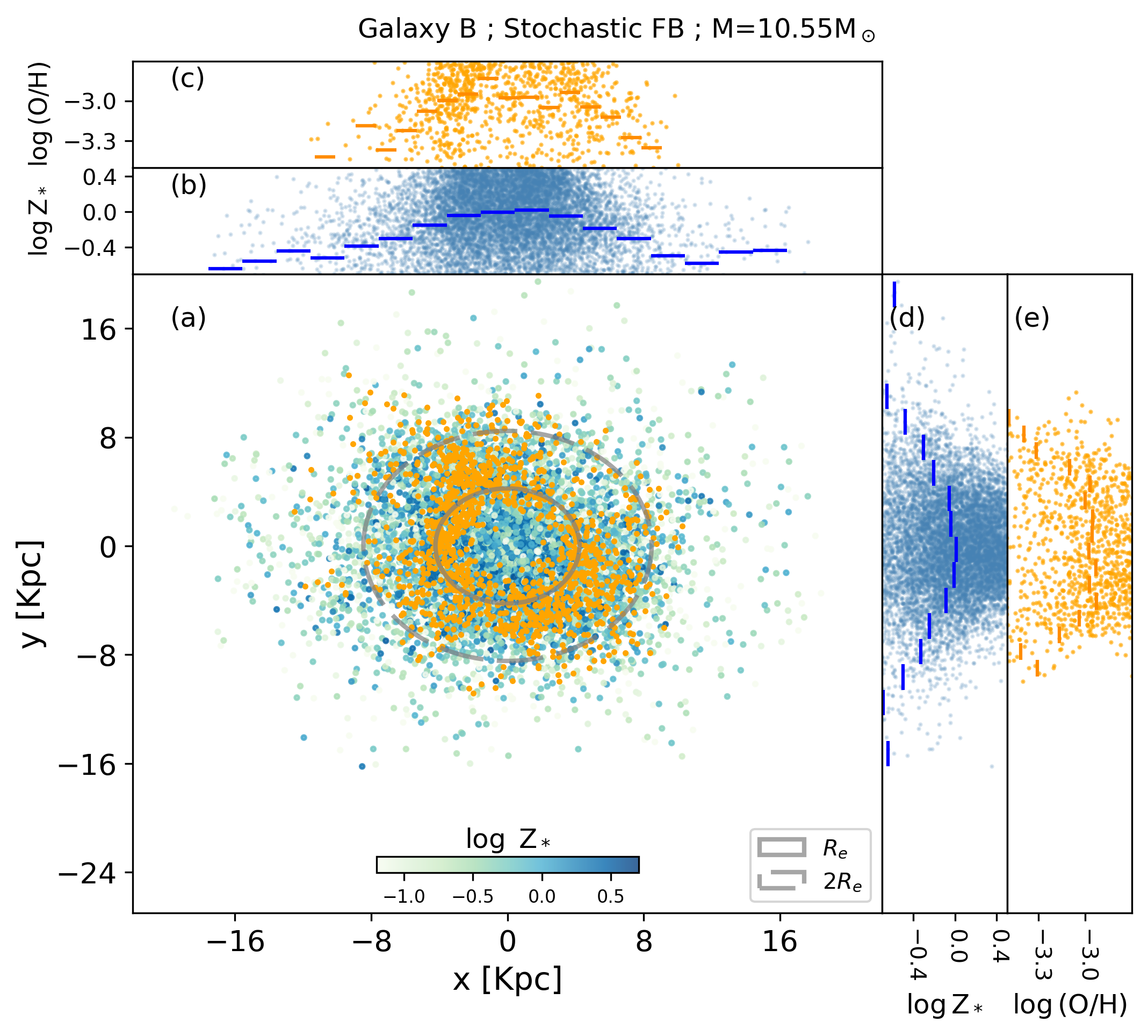}
    \hfill
    \includegraphics[width=0.33\textwidth]{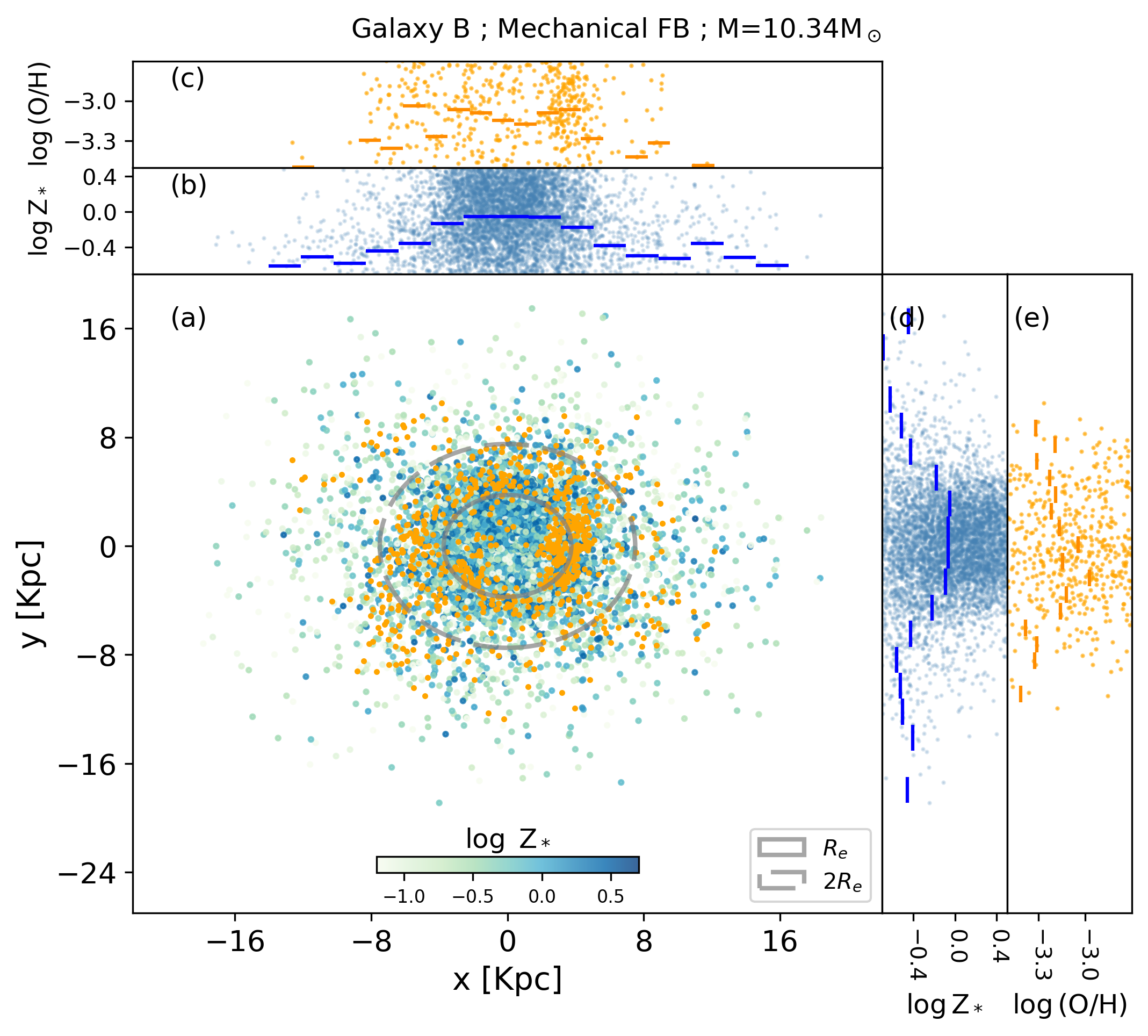}
    \caption{Same as Fig. \ref{map_highMass} but for an intermediate-mass galaxy B ($M_*\sim 10^{10}M_\odot$).}\label{map_lowMass}
\end{figure*}

\begin{figure}
    \includegraphics[width=0.48\textwidth]{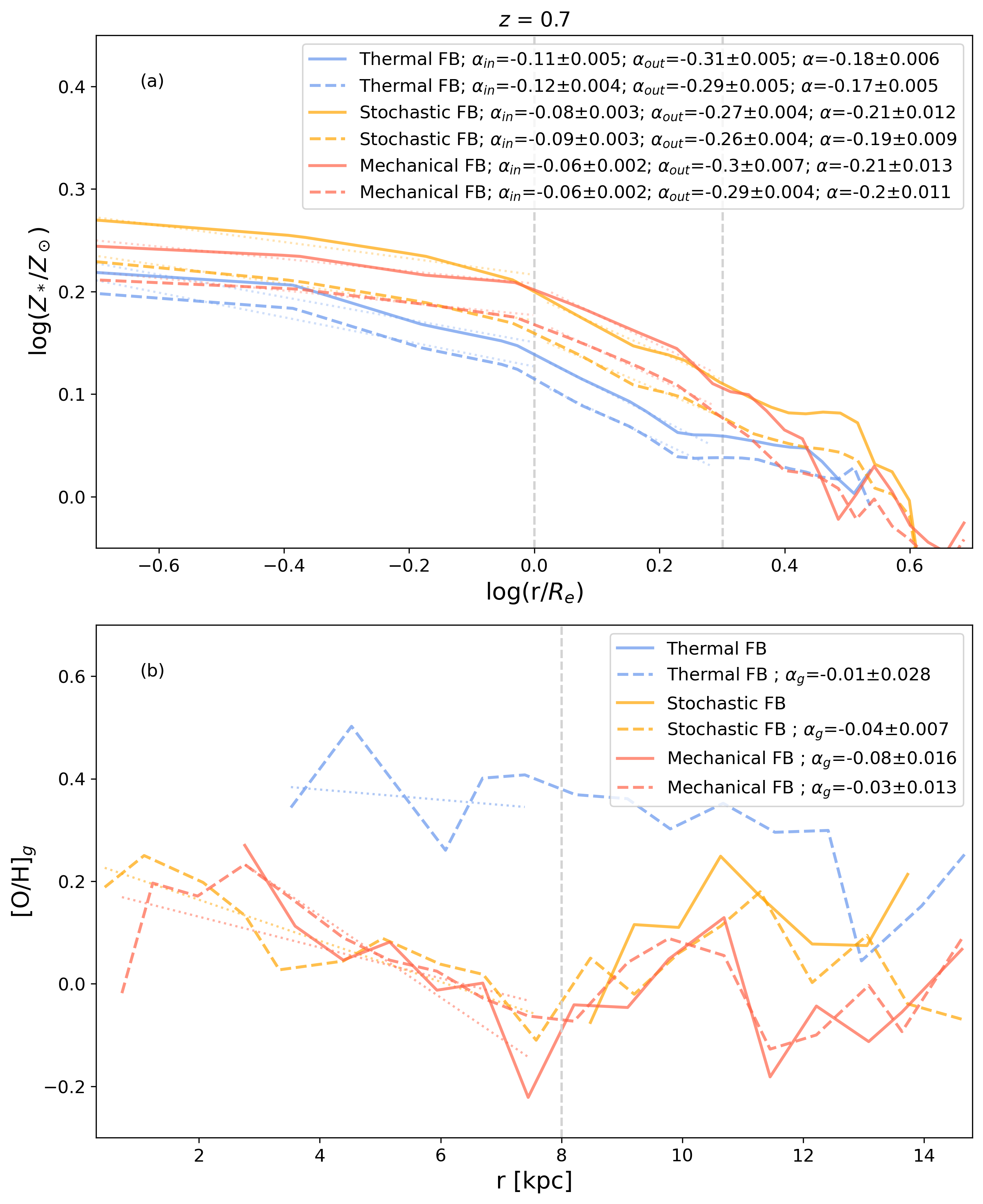}
\caption{(a) V-band luminosity-weighted (solid lines) and mass-weighted (dashed lines) stellar metallicity profiles for Galaxy A at $z=0.7$ for the thermal (blue), stochastic (orange) and mechanical (red) feedback models. We show the profiles with measurable gradients along the total projected radius $\alpha$, the inner gradient $\alpha_{*,\rm in}$ within $R_{\rm e}={4.59}$ kpc, and the outer gradient $\alpha_{*, \rm out}$ between $R_{\rm e}$ and 2$R_{\rm e}$ (vertical dashed grey lines).
(b) SFR-weighted (solid lines) and mass-weighted (dashed lines) gas-phase metallicity profiles for Galaxy A at $z=0.7$ with measurable gradients $\alpha_{\rm g, in}$ in 8 kpc (vertical dashed grey lines). The dotted lines show the best linear regression fits.}\label{Z_R_galA}
\end{figure}

\begin{figure}
    \includegraphics[width=0.48\textwidth]{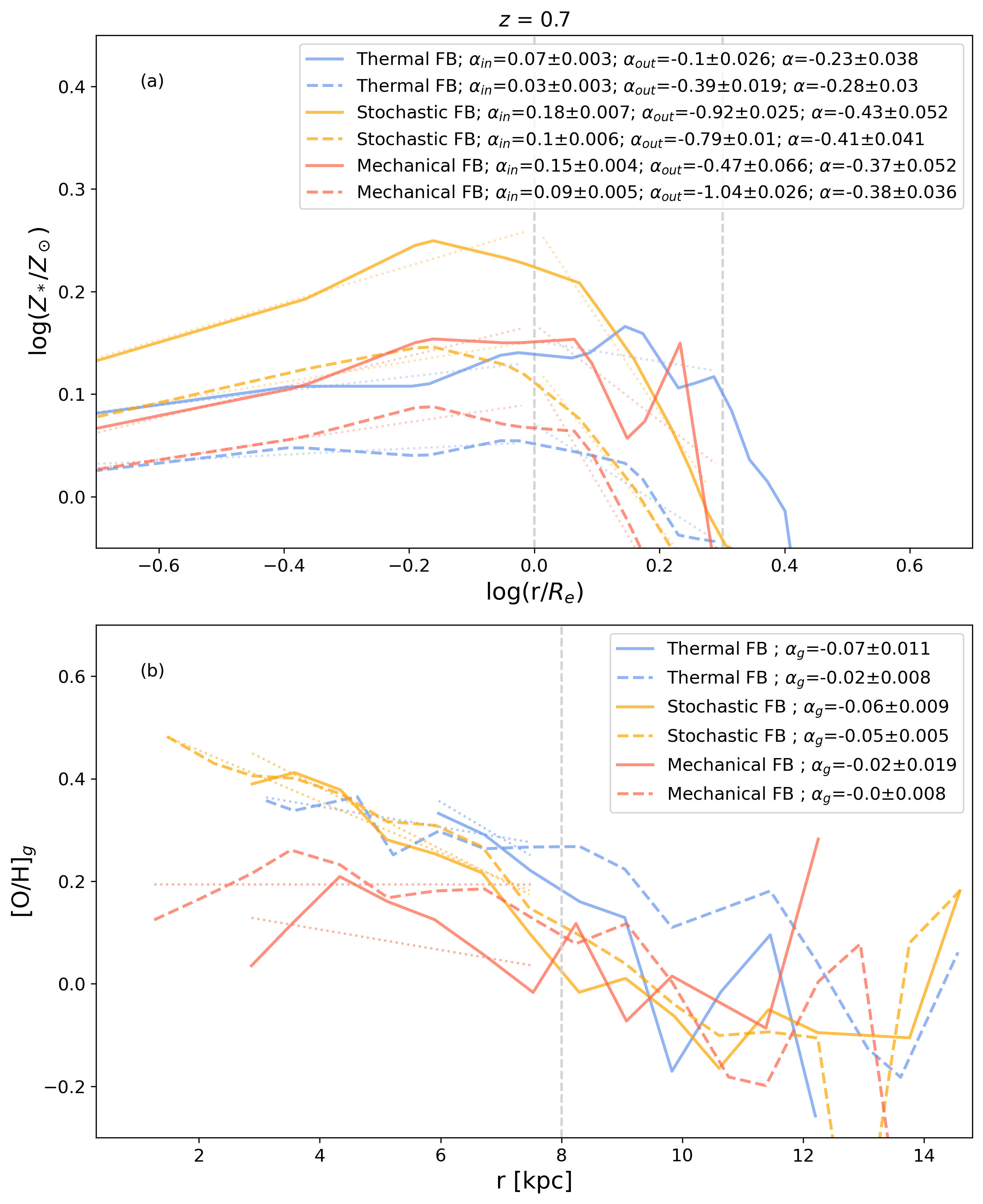}
\caption{Same as Figure \ref{Z_R_galA}, but for Galaxy B with $R_{\rm e}={3.74}$ kpc.}\label{Z_R_GalB}
\end{figure}

\subsubsection*{Galaxy A}
Panels (a) of Figure \ref{map_highMass} show the stellar (blue) and gas-phase (orange) distributions for an example massive galaxy (Galaxy A) with $M_*\sim 10^{11}$M$_\odot$ in our cosmological simulations at $z=0.7$, with the same initial conditions for the thermal (left panels), stochastic (middle panels), and mechanical (right panels) feedback models. The individual stellar particles on the map are colour-coded with the stellar metallicity $\log Z_{*}$.
The grey solid and dashed circles on each map represent $1 R_{\rm e}$ and $2 R_{\rm e}$, respectively.
The upper histograms on each map represent the radial metallicity profiles along the $x$-axis for stellar (blue) and gas-phase (orange) metallicities in panels (b) and (c), respectively. The right-side histograms (d) and (e) represent the radial metallicity profiles along the $y$-axis for stars and gas, respectively. The solid lines in the histograms are the median metallicity in each bin of $x$ or $y$ (20 bins within the total diameter of 40 kpc). 

The stellar distribution in this massive galaxy is similar among the three feedback models. No specific pattern is distinguished from the stellar metallicity map either. However, the gas phase distribution is considerably impacted by the supernova feedback.
The gas particles are non-uniformly distributed in all models, but with the thermal model (left), they are concentrated in one region at the bottom, and with the mechanical model (right), they are more centred around the centre of the stellar component. The stellar and gas-phase metallicity histograms show the presence of a gradient along both the $x$ and $y$ axes. Gas gradients show more azimuthal variations than stellar gradients.

To compare with observations, we use the V-band luminosity-weighted (solid lines) and mass-weighted (dashed lines) stellar metallicity profiles against projected radius $r$, as shown in Figure \ref{Z_R_galA}\textcolor{blue}{(a)}, and calculate the slopes as {defined} in Section \ref{sect_metallicity_grad}. The profiles can be well fitted with a linear line (Eq. \ref{grad_str_Weighted}) at $r<R_{\rm e}$ and $r=1-2R_{\rm e}$ separately, as shown with the faint dotted lines. For all feedback models, the luminosity-weighted and mass-weighted stellar metallicity give {almost} the same inner gradients $\alpha_{*,\rm in}$. The luminosity-weighted gradients are steeper by $\sim$0.01 than the mass-weighted ones for the outer and total gradients ($\alpha_{*,\rm out}$ and  $\alpha_{*}$, respectively) due to slight positive age gradients. 
For luminosity-weighted stellar metallicities, the total gradients  $\alpha_*$ for this massive galaxy are not so much affected by the supernova feedback model. The gradient is only slightly flatter in the thermal (blue) case by $0.03\pm0.013$ and $0.03\pm0.014$ than in the stochastic (orange) and mechanical (red) models, respectively. The outer gradient $\alpha_{*,\rm out}$ is flatter in the stochastic case by $0.03\pm0.008$ and $0.04\pm0.006$ than in the mechanical and thermal cases, respectively. However, the inner gradient $\alpha_{*,\rm in}$ is significantly steeper with the thermal feedback by $0.05\pm0.005$ than in the mechanical case. This means it is important to measure the inner and outer gradients separately to see the impact of stellar feedback on the stellar metallicity gradients. All the stellar gradients are negative independently from the feedback model or weighting, meaning that Galaxy A is more metal-rich in the central part, as already seen in the stellar histograms of Figure \ref{map_highMass}, which indicates an inside-out growth \citep[e.g.,][]{Vincenzo2020}.

The SFR-weighted (solid lines) and mass-weighted (dashed lines) gas-phase oxygen abundance gradients, $\alpha_{\rm g, in}$, are quantified in Figure \ref{Z_R_galA}\textcolor{blue}{(b)}, where the SFR-weighted profiles start at $r>2$ kpc (for mechanical) and $r>8$ kpc (for stochastic) 
due to the SFR-weighting as the central part of the galaxy has very few star-forming gas particles. The gas is not star-forming at all in the thermal case, resulting in the absence of the SFR-weighted metallicity profile. Although some enhancement is seen in the outskirts with the stochastic and mechanical feedback,  the gradients can be fitted with a linear line within $8$ kpc as shown with the dotted line (Eq. \ref{eq_grad_g_w}). 
Considering the fitting error of the thermal model, we do not find a significant difference in the mass-weighted gradients.
Within 8kpc, the SFR-weighted gradient for this galaxy is only available for the mechanical feedback and is $\alpha_{\rm g, in}=-0.08\pm0.016$ dex/kpc, which is steeper than the mass-weighted one by $0.05\pm{0.021}$ dex/kpc.

To understand the impact of stellar feedback on the gas distribution further, we study the kinematics of this galaxy in Appendix \ref{sect_kinematics} using velocity maps. Galaxy A shows no significant rotation for stars and gas. The stellar feedback models do not seem to impact the kinematics so much either for this massive galaxy.

\subsubsection*{Galaxy B}
Figure \ref{map_lowMass} is the same as Figure \ref{map_highMass} but for an example intermediate-mass \footnote{\dii{Regarding the inclusion of a representative low-mass galaxy ($M_*\sim$ $10^{9}$ M$_\odot$), 
we have carefully examined our sample and found that most galaxies in this mass regime contain insufficient cold gas to robustly measure spatial metallicity gradients or produce meaningful maps. Moreover, the resolution limits of our simulations at these mass scales make interpretation of the gas-phase structures more uncertain. For this reason, we have chosen not to include a low-mass galaxies in this paper.}}

galaxy ($M_*\sim$ $10^{10}$ M$_\odot$), that we will refer to as Galaxy B. The stellar distribution is similar for all feedback models, although it is slightly more elongated with the stochastic feedback and more concentrated in the galactic centre with the mechanical feedback.
Supernova feedback has a larger impact on gas-phase distributions for intermediate-mass galaxies {than for massive galaxies}. In this example, the gas particles are pushed away from the centre of the galaxy {mainly} due to AGN feedback, leading to the absence of gas particles in the centre. This ``hole'' feature is most clearly seen in the thermal feedback at $r \sim 6$ kpc, while the feature is less visible with the mechanical feedback.

The stellar and gas-phase metallicity gradients for this galaxy are shown in Figure \ref{Z_R_GalB}\textcolor{blue}{(a)} and \ref{Z_R_GalB}\textcolor{blue}{(b)}, respectively, for the thermal (blue), stochastic (orange), and mechanical (red) feedback. The stellar metallicity profiles change around $1R_{\rm e}$, so we also measure inner and outer gradients. The luminosity-weighted (solid) and mass-weighted (dashed) stellar gradients are similar for the total gradients $\alpha_{*}$. However, the inner gradients $\alpha_{*,\rm in}$ are positive and steeper by $\sim$0.04-0.08 with luminosity-weighted, possibly due to negative age gradients of inside-out quenching. The outer gradients $\alpha_{*,\rm out}$ significantly depend on the weighting, where the luminosity-weighted gradients are steeper by $0.57\pm0.07$ for the mechanical, and flatter by $0.13\pm0.02$ for the stochastic model. 
For luminosity-weighted stellar metallicities, the total gradient $\alpha_{*}$ is steeper for the stochastic feedback by $0.20\pm$0.064 and $0.06\pm 0.07$ than the thermal and mechanical, respectively. The outer gradients $\alpha_{*,\rm out}$ are considerably steeper in the stochastic models ($\alpha_{*,\rm out}=-0.92\pm0.03$) than in the thermal and mechanical feedbacks ($\alpha_{*,\rm out}=-0.10\pm0.03$ and $\alpha_{*,\rm out}=-0.47\pm0.06$, respectively), this may be caused by the limited number of particles at the outskirts {(Fig. \ref{map_lowMass})}. 
With all feedback models, the inner gradients $\alpha_{*,\rm in}$ for Galaxy B is {\it positive} (inverted), meaning the star particles in the galactic centre are more metal-poor. As shown in the following sections, the inverse metallicity gradients are unusual, which might be because this example galaxy was selected from the unique ring structure of the gas.

In Figure \ref{Z_R_GalB}\textcolor{blue}{(b)} we compare the SFR-weighted (solid) and mass-weighted (dashed) gas-phase oxygen abundance gradients, $\alpha_{\rm g, in}$, again within $8$ kpc. The SFR-weighted gradient is steeper than the mass-weighted one by $0.05\pm0.01$ and $0.01\pm0.01$ dex/kpc for the thermal and stochastic feedback models, respectively. The SFR-weighted gas-phase metallicity profile of this galaxy in the thermal case starts at $r>6$ kpc because the gas particles are not star-forming at $r<6$ kpc, which may cause the steepness of the gradient by $\sim$0.05 dex{/kpc} compared the mass-weighted one.

We study the kinematics of Galaxy B in Appendix \ref{sect_kinematics} and find stellar rotation in the $x-y$ plane. The gas phase particles are also rotating but on a perpendicular angle. The rotation seems stronger with the stochastic feedback. 

\subsection{Radial stellar metallicity profiles}\label{sect_Z_profiles}
We extend our analysis of radial metallicity gradients to all galaxies in our simulations. We {find it interesting to show} the metallicity profiles of massive and lower-mass galaxies separately at the threshold stellar mass $M_*\sim 10^{10}$M$_\odot$. The metallicity profiles for lower-mass galaxies ($M_* <10^{10}$ M$_\odot$) in our simulations at $z=0.7$ are shown in Figure \ref{Z_R_M<10} for V-band luminosity-weighted stellar metallicities (the upper panel a) and SFR-weighted gas-phase oxygen abundances (the lower panel b).

For stars, the solid lines are the medians calculated for 20 bins along {the projected radius} $\log(r/R_{\rm e})$ of $631$, $794$, and $434$ lower-mass galaxies respectively for the thermal (blue), stochastic (orange), and mechanical (red) feedback models. The shaded areas are for 1$\sigma$ scatter. The dotted lines show the best linear regression fits of these medians. 
As in Figs. \ref{Z_R_galA}\textcolor{blue}{(a)} and \ref{Z_R_GalB}\textcolor{blue}{(a)}, we fit with a broken power-law to calculate the inner gradient $\alpha_{*,\rm in}$ {between {$0.73$} {kpc} (which corresponds to our spatial resolution limit {at $z=0.7$}) and} $1R_{\rm e}$, and the outer gradient $\alpha_{*,\rm out}$ between $1R_{\rm e}$ and $2R_{\rm e}$. 
We also fit with a single slope $\alpha_{*}$ along the total radius. These slope values are presented on each figure for each model. 

As shown in \citet{Ibrahim_Kobayashi_2024}, the mechanical feedback produces a lower stellar metallicity. Here, we find that all median gradients are negative independent of the feedback models. The single-slope gradient $\alpha_{*}$ for the mechanical feedback model is $\alpha_{*}=-0.45\pm0.041$, which is significantly flatter by $\sim$0.3 than for the other models; the stochastic feedback gives slightly steeper gradient than the thermal feedback. Also, with the broken power-law fit, both the inner and outer gradients are flatter with the mechanical feedback. For all feedback models, the inner gradients are much flatter than the outer gradients.

Figure \ref{Z_R_M>10}\textcolor{blue}{(a)} is the same as Figure \ref{Z_R_M<10}\textcolor{blue}{(a)}, but for massive galaxies ($M_*>10^{10}$ M$_\odot$) in our simulations, which have $207$, $325$, and $157$ massive galaxies for the thermal (blue), stochastic (orange), and mechanical (red) feedback models, respectively. The inner gradients are considerably flatter than the outer gradients. Unlike for lower-mass galaxies, the median stellar metallicity gradient of massive galaxies in our simulations is not highly impacted by the stellar feedback for both the total and outer gradients. Nevertheless, the inner gradient $\alpha_{*,\rm in}=0.05\pm0.03$ is flatter with the thermal model by $0.17\pm0.03$ than in the stochastic model, and by $0.09\pm0.04$ than in the mechanical model.

For all mass ranges, we find that the median stellar gradient in our simulations is always negative, making most galaxies more metal-rich in the central regions. For massive galaxies, the impact of stellar feedback on the gradients is subtle; the thermal feedback results in a flatter inner gradient for the median but not for Galaxy A (\S \ref{sect_met_maps}).
For lower-mass galaxies, however, both the inner and outer gradients are flatter with the mechanical feedback, which can be explained by the suppression of star formation at the centre and the ejection of metals to the outskirts.

\subsection{Radial gas-phase metallicity profiles}\label{sect_gas_profiles}
\begin{figure}
	\includegraphics[width=0.48\textwidth]{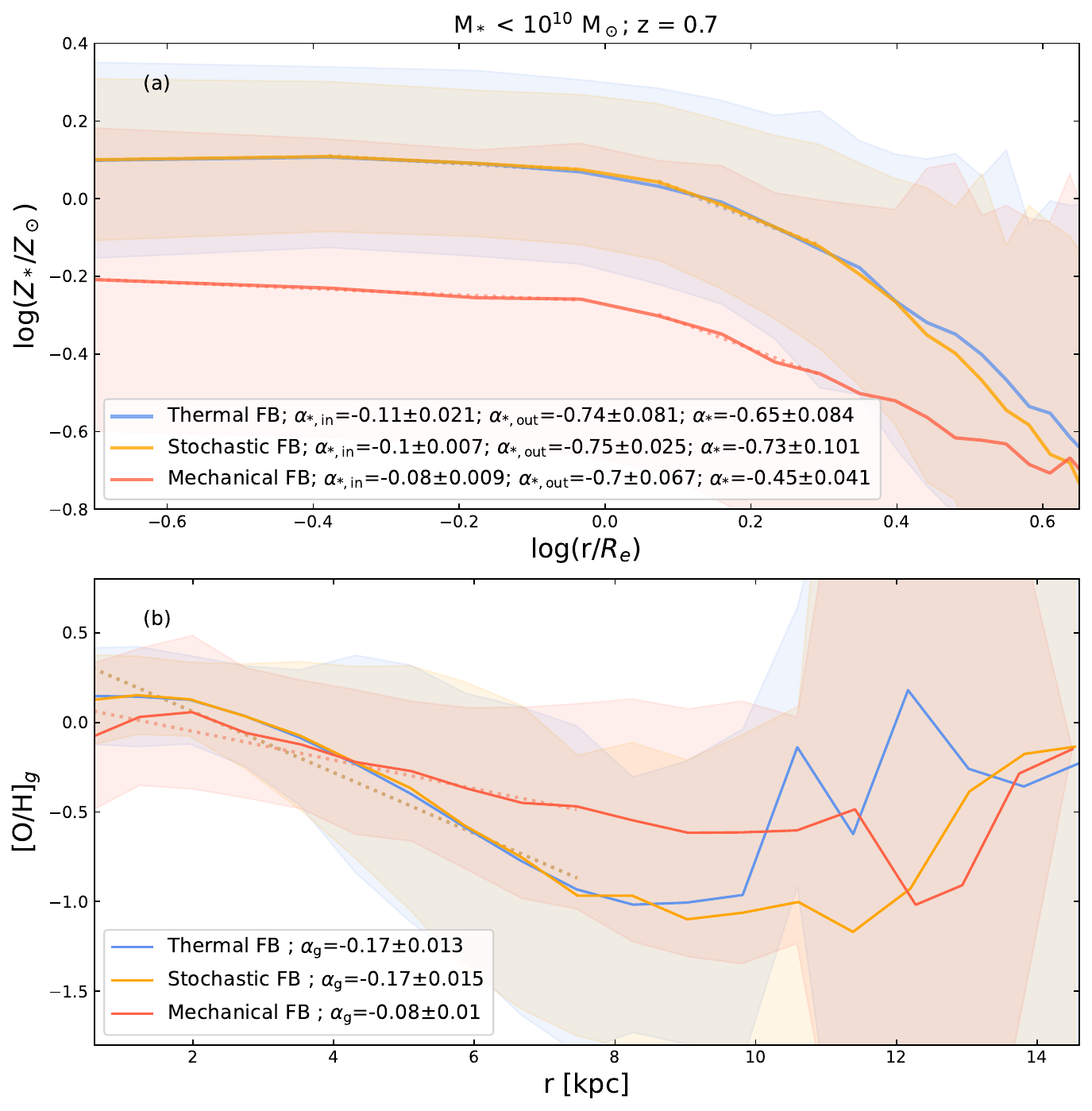}
    \caption{(a) V-band luminosity-weighted stellar metallicity profiles for lower-mass galaxies ($M_*<10^{10} M_\odot$)  with measurable gradients at $z=0.7$ in our simulations. The solid lines are the median for the thermal (blue), stochastic (orange), and mechanical (red) feedback models. The shaded areas are $1 \sigma$ scatter. The dotted lines are the best linear fit of the medians for a single slope $\alpha_{*}$, the inner gradient $\alpha_{*,\rm in}$ within $1R_{\rm e}$, and the outer gradient $\alpha_{*,\rm out}$ between $1R_{\rm e}$ and $2R_{\rm e}$. 
    (b) The same as (a), but for SFR-weighted gas-phase oxygen abundance profiles. \diii{The medians are fitted with a single slope (dotted line), such that the gradients $\alpha_{\rm g,in}$ are measured within $r < 8$ kpc.}
    }\label{Z_R_M<10}
\end{figure}

\begin{figure}    
    \includegraphics[width=0.48\textwidth]{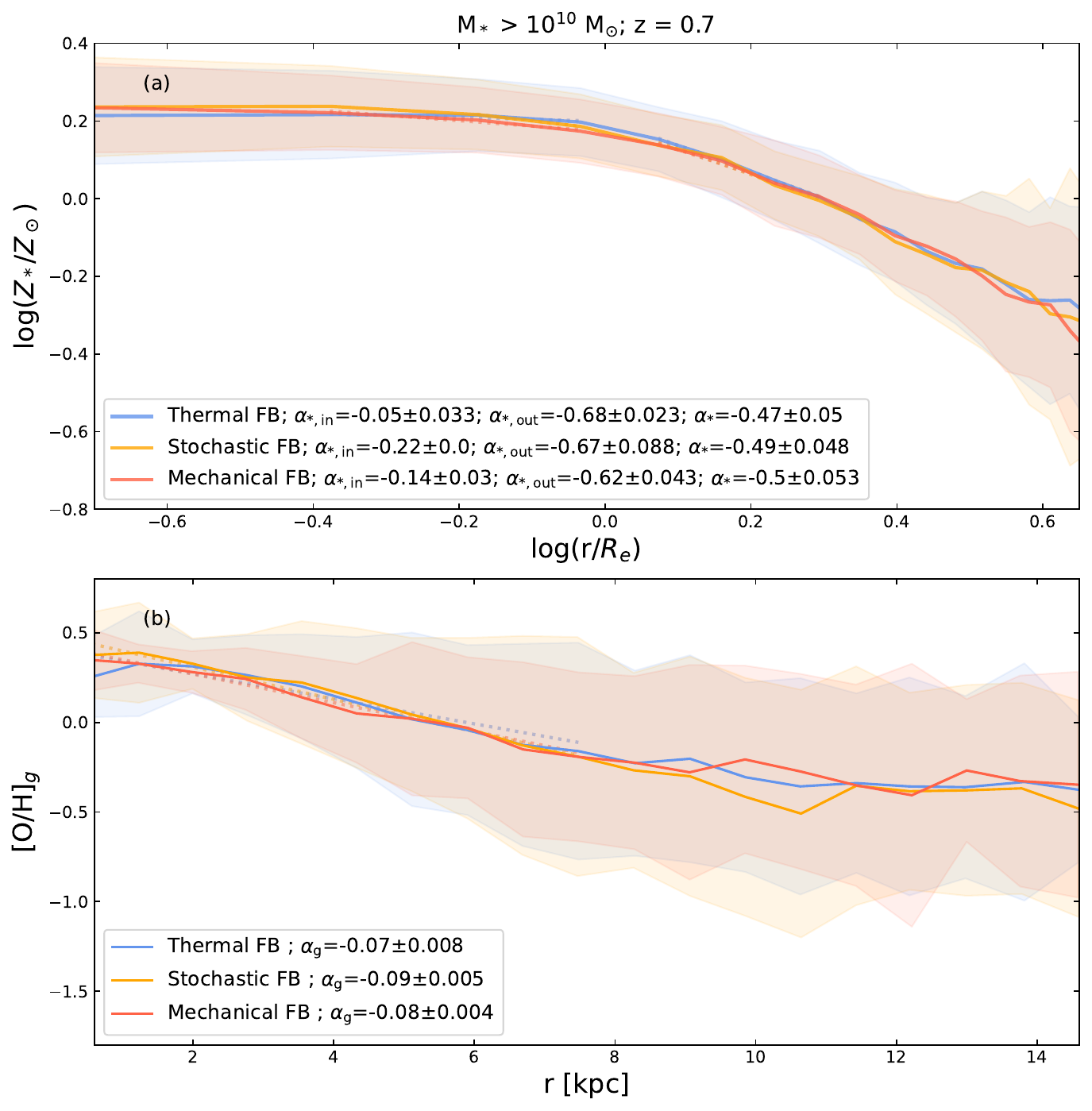}
    \caption{Same as Figure \ref{Z_R_M<10}, but for massive galaxies with $M_*>10^{10} M_\odot$. }\label{Z_R_M>10}
\end{figure}

Figures \ref{Z_R_M<10}\textcolor{blue}{(b)} and \ref{Z_R_M>10}\textcolor{blue}{(b)} show the SFR-weighted gas-phase oxygen abundance profiles for our lower-mass ($M_*<10^{10}$ M$_\odot$) and massive ($M_*>10^{10}$ M$_\odot$) galaxies, respectively. 
The blue, orange, and red solid lines are the best linear fits of medians for galaxies obtained with the thermal, stochastic and mechanical feedback models, respectively. The shaded area is for 1$\sigma$ scatter. As in Figs. \ref{Z_R_galA}\textcolor{blue}{(b)} and \ref{Z_R_GalB}\textcolor{blue}{(b)}, the gas-phase metallicity gradients $\alpha_{\rm g, in}$ are calculated within $8$ kpc,  
which approximately corresponds to $2 R_{\rm e}$ \citep{sanchez2014}.

Figure \ref{Z_R_M<10}\textcolor{blue}{(b)} shows that the stellar feedback significantly impacts the gas-phase metallicity gradients for lower-mass galaxies; although the central gas-phase metallicities are not so different, a clear difference is seen at $r>5$ kpc. The gradient is flatter for the mechanical case by $\sim0.1$ dex/kpc than for the other models. For massive galaxies, however, we do not see a significant difference for gas-phase metallicity gradients, as shown in Figure \ref{Z_R_M>10}\textcolor{blue}{(b)}, which is also the case for stellar gradients in Figure \ref{Z_R_M>10}\textcolor{blue}{(a)}.

For all mass ranges of galaxies, the median gas-phase metallicity gradient is always negative. This could be due to the inflow of metal-rich gas, but for galaxies with low gas density at the centre (\S \ref{sect_met_maps}), it is more likely to be caused by stellar mass-loss (\S \ref{sect_Z_profiles}). For lower-mass galaxies, the median gradient is flatter with the mechanical feedback, possibly due to outflows driving metal-rich gas to the outer regions of the galaxies. As for stellar gradients, the supernova feedback model has no significant impact on the gradient of massive galaxies.

\subsection{Present-day gradients vs mass}\label{sect_z0}
\begin{figure}
    \includegraphics[width=8.7cm]{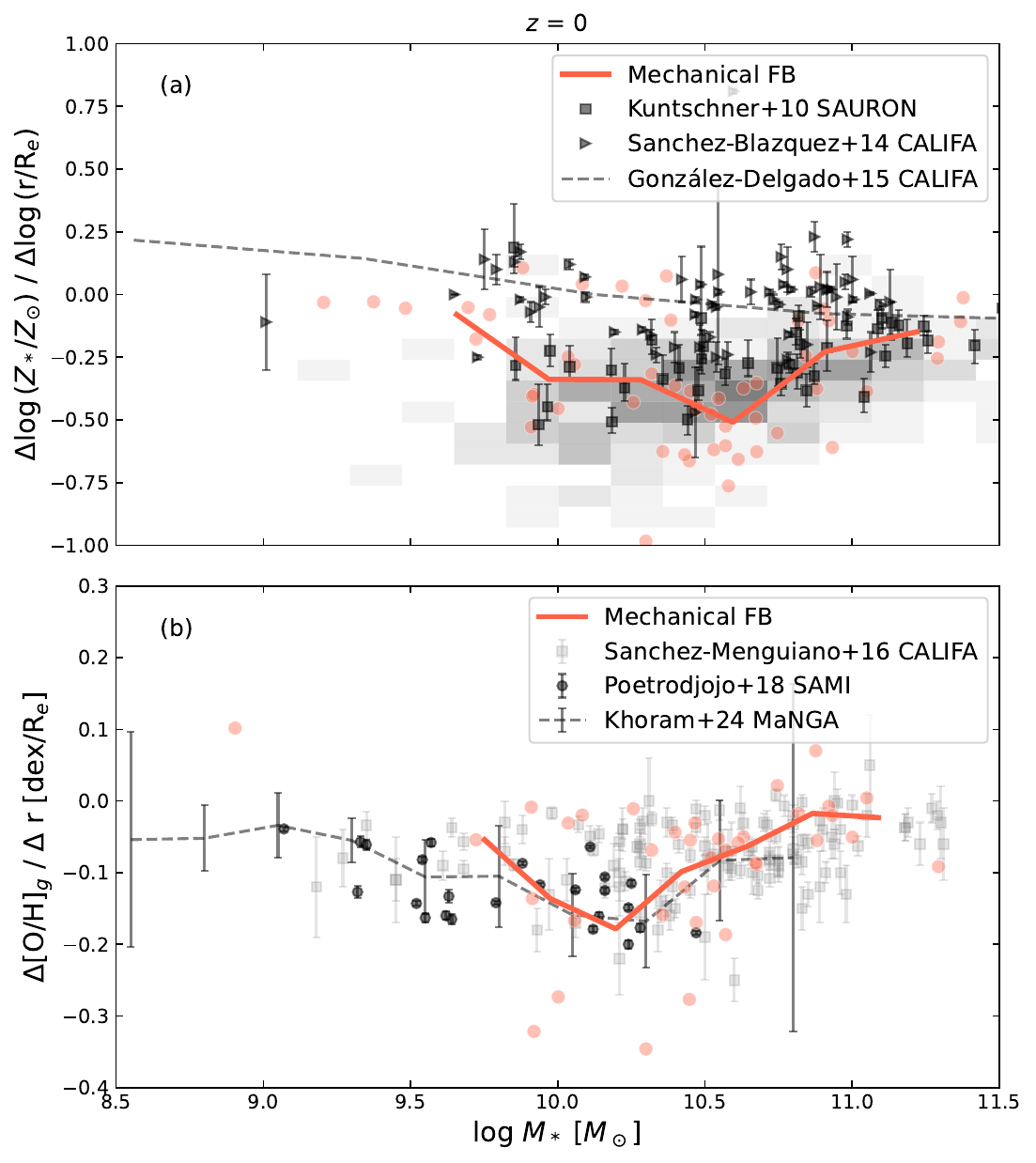}
    \caption{{\it Top panel:} V-band luminosity weighted stellar metallicity gradients \di{measured within 1.5 $R_{\rm e}$} as a function of the galaxy's total stellar mass at $z=0$ in our simulation with the mechanical feedback model (red circles). The red line shows the {median} value {at a given mass}. Observational data are taken from the SAURON survey (\citealt{Kuntschner_2010}, black squares), the CALIFA survey (\citealt{Sanchez_Blazquez_2014}, black triangles; and \citealt{gonzalez_delgado_2015}, black dashed line), and from the ATLAS$^{\rm 3D}$ survey (grey shading, Kuntschner et al., priv. comm.).
    {\it Bottom panel:} Same, but for the SFR-weighted gas-phase oxygen abundance gradients \di{measured within $8$ kpc}. Observational data are taken from \di{\citet[][CALIFA, black square]{sanchez_Menguiano2016}}, 
    \citet[][SAMI, black circle]{Poetrodjojo2018} and \citet[][MaNGA, black line]{Khoram2024}.
    }\label{Grad_M_z0}
\end{figure}

In Figure \ref{Grad_M_z0}, we show the dependence of the metallicity gradients on stellar mass at $z=0$ specifically with the mechanical feedback model, which gave the best matches to the observed MZRs in \citet{Ibrahim_Kobayashi_2024}. The upper panel is for V-band luminosity-weighted stellar metallicity gradients for all galaxies in our simulation. The red circles represent the gradients of individual galaxies, here within \di{
${1.23}$ kpc$<r<{1.5} R_{\rm e}$} 
to compare with the observational data (black symbols and lines, and grey shading).
The red solid line connects the median values at a given stellar mass of our simulated galaxies. The median stellar metallicity gradient is steeper for intermediate-mass galaxies ($M_* \sim 10^{10}$M$_\odot$) by $\sim 0.2$ dex than for massive galaxies, which is in excellent agreement with observational data from the ATLAS3D sample of 253 galaxies (grey shading, Kuntschner et al., priv. comm.), the SAURON sample of 48 early-type galaxies \citep[black squares with errorbars]{Kuntschner_2010}, \dii{and in good agreement with} the CALIFA sample of 62 nearly face-on, spiral galaxies \citep[black triangles with errorbars, \di{measured in $1.5R_{\rm e}$}]{Sanchez_Blazquez_2014} and the CALIFA sample of 300 from spheroids to spiral galaxies \citep[dashed black line, \di{measured in $\sim1R_{\rm e}$}]{gonzalez_delgado_2015}. 

This flattening of gradients toward the massive end is found to be due to major mergers in \cite{Kobayashi2004} and \cite{Taylor_2017}. Then our gradients become flatter again toward the low-mass end, which is possibly due to supernova feedback (Fig. \ref{Z_R_M<10}\textcolor{blue}{(a)}). A similar V-shape trend (against the central velocity dispersion) is found in the observations by \citet{Spolaor_Kobayashi_2010}.

The lower panel is the same but for SFR-weighted gas-phase oxygen abundance gradients within \di{8 kpc}; \di{the fitting range does not change the results due to the SFR weighting (see also Figs.\, \ref{grad_z_allM} and \ref{grad_z_allM_Re}).} To compare with observational data, we convert our gradients such that our units become [dex/$R_{\rm e}$].
Galaxies around $M_* \sim 10^{10}$M$_\odot$ have steepest gradients by $\sim 0.15$ {dex/$R_{\rm e}$}. This is also in excellent agreement with recent observational data from direct method abundances of stacked spectra of 4140 star-forming galaxies from the MaNGA survey (\citealt{Khoram2024}, black line) and from 25 star-forming face-on spiral galaxies from the SAMI survey (\citealt{Poetrodjojo2018}, black points with tiny errorbars).

\di{\cite{Poetrodjojo2018} originally measured metallicity gradients in SAMI galaxies using the R$_{23}$ diagnostic, but later noted that the weak [OII] emission line limited the reliability of those measurements. In \cite{Poetrodjojo2021}, they improved upon this analysis {by using} more robust diagnostics. They concluded that {their} `Rcal' calibration provides the most robust and consistent metallicity gradients and recommended it for comparisons with theoretical predictions. Using this method, they found a clear correlation between stellar mass and metallicity gradient: more massive galaxies tend to have shallower gradients, while lower-mass systems show steeper and more scattered gradients.}
\di{Our gradients seem to have comparable scatter to this observation with Rcal metallicity diagnostics, but larger scatter than the Scal one.}

\di{The steep gradients at intermediate stellar masses are also seen in \citet{Sharda2021}, who used semi-analytic scaling models to reproduce observed trends. Their models {also} predict the steepest gradients at $M_* \sim 10^{10}$–$10^{10.5},M_\odot$, with a flattening at both lower and higher masses. This curvature is interpreted as a transition from advection-dominated to accretion-dominated regimes in galaxy evolution. 
Our simulations indicate that {their} effective yield reduction factor, $\phi$, varies with stellar mass: $\phi = {0.4}$ at $M_* = 10^{10}M_\odot$, $\phi = {0.2}$ at $10^{11}M_\odot$.
Our gradients are quantitatively flatter than those predicted by the two empirical scalings adopted in their models. However, the qualitative shape of the relation remains similar, and our results fall within the range spanned by their model predictions for different $\phi$ values.
}

\section{Redshift Evolution}\label{sect_redshift_evo}

\subsection{Stellar Gradients vs Mass}
\begin{figure*}
	\includegraphics[width=\textwidth]{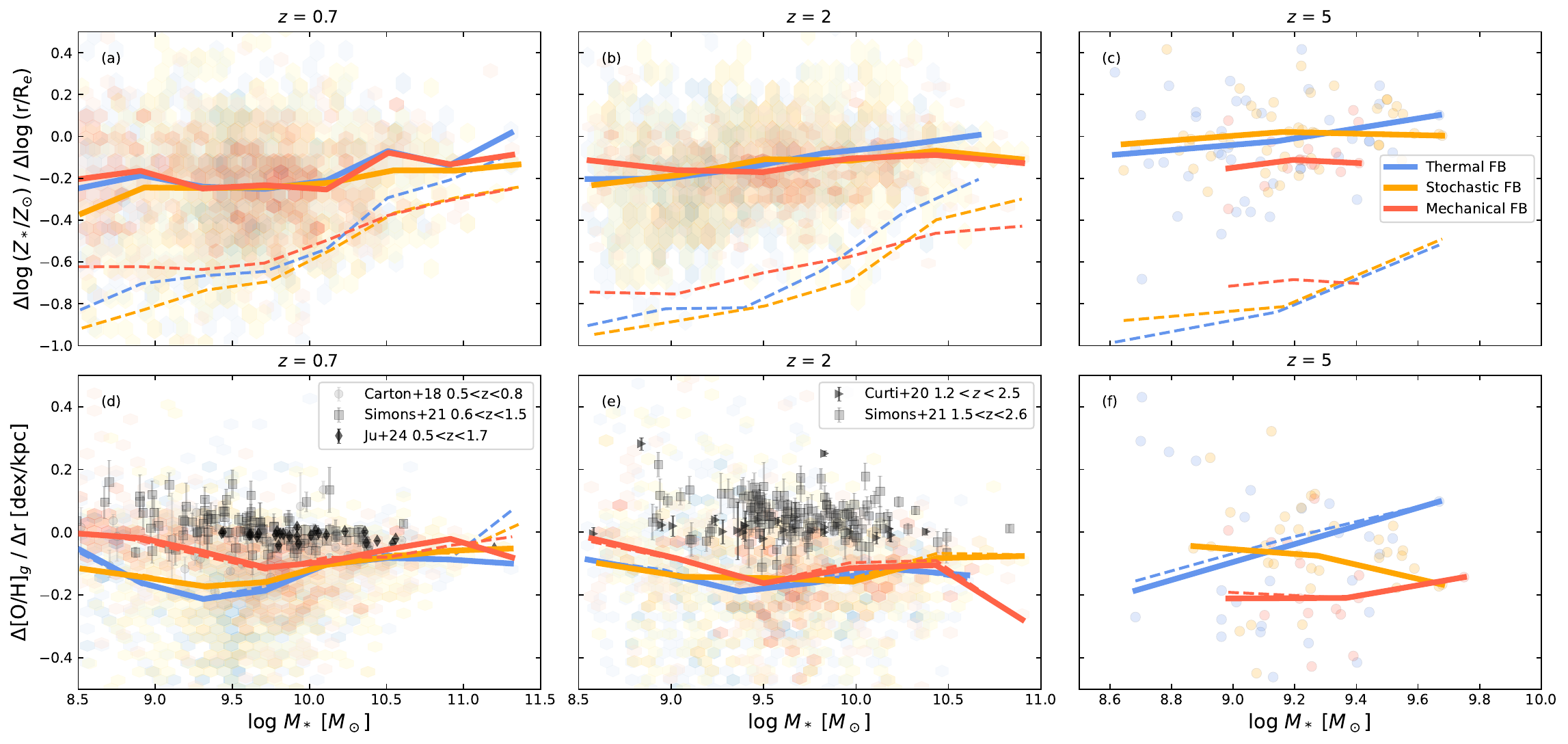}
    \caption{Upper panels: The V-band luminosity weighted stellar metallicity inner gradients $\alpha_{*,\rm in}$ within 1.5 $R_{\rm e}$ as a function of the galaxy total stellar mass at $z=0.7$ (a), $z=2$ (b), and \di{$z=5$} (c) for all galaxies in our simulations with the thermal (blue), stochastic (orange), and mechanical (red) feedback models {(see main text)}. The solid lines are the median at a given mass. The dashed lines are the metallicity gradients $\alpha_{*}$ along the total projected radius. 
    The lower panels (d, e, and f) are the same, but for the SFR-weighted gas-phase oxygen abundance gradients $\alpha_{\rm g}$ within $8$ kpc (solid lines) and total gradients (dashed lines).
    \di{The distributions of simulated galaxies are shown with hexagonally binned density maps at $z=0.7$ and 2, while with scattered points at $z=5$, with the same colour.}
    \di{The black/gray symbols are} observational data from \citet{Carton2018} at $0.5<z<0.8$ (circles),
    \citet{Simons2021} at $0.6<z<1.5$ and  $1.2<z<2.6$ (squares), \citet{Curti2020} at  $1.2<z<2.5$ (triangles), and \citet{Ju2024} at $0.5<z<1.7$ (diamonds). 
    }\label{Grad_M}
\end{figure*}

The top panels in Figure \ref{Grad_M} show the V-band luminosity-weighted stellar metallicity gradients with respect to galaxy total stellar mass at $z=0.7$ (panel a), $z=2$ (b), and $z=5$ (c) in our simulations, with the thermal (blue triangles), stochastic (orange diamonds), and mechanical (red circles) feedback models.
As in Fig. \ref{Grad_M_z0}, these symbols represent the inner gradients $\alpha_{*, \rm in}$ within 1.5 $R_{\rm e}$ for individual galaxies, and the solid lines represent the median metallicity gradients at a given mass in each simulation. As already shown in Figs. \ref{Z_R_M<10}\textcolor{blue}{(a)} and \ref{Z_R_M>10}\textcolor{blue}{(a)}, the medians of inner gradients are much flatter than the medians of the total metallicity gradients ($\alpha_*$, dashed lines) along the total radius.
The distribution of simulated galaxies are also shown. To reduce overcrowding from individual data points at $z=$0.7 and 2 (a and b), \di{we use hexagonally binned density maps in the background with a consistent color scheme. This hexbin visualization highlights the data distribution, especially in regions of high density points.}
\di{At $z=5$ (panel c) , we plot all individual galaxy points (circles) for the thermal (blue), stochastic (orange), and mechanical (red) feedback models.} \di{Some points lie outside the gradient range shown, which explains why, for example, the blue fit at the low-mass end in the bottom panel does not align with the displayed blue points.}

At $z=0.7$, the inner gradients of individual galaxies (points) are clustered between $\alpha_{*, \rm in} \sim -0.5$ and $\sim 0.0$ for all models. At the low-mass end ($M_*<10^{9}$ M$_\odot$), a significant fraction of galaxies with the mechanical ($22\%$) and thermal ($24\%$) and stochastic ($14\%$) feedback show inverted inner gradients ($>0$). At the intermediate mass ($M_* \sim 10^{10}$ M$_\odot$) quite a few individual galaxies with the stochastic feedback have very steep gradients ($\sim -0.9$). At the massive end ($M_*>10^{11}$ M$_\odot$) the gradients are close to $0$ dex (i.e. flat). This flattening of stellar metallicity gradients for massive galaxies was explained by the occurrence of major mergers \citep{Kobayashi2004, Taylor_2017}, which perturb the spatial distribution of stars (see Fig.12 of \citealt{Kobayashi2004}). Hence, we do not expect to see a significant difference among feedback models, as in Figure \ref{Z_R_M>10}\textcolor{blue}{(a)}.
The medians (solid lines) show only a small ($\sim 0.1$) difference among the feedback models depending on the mass. Mechanical feedback gives slightly flatter gradients for lower-mass galaxies and shows no significant difference at higher-mass galaxies compared with the other feedback models. These are consistent with what we find in Section \ref{sect_Z_profiles}. From low-mass to high-mass galaxies, the median inner gradient becomes flatter, with a maximum difference of $\sim 0.2$ in the plotted mass range. On the other hand, the median total gradient $\alpha_{*}$ (dashed lines) becomes much flatter by $\sim 0.4$ and $\sim 0.8$ for the mechanical and thermal feedback models, respectively. This stronger mass dependence should be caused by outer gradients.

At $z=2$ (panel b), the median of inner gradients $\alpha_{*,\rm in}$ (solid lines) is not so much impacted by the stellar feedback models, while at $M_*<10^{9.5}$ M$_\odot$ it is flatter with the mechanical model by $\sim 0.1$, and at $M_*>10^{9.5}$ M$_\odot$ it is flatter with the thermal model by $\sim 0.1$, than the other models. 
The median total gradients $\alpha_{*}$, however, greatly depend on the feedback model at all mass ranges. At $M_*<10^{10}$ M$_\odot$, the mechanical gradient is flatter by $\sim 0.2$ than the stochastic one, while at $M_*>10^{9.7}$ M$_\odot$, the thermal one is flatter by $\sim 0.3$. 

We find a similar gradient--mass relation and feedback dependence at $z=4$. However, we find an interesting transition at $z=5$.
At $z=5$ (panel c),the inner gradient becomes much steeper ($\alpha_{*,\rm in}\sim -0.2$) with the mechanical feedback than for the other models. There is no clear mass dependence of gradients with the stochastic and mechanical models, although there is with the thermal feedback model.

This feedback dependence for the total gradient (including $>1.5R_{\rm e}$) is remarkable, and it would be very useful if stellar gradient evolution could be measured for a wide range of galaxy mass.
For all shown redshifts and with all feedback models, the stellar metallicity total gradient is steeper for lower stellar mass, on average. However, the gradients increase from the low-mass to high-mass end is larger for the stochastic and thermal feedback models and smallest for the mechanical one, which tends to give flatter gradients at $M_*<10^{10}$M$_\odot$. This mass dependence is not as much retrieved for the inner gradients.

\subsection{Gas-phase Gradients vs Mass}
The bottom panels in Figure \ref{Grad_M} are the same as the top panels, but for the SFR-weighted gas-phase oxygen abundance gradients at $z=0.7$, $z=2$, and $z=5$ (in panels (d), (e), and (f), respectively) for the three difference feedback models. The solid lines represent the medians inner gradients $\alpha_{\rm g, in}$ (within $8$ kpc) at a given mass of simulated galaxies (\di{hexbin or} points). The dashed lines are the total gradients along the total radius, which are very similar to the inner gradients except for the massive end. 

At $z=0.7$, the median gradient is significantly flatter with the mechanical feedback for low-mass galaxies ($M_*<10^{10}$ M$_\odot$). It is less impacted by the supernova feedback for massive galaxies ($M_*>10^{10}$ M$_\odot$), as previously discussed in Section \ref{sect_gas_profiles}. By looking at individual galaxies (points) in all feedback models, most galaxies have gas-phase gradients clustered between $\alpha_{\rm g, in} \sim -0.3$ and $\sim 0$ dex/kpc. However, certain galaxies with $M_*\sim10^{9.5}$ M$_\odot$ in the thermal and stochastic models have very steep gradients reaching $\alpha_{\rm g, in} \sim-0.5 $ dex/kpc. Massive galaxies ($M_*>10^{10}$ M$_\odot$) have gradients $\alpha_{\rm g, in} \sim -0.1 $ dex/kpc for all feedback models.
At the low-mass end, the galaxies with the mechanical feedback (red) are clustered around $\alpha_{\rm g, in} = 0$ dex/kpc, again showing flatter gradients in this model. This is likely due to outflows driving metal-rich gas to the outer regions of the galaxies (\S \ref{sect_gas_profiles}). In addition, $46\%$ (mechanical), $22.5\%$ (thermal), and $15.8\%$ (stochastic) of low-mass galaxies with $M_*<10^{9}$ M$_\odot$ show a positive gradient in the three models, meaning they have more metal-rich gas in the outskirt. This requires an additional effect, such as the inflow of pristine gas. The mechanical feedback gradients are closer to observational data from \di{MUSE/VLT \citep[light gray {circles}]{Carton2018} at $0.5\!<\!z\!<\!0.8$}, the CANDELS CLEAR survey with HST \citep[gray squares]{Simons2021} at $0.6\!<\!z\!<\!1.5$, and the MSA-3D survey with JWST/NIRSpec of 26 galaxies at $0.5\!<\!z\!<\!1.7$ \citep[black diamonds]{Ju2024}, \di{
as well as other observational datasets \citep[e.g.,][]{Stott2014, Schreiber2018},  that are not included in the figure to avoid overcrowding.}

At $z=2$ (panel e), the median gradient with the mechanical feedback is slightly steeper than at $z=0.7$. We can also see the feedback dependence of the gradients here. The stochastic and thermal gradients are clustered near $\sim-0.1$ dex/kpc at $M_*<10^{10}$ M$_\odot$, while the mechanical feedback gives slightly flatter gradients. At $M_*>10^{10}$ M$_\odot$, the mechanical feedback gives steeper gradients reaching $\alpha_{\rm g, in} \sim -0.2$ dex/kpc, however, this may be due to the small number of massive galaxies at this redshift. All feedback models give gradients steeper by $\sim 0.1$ dex/kpc compared to observational data from {the KLEVER survey with KMOS} at $1.2\!<\!z\!<\!2.5$ \citep[{black triangles}]{Curti2020}  and from {the CLEAR survey} at $1.5\!<\!z\!<\!2.6$ \citep[{gray squares}]{Simons2021}.

Up to $z=4$, specifically with the mechanical feedback, the gas-phase oxygen abundance gradients tend to become flat at the low-mass end, which is consistent with the observational data. 
At $10^9M_\odot < M_* < 10^{10}M_\odot$, the predicted gradients are steeper than observed, which might indicate that metal outflow, pristine gas inflow, and/or gas mixing is {still} inefficient in our simulations. 

As for stellar gradients, the evolutionary transition is seen not at $z=4$ but at $z=5$.
At $z=5$ (panel f), the mechanical model gives much steeper median gradients ($\alpha_{\rm g,in} \sim -0.2$ dex/kpc) with no clear mass dependence. 
In our future work, we will extend this study for $z>5$, to have a more complete sample of gas-phase oxygen abundance gradients in our hydrodynamical simulations at high redshifts.

\subsection{Time evolution of metallicity gradient}
\begin{figure}
    \includegraphics[height=5cm]{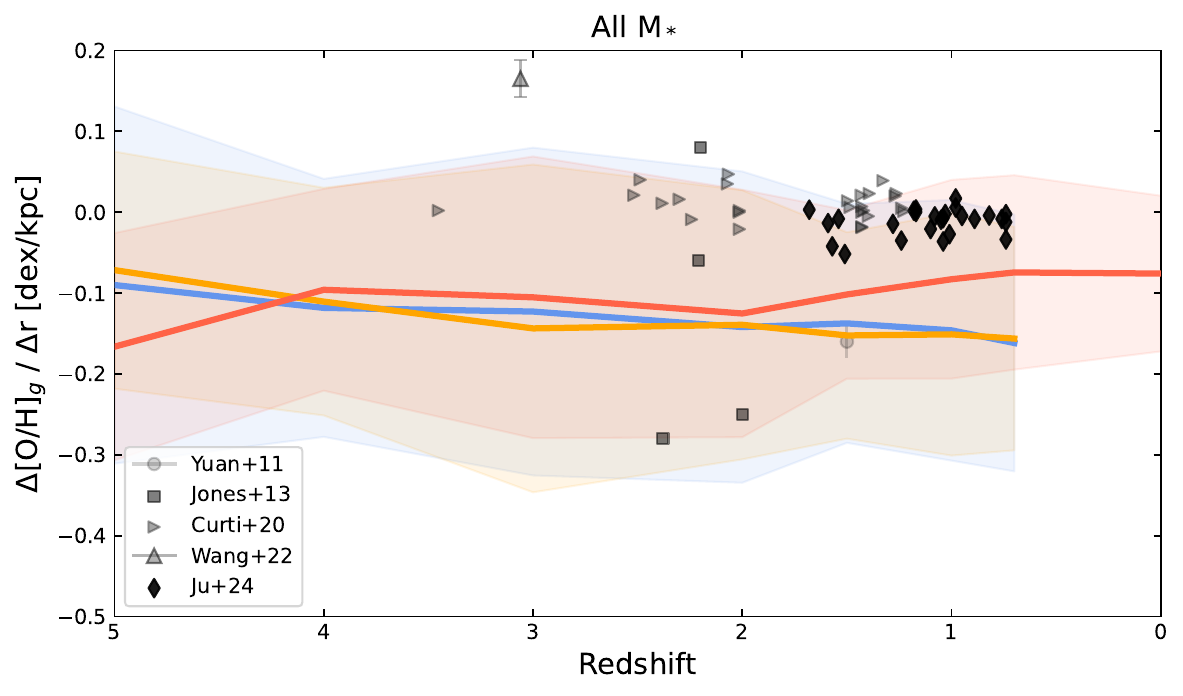}
    \includegraphics[height=5cm]{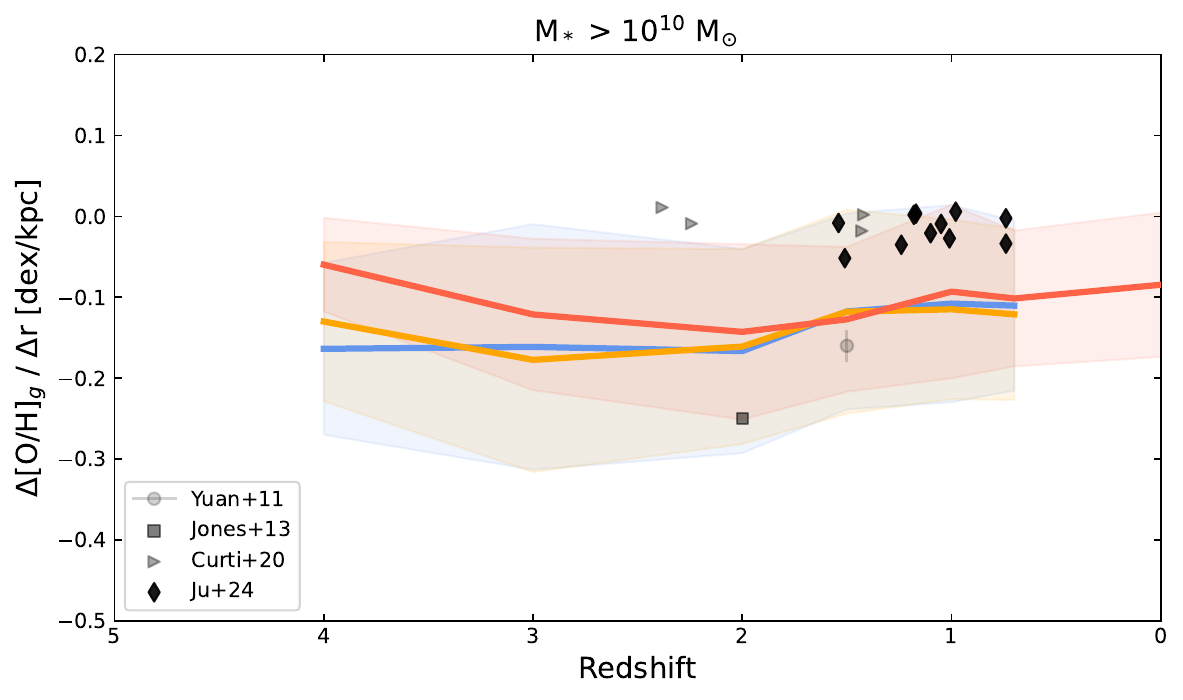}
    \caption{{\it Top panel:} SFR-weighted gas-phase metallicity gradient \di{in [dex/kpc]} as a function of redshift for all galaxies in our simulations with the thermal (blue solid line), stochastic (orange), and mechanical (red) feedback models. The shaded areas are 1$\sigma$ scatter. The {black/gray} symbols  
    are observational data from \citet[][circle]{Yuan2011} using AO-assisted spectroscopy OSIRIS on Keck II on a face-on spiral galaxy at $z \sim 1.5$, \citet[][Square]{Jones2013} using AO-assisted spectroscopy OSIRIS on Keck on gravitationally lensed systems, \citet[][right-pointing triangles]{Curti2020_klever} using KMOS KLEVER survey on 42 gravitationally lensed galaxies, 
    \citet[][upward-pointing triangle]{Wang2022} with NIRISS, early result from GLASS-JWST and \citet[][diamonds]{Ju2024} with the JWST/NIRSpec Slit-stepping Spectroscopy.
    {\it Bottom panel:} Same as the top panel but for massive galaxies ($M_*>10^{10} M_\odot$) {only}. 
    }\label{grad_z_allM}
\end{figure}

\begin{figure}
    \includegraphics[height=5cm]{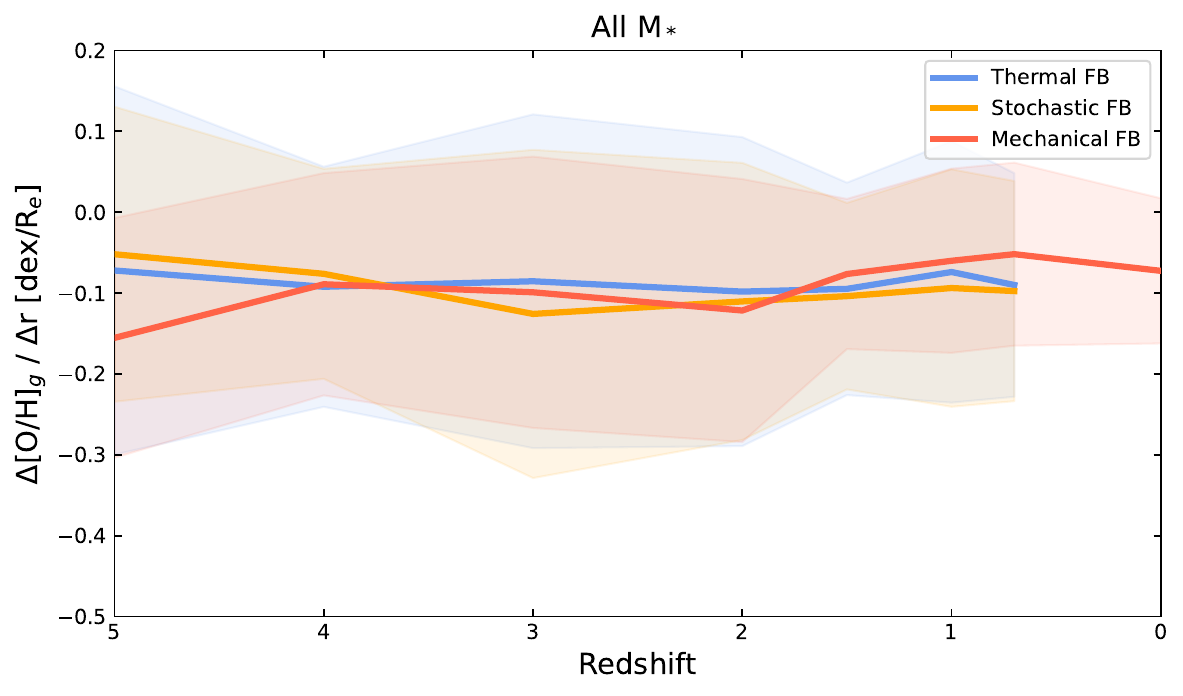}
    \includegraphics[height=5cm]{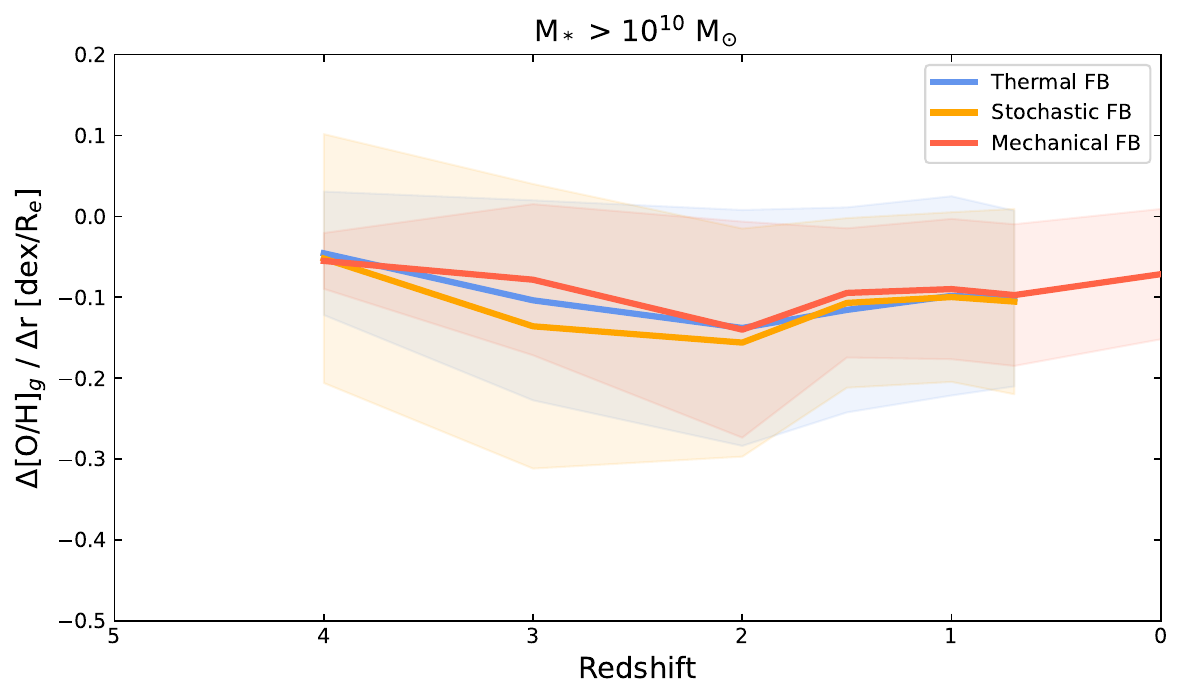}
    \caption{\di{Same as Figure \ref{grad_z_allM}, but in [dex/$R_{\rm e}$] {measured within 2.5 $R_{\rm e}$}.}}\label{grad_z_allM_Re}
\end{figure}

At high redshifts, available observational data are for gas-phase oxygen abundances {only} (e.g. \citealt{Cresci2010}; \citealt{Jones2010}; \citealt{Yuan2011}; \citealt{Curti2020_klever}), thus we focus on gas-phase gradients, but we also show our predictions of stellar metallicity gradients for future. The medians and one sigma scatter are also listed as a function of redshift in Table \ref{table_z_evolution}. 

The top panel of Figure \ref{grad_z_allM} shows the redshift evolution of the SFR-weighted gas-phase oxygen abundance gradients for all galaxies in our simulations with the thermal (blue), stochastic (orange), and mechanical (red) feedback models. The gradients are measured along the projected radius for each galaxy, in a fixed range $r<8$ kpc, and then the medians at a given redshift are calculated (solid lines). The shaded areas are $1\sigma$ scatter. Note that we show our simulations wherever available, i.e., down to $z=0.7$ for the stochastic and thermal models and down to $z=0$ for the mechanical model.
\di{This is due to overheating in the feedback runs at $z < 0.7$, which leads to suppressed star formation and unrealistic galaxy properties. These effects are discussed in detail in \citet{Ibrahim_Kobayashi_2024}. For $z = 0$, our main conclusions rely on the mechanical feedback model, which remains consistent with observations down to $z = 0$.}

There is only a mild evolution in the metallicity gradients from $z=0$ to $z=4$ for all feedback models.
The mechanical feedback has the flattest gradient ($\alpha_{\rm g,in} = -0.07\pm0.12$ dex/kpc at $z=0.7$) up to $z=4$, as already shown in Fig. \ref{Grad_M}. The thermal and stochastic models produce similar gradients up to $z=4$, where the gradient becomes steeper in the thermal case ($-0.12\pm0.15$ dex/kpc at $z=4$).
With all three supernova feedback models, the gradients always remain negative at all redshifts, making it steeper than observational data (e.g., \citealt{Curti2020_klever}, \citealt{Ju2024}). 

Since there is a gradient--mass relation {(Fig. \ref{Grad_M})}, this figure looks different depending on the galaxy mass. The bottom panel of Figure \ref{grad_z_allM} is the same as the top panel, but only for the massive galaxies ($M_* \ge 10^{10} M_\odot$) in our simulations. This figure shows a clear impact of the feedback models on the gradients at $z>2$, where the mechanical feedback gives significantly flatter gradients, although there is no impact at $z<2$ for these massive galaxies. Even so, our predicted gradients are still steeper than observed.

\di{As galaxy sizes evolve with redshift, one might think that it would be preferable to measure the gradients differently.
Figure \ref{grad_z_allM_Re} is the same as Figure \ref{grad_z_allM} but with 
units of dex/$R_{\rm e}$
\di{and measured in 2.5 $R_{\rm e}$ instead of 8 kpc using the effective radius $R_{\rm e}$ of each galaxy.} 
\di{Since these metallicities are weighted by the SFR, the fitting range is not so important, as long as a sufficiently large range is covered.}
While \di{the gradient in} dex/kpc is more straightforward to measure observationally, this scale does not account for galaxy size evolution and can be misleading when comparing across redshifts. Galaxies at high redshifts are typically more compact for a given stellar mass than their low-redshift counterparts \dii{\citep{VanDerWel2014}}.
This size growth is particularly pronounced for massive galaxies, which experience stronger inside-out growth and feedback driven expansion of their stellar and gas distributions.
This normalization by effective radius {provides a comparison of metallicity gradients across galaxies taking account of their size evolution}. The trend {in Figure \ref{grad_z_allM_Re} is very similar to that in Figure \ref{grad_z_allM}, and both} show that the metallicity gradients remain relatively flat across cosmic time, with mild fluctuations.
\di{For the massive galaxies}, however, the unit change tends to flatten the gradients, and there is no significant difference among feedback models at $z=4$.
}

\begin{figure}
    \includegraphics[height=5cm]{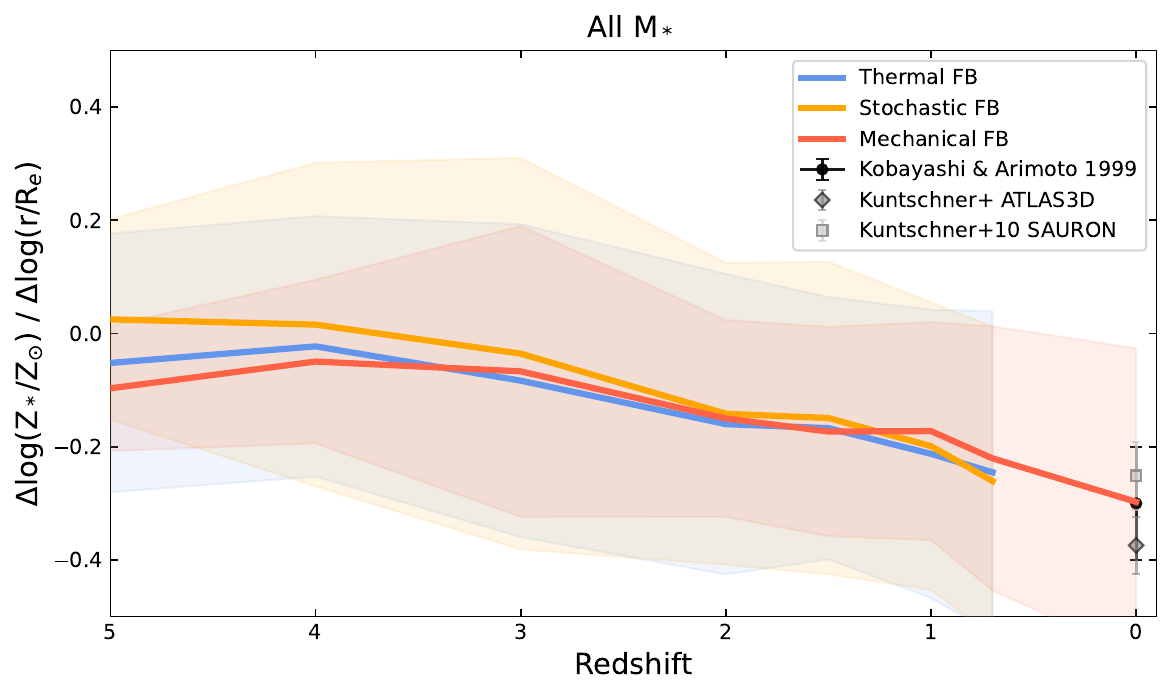} 
    \includegraphics[height=5cm]{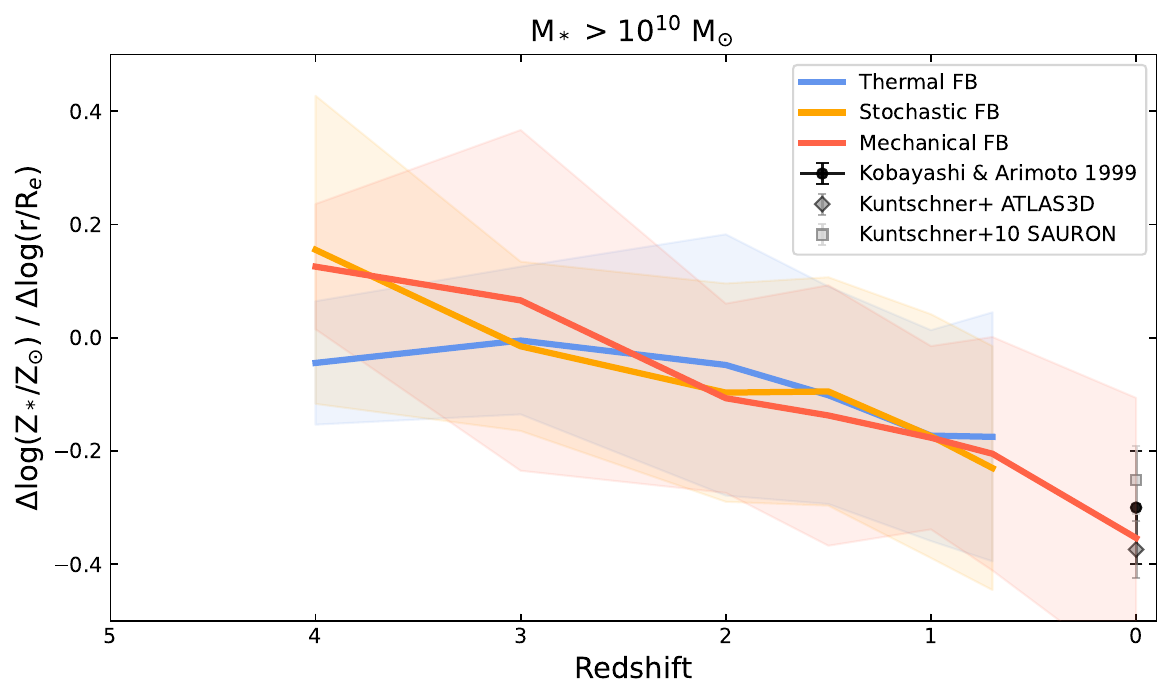}
    \caption{{\it Top panel:} V-band luminosity weighted stellar metallicity gradients {measured} within 1.5 $R_{\rm e}$ as a function of redshift for all galaxies in our simulations with the thermal (blue), stochastic (orange), and mechanical (red) feedback models. Gradients are measured for each galaxy, normalized by the effective radius $R_{\rm e}$, and then the median is calculated at a given redshift (solid lines). The shaded areas are 1$\sigma$ scatter. Observational data are taken from \citet[][black point with error bar]{Kobayashi1999}, the SAURON survey (\citet[][light gray square with error bar]{Kuntschner_2010}; and the ATLAS$^{\rm 3D}$ survey (Kuntschner et al., priv. comm; gray diamond with error bar).
    {\it Bottom panel:} Same as top panel but for massive galaxies with $M_*>10^{10} M_\odot$. }\label{grad_z_allM_str}
\end{figure}

Figure \ref{grad_z_allM_str} is the same as Figure \ref{grad_z_allM} but for the V-band luminosity weighted stellar metallicity gradients along the projected radius within $1.5 R_{\rm e}$. Gradients are measured against $\log(r/R_{\rm e}$) of each galaxy, 
then medians are calculated at a given redshift (solid lines).
The top panel includes all galaxies in our simulations and shows that the stellar metallicity gradient is not highly impacted by supernova feedback models, although the stochastic model gives flatter gradients (by $\sim 0.1$ than the mechanical model) at $z>2$. 

As seen in the gradient--mass relations (Fig. \ref{Z_R_M>10}), massive galaxies show a clear impact of the feedback models.
The bottom panel shows that, for massive galaxies ($M_* \ge 10^{10} M_\odot$), the stellar gradients are more impacted by the feedback model at $z>2$, where the mechanical feedback \di{also} gives flatter gradients (by $\sim 0.2$ than the thermal feedback at $z=4$). \di{The {\it positive} gradients are caused by slower chemical enrichment at the centre, which is surprising and could be due to inflow of low-metal gas and/or outflow of metal-rich gas (see Section \ref{sect_discussion} for more discussion).}

Although there are no observational data at high redshifts, at $z=0$, we find that our gradient with the mechanical feedback shows a great match with observational data of nearby galaxies \citep[e.g.][]{Kobayashi1999, Kuntschner_2010}. \dii{
By contrast, our predicted gradients 
are more negative by $\sim0.3$~dex than CALIFA disc galaxies \citep{Sanchez_Blazquez_2014}}. 

\begin{table}
\centering
\resizebox{\columnwidth}{!}{%
\begin{tabular}{|l|c c c r|} 
\hline
$z$         & Thermal     & Stochastic  & Mechanical   \\
\hline
\hline
     &     &   $\alpha_{\rm g,{in}}\equiv{\Delta {\rm [O/H]}_{\rm g} }/{\Delta r}$ [dex/kpc]   &      \\
\hline
0          & $--$                   &     $--$              &  -0.07 $\pm$ 0.09 \\
0.7       &  -0.16$\pm$ 0.16      &  -0.16 $\pm$ 0.13    &   -0.07 $\pm$ 0.12          \\
1       &   -0.15 $\pm$  0.16     &   -0.15$\pm$0.15         &  -0.08  $\pm$0.12         \\
1.5       &   0.14 $\pm$0.15     &   -0.15$\pm$0.13        &   -0.10   $\pm$ 0.10     \\
2        &     -0.14 $\pm$0.19 &   -0.14 $\pm$0.17         &   -0.13$\pm$0.15         \\
3         &    -0.12 $\pm$  0.20    &    -0.14$\pm$ 0.20      &   -0.11 $\pm$ 0.17      \\
4        &     -0.12 $\pm$ 0.15   &   -0.11 $\pm$ 0.14   &   -0.10 $\pm$ 0.12           \\
5       &       -0.09 $\pm$ 0.22        & -0.07 $\pm$ 0.15  &           -0.17 $\pm$ -0.14  \\
\hline
     &     &   
     \di{{$\Delta {\rm [O/H]}_{\rm g} $}/{$\Delta r$} [dex/$R_{\rm e}$]}   &      \\
\hline
0          &      $--$           &        $--$            &  -0.07 $\pm$ 0.06 \\
0.7       &  -0.09$\pm$ 0.14      &  -0.10 $\pm$ 0.13        &   -0.05 $\pm$ 0.11          \\
1       &   -0.07 $\pm$  0.16     &   -0.9$\pm$0.14         &  -0.06  $\pm$0.11         \\
1.5       &   -0.09 $\pm$0.13      &   -0.10$\pm$0.11        &   -0.08   $\pm$ 0.09     \\
2        &     -0.10 $\pm$0.19     &   -0.11 $\pm$0.17         &   -0.12 $\pm$0.16         \\
3         &    -0.09 $\pm$  0.20    &    -0.13 $\pm$ 0.20      &   -0.09 $\pm$ 0.17      \\
4        &     -0.09 $\pm$ 0.15      &   -0.07 $\pm$ 0.12   &   -0.08 $\pm$ 0.13           \\
5       &       -0.07 $\pm$ 0.22        & -0.05 $\pm$ 0.18  &           -0.16 $\pm$ 0.15  \\
\hline
            &             &   $\alpha_{*,{\rm in}}\equiv\Delta \log(Z_*/Z_\odot) / \Delta \log(r/R_{\rm e}) $   &   \\
\hline
0          &  $--$      &      $--$            &  -0.25 $\pm$ 0.23 \\    
0.7       &  -0.19  $\pm$  0.20   &     -0.19$\pm$0.20       &     -0.17 $\pm$ 0.18     \\
1       &     -0.18 $\pm$ 0.23        &    -0.18$\pm$ 0.22  &      -0.16 $\pm$  0.18   \\
1.5       &   -0.15 $\pm$ 0.22        &  -0.13$\pm$0.28       &     -0.16 $\pm$ 0.18        \\
2        &      -0.10 $\pm$ 0.23        & -0.08 $\pm$ 0.23  &    -0.11   $\pm$0.15    \\
3         &    -0.05 $\pm$ 0.25        &   -0.006 $\pm$ 0.32          &    -0.05 $\pm$ 0.25         \\
4        &    0.02 $\pm$ 0.22         &     0.01 $\pm$0.28     &   -0.05  $\pm$  0.14       \\
5        &     -0.05$\pm$ 0.23         &    0.03$\pm$ 0.17       &   -0.10  $\pm$ 0.11      \\
 \bottomrule
\end{tabular}
}
\caption{The median of the SFR-weighted gas-phase oxygen abundance gradients in \di{dex/kpc (upper table) and dex/$R_{\rm e}$ (middle table)}, and the V-band luminosity-weighted stellar metallicity gradients (bottom table), as a function of redshift $z$ for all galaxies in our simulations with the thermal, stochastic, and mechanical feedback models. The 1$\sigma$ scatters are also given.}\label{table_z_evolution}
\end{table}

\section{Galaxy type dependence}\label{gal_type}

To understand the offset between the simulations and observations, one may ask if observational data are biased toward star-forming galaxies. In this section, we study the dependence of the metallicity gradient on the galaxy ``type''. All simulated galaxies for which we measured gas-phase metallicity gradients are star-forming. Thus, we split our sample into lower star formation earlier-type galaxies (ETGs) and higher star formation later-type galaxies (LTGs), using the star-formation main sequence (SFMS) of galaxies that are commonly used in observations (e.g. \citealt{Renzini_Peng_2015}) and in simulations (e.g. \citealt{Taylor2016}). 

The SFMS of our simulated galaxies are shown in Figure \ref{SFMS} in Appendix \ref{sect_SFMS}. To select ``ETGs'' and ``LTGs'', we use the same method as in \citet{Taylor_2017}, where we take the best linear fit of our simulated SFMS and calculate the perpendicular distance ($\Delta$SFMS) of this fit from the data. We then use this quantity to define the galaxies such that ``ETGs'' have $\Delta$SFMS < $-0.5$, and ``LTGs'' have $\Delta$SFMS $\geq$ $-0.5$. 

\begin{figure}
    \includegraphics[width=0.48\textwidth]{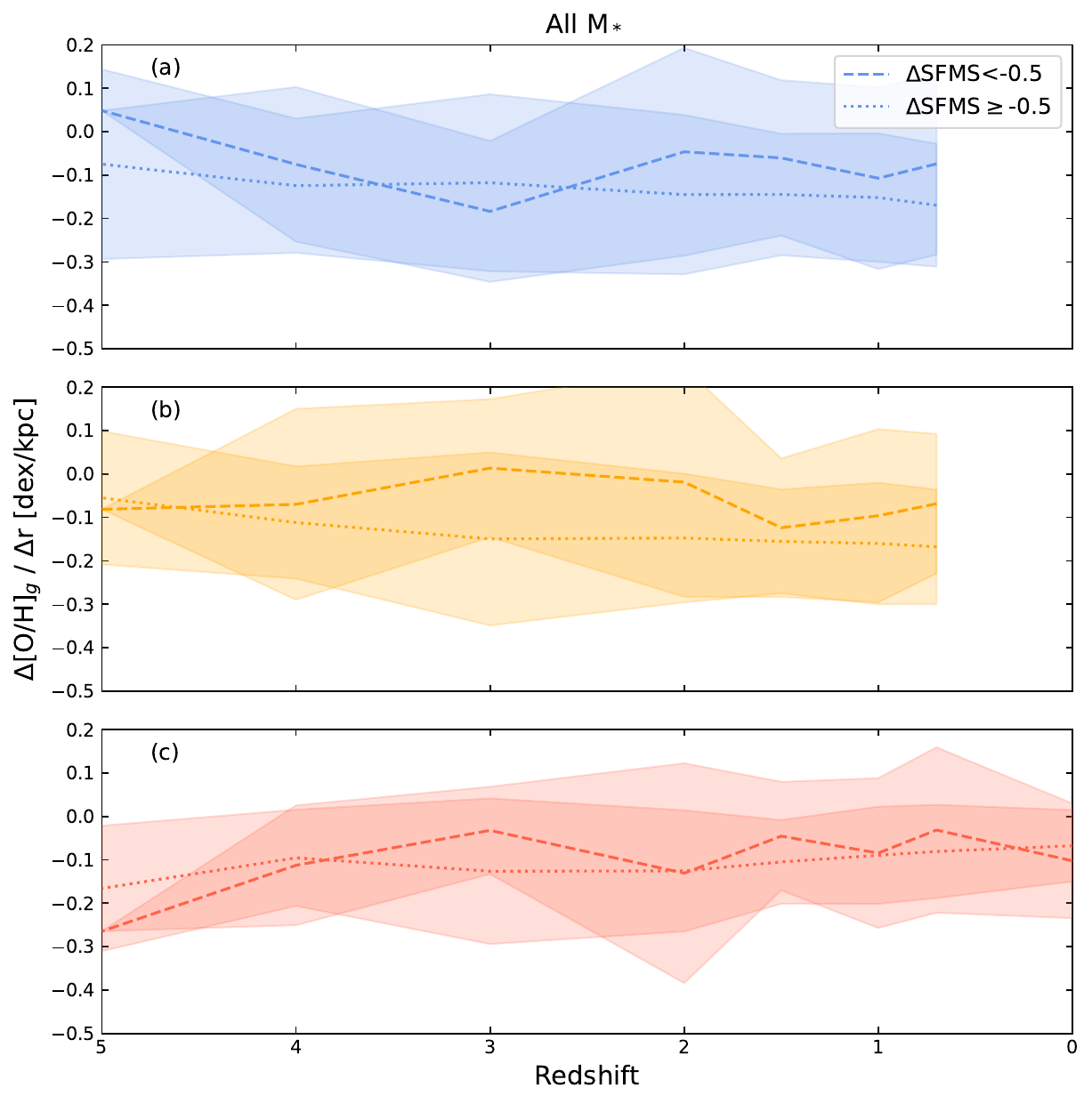}
    \caption{The SFR-weighted gas-phase oxygen abundance gradients as a function of redshift for all galaxies in our simulations with the thermal (blue, top panel), stochastic (orange, middle panel), and mechanical (red, bottom panel) feedback models. The dashed and dotted lines show the medians of earlier-type galaxies ($\Delta$SFMS < $-0.5$) and later-type galaxies ($\Delta$SFMS > $-0.5$), respectively. The shaded areas are 1$\sigma$ scatter.
    }\label{z_evo_type}
\end{figure}

Figure \ref{z_evo_type} shows the redshift evolution of the SFR-weighted gas-phase metallicity gradients for ETGs (dashed lines) and LTGs (dotted lines) for thermal (blue top panel), stochastic (orange, middle panel), and mechanical (red, bottom panel) feedback models. Overall, ETGs have a flatter gradient at all redshifts with all feedback models. Hence, the offset between observations and simulations is not due to the selection bias. At $z\sim5$, ETGs seem to show a steep gradient with mechanical feedback.

\section{\di{Discussion}}\label{sect_discussion}

\di{In hydrodynamical simulations, metallicity gradients are affected by several physical processes: inside-out growth, metal flows, stellar migration, and galaxy mergers. These processes interact with one another, and it is beyond the scope of this work to disentangle their individual impacts. However, we wish to discuss our findings in relation to previous studies.}

\subsection{Present-day gradients}

\di{
At $z = 0$, both observations and theory consistently find that gas‐phase metallicity gradients are predominantly negative.  Surveys such as CALIFA, MaNGA, and SAMI measure typical slopes of $-0.05$ to $-0.1\;\mathrm{dex}/R_{\rm e}$ \citep{sanchez2014, Belfiore2017, Poetrodjojo2021}, while cosmological simulations (Illustris/TNG50, EAGLE, FIRE, SIMBA) predict similarly mild negative values of
$-0.01$ to $-0.06\;\mathrm{dex}/\mathrm{kpc}$
 \citep{Hemler2021, Ma_2017, Dave2019, Garcia_2025}. 
 Analytic inside‐out and ``bathtub'' models {can create} negative gradients,  
 depending on the assumed gas accretion and star‐formation efficiencies \citep[e.g.][]{Chiappini2001, Molla_Diaz2005, Sharda2021}.
We find $\sim-0.07$ both for $\mathrm{dex}/\mathrm{kpc}$ and $\mathrm{dex}/R_{\rm e}$ (Table \ref{table_z_evolution}) for gas with our preferred, mechanical feedback model.}

\begin{figure}
	\includegraphics[width=0.5\textwidth]{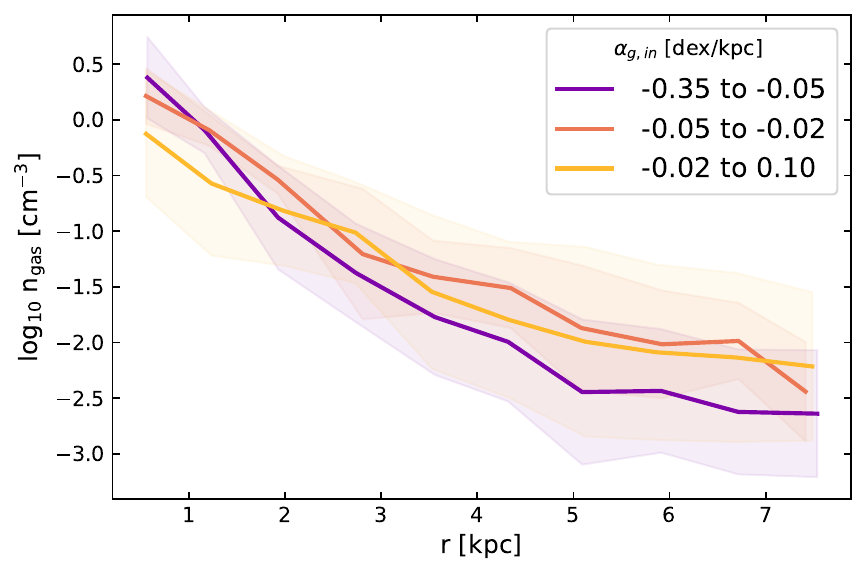}
    \caption{\di{Radial gas number density profiles for galaxies grouped by different bins of gas-phase metallicity gradients $\alpha_{\rm g, in}$ within 8 kpc. Each line represents the median profile within a given gradient range. The shaded area is the 1 $\sigma$ scatter. {Results are shown for $z=0$ with the mechanical feedback}.}}\label{density_profile}
\end{figure}

\begin{figure}
	\includegraphics[width=0.5\textwidth]{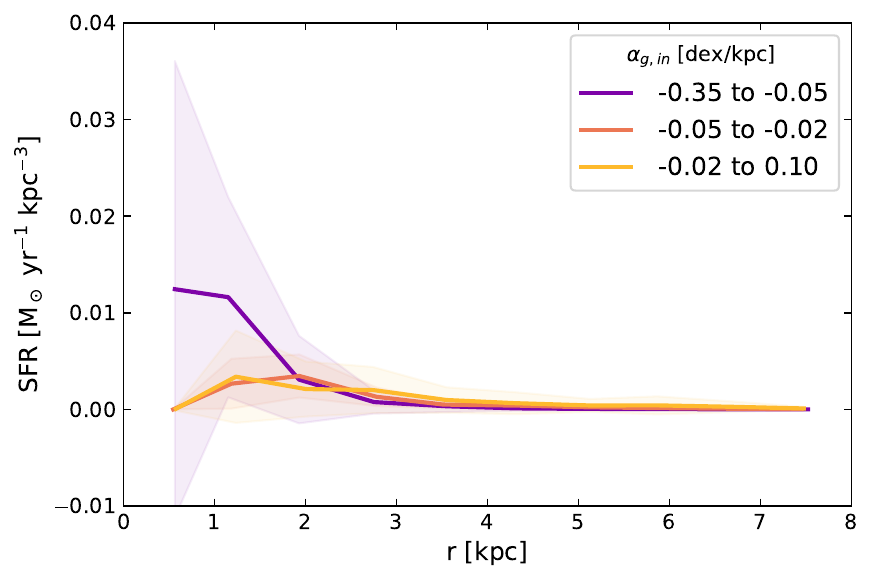}\caption{
    \di{{Same as Fig.\,\ref{density_profile}, but for}
    radial star formation rate profiles. 
    }}\label{sfr_profil}
\end{figure}

\di{
Moreover, both for observations and simulations, there is a {galaxy} mass dependence {(Fig. \ref{Grad_M_z0}b)}, where massive galaxies ($M_* \gtrsim 10^{10.5}\,M_\odot$) exhibit flatter gradients ($-0.06$ to $-0.03\;\mathrm{dex}/R_{\rm e}$).
This could be due to `equilibrium' reached with low gas density at the centre \citep{Belfiore2017}. 
Is is also important to note that we find a similar trend also for stellar gradients (Fig. \ref{Grad_M_z0}a), which is predominantly caused by galaxy mergers \citep{Kobayashi2004}. 
{This leads to a radial gradient of metals in stellar mass-loss}, which keep supplying metals in the ISM.
Therefore, the flattening of gas gradients at the massive end is also due to galaxy mergers. 
On the other hand, \di{intermediate}‐mass systems ($M_* \sim 10^{10.5}\,M_\odot$) show steeper gas-phase gradients (down to $-0.3\;\mathrm{dex}/R_{\rm e}$), which we also see in stellar gradients. 
{We find that the steep gas-phase gradients are caused by on-going star formation and higher gas density at the centre (Figs. \ref{density_profile} and \ref{sfr_profil}).}
However, lower‐mass galaxies ($M_* \sim 10^{9}\,M_\odot$) show flat gradients, which could be explained by enhanced radial mixing likely driven by feedback in shallow potentials.  
\dii{This interpretation is supported by our finding that the flattening at the low-mass end is weaker with the other feedback models (Fig. \ref{Grad_M}).}
}


\di{
To better understand the physical origin of the {\it steep} metallicity gradients in our simulated galaxies, we examined the radial profiles of gas surface density (Figure \ref{density_profile}) and star formation rate (Figure \ref{sfr_profil}) for the mechanical feedback model at $z=0$.  
}
\di{
We compute the radial hydrogen number density profiles (in 3D) by binning the gas particles in concentric spherical shells centred on each galaxy. For each shell between inner radius $R_1$ and outer radius $R_2$, thesgas mass is measured by summing the masses of all particles whose radii satisfied $R_1 \leq r < R_2$. The shell volume was computed as $V = \frac{4}{3}\pi (R_2^3 - R_1^3)$, yielding a mass density $n_{\mathrm{gas}} = M_{\mathrm{gas}} / V$ in units of $M_\odot\,\mathrm{kpc^{-3}}$. We then converted this to a number density in $\mathrm{cm^{-3}}$ and dividing by the mean particle mass $\mu m_{\mathrm{H}}$. We adopted $\mu = 1.4$ to account for the presence of helium in addition to hydrogen. The resulting profile therefore represents the median hydrogen number density of gas as a function of radius.}
\di{We also compute the radial SFR profiles (in 3D) using the same concentric spherical shells defined for the gas density profiles. For each shell between inner radius $R_1$ and outer radius $R_2$, the SFRs of all gas particles are added, then divided by the shell volume to obtain the SFR volume density in units of $M_\odot\,\mathrm{yr^{-1}\,kpc^{-3}}$. This approach yields the SFR density profiles of individual galaxies, which were then stacked to compute the median profiles for each subsample.
}

\di{
Figure~\ref{density_profile} shows the hydrogen number density profiles for different bins of gas-phase metallicity gradients, defined such that each bin contains the same number of galaxies.
Galaxies with steeper metallicity gradients tend to have more centrally concentrated gas distributions. In the steepest-gradient bin ($-0.35$ to $-0.05$ dex/kpc; purple), the gas density peaks sharply at $r < 1$ kpc 
and declines rapidly beyond $\sim$1.5 kpc. Galaxies with intermediate gradients ($-0.05$ to $-0.02$ dex/kpc; orange) {show a similar but less pronounced central concentration with lower maximum densities}. 
Meanwhile, galaxies with flatter or inverted gradients ($-0.02$ to $+0.10$ dex/kpc; yellow) have the most diffuse central gas, with comparatively low central densities.}

\di{
Similarly, Figure \ref{sfr_profil} shows that steep-gradient galaxies ($-0.35$ to $-0.05$ dex/kpc; purple) exhibit a SFR peak 
{($\sim$0.013 M$_\odot$ yr$^{-1}$ kpc$^{-3}$) at $r < 1$ kpc followed by a rapid decline beyond $r \approx 1.5$ kpc, indicating highly concentrated star formation. Galaxies with flatter or positive gradients have lower SFRs with a broad maximum of $\sim$0.004 M$_\odot$ yr$^{-1}$ kpc$^{-3}$ that falls off at $r \sim 1.5$--2 kpc and sustains activity out to $\sim 3.5$ kpc.}
These profiles show that the radial distribution of star formation is a driver of the overall metallicity gradient. 
}

\di{
{From these two figures, we can conclude} that steeper metallicity gradients are caused in galaxies that are more centrally concentrated in both gas density and star formation, and thus there is a small, dense star-forming core that rapidly enriches its inner few kpc. On the other hand, flatter (or inverted) gradients arise in systems where the gas reservoir and star formation is  spread out to larger radii,  
diluting any strong inner metallicity peak. }


\subsection{Overall evolution of gradients}

\di{
At high redshifts ($z \gtrsim 1$), the picture of gas-phase metallicity‐gradient evolution becomes notably more complex, with some studies finding significantly steeper gradients while others report flat or even inverted profiles.  Early, small‐sample AO observations {of lensed galaxies} (e.g. \citealt{Jones2010, Jones2013, Yuan2011}) predict that gradients steepen with redshift, {which were in excellent agreement with zoom-in simulations of a Milky-Way type galaxy \citep{Kobayashi_Nakasato_2011,Pilkington2012} and some cosmological simulations} (e.g.\ TNG50, with a steepening rate of $\sim0.02\;\mathrm{dex}/\mathrm{kpc}\,\mathrm{per}\;z$; \citealt{Hemler2021}; MUGS with ``normal'' feedback; \citealt{Gibson_Pilkington_2013}).  Our mechanical feedback {(Fig. \ref{grad_z_allM})} runs likewise steepen by $\sim0.1\;\mathrm{dex}/\mathrm{kpc}$ from $z=0$ to 5, although for $M_* > 10^{10}\,M_\odot$ systems the gradient peaks around $z\approx2$ before flattening again by $z\sim4$ {(see below for more discussion)}.}

\di{
{In contrast}, other works, both observational (e.g.\ \citealt{Swinbank2012, Cresci2010, Leethochawalit2016, Wuyts_2016, Wang_2017, Schreiber2018, Curti2020_klever})
and theoretical (enhanced‐feedback MaGICC; \citealt{Gibson_Pilkington_2013}; FIRE; \citealt{Ma_2017}; EAGLE; \citealt{Tissera_2022}; analytic Milky Way models; \citealt{Mott_2013, Sharda2021}), find no evolution or flatter (or inverted) gradients at $z>1$ than $z=0$, often with large scatter.  
Our stochastic and thermal feedback models also produce {flatter (or inverted) gradients in some galaxies (Fig. \ref{Grad_M})}.}

\di{
\dii{As galaxies evolve, mergers tend to flatten metallicity gradients \citep{Kobayashi2004, Rupke2010, Taylor_2017}}, as does inside-out growth \citep{Vincenzo2020}. On the other hand, dilution and mixing of metals driven by strong feedback can produce flat gradients at any stage of a galaxy’s history.
In addition to these gas-phase processes, physical mechanisms such as stellar migration can also influence the evolution of metallicity gradients, particularly in the stellar component. 
Radial migration flattens the stellar metallicity gradient over time by redistributing older stars outward from the chemically enriched inner regions \citep{Vincenzo2020}. 
This process is expected to be especially important in massive, well-structured disks. 
However, {for massive galaxies,} we find stellar gradients becoming steeper as time (Fig. \ref{grad_z_allM_str})}.
\dii{We do not see the impact of stellar migration, although this could be due to insufficient resolution.}

\di{{Instead, we find that the inverse stellar gradients of massive galaxies are caused by off-center star formation (in a similar SFR profile as in Fig. \ref{sfr_profil}).}
 In these galaxies, early central star formation occurred from relatively metal-poor gas, while later star formation in the outskirts used gas that had already been enriched. 
This combination can explain the positive stellar gradients at early times, even in systems whose gas-phase metallicity decreases with radius.
{We have seen this in Galaxy B at $z=0$ with a comparable stellar mass (Fig.\,\ref{map_lowMass}), and the stellar gradient was indeed positive (Fig.\,\ref{Z_R_GalB}).}}

\di{It is important to acknowledge other small-scale mixing processes such as turbulence driven diffusion and superbubble evolution, which {can} impact metallicity gradients (e.g. \citealt{Sharda2021}), are not explicitly captured. This may affect the fine structure of metallicity gradients, but we expect the global trends with mass and redshift to remain robust. Future work with higher-resolution simulations and explicit mixing models will help quantify these effects.}

\subsection{Gradient evolution in an individual galaxy}

\di{
In cosmological simulations, tracing the detailed evolutionary history of individual galaxies is challenging. Here, we present the redshift evolution of gas-phase and stellar metallicity gradients for a representative system, Galaxy A.}
\di{
We track Galaxy A across snapshots using its unique catalog ID to link the same system between redshifts. At each snapshot all candidate star particles associated with that ID (including post mergers fragments; see Figure \ref{galA_map_evo}) are collected to assign the most massive bound system as the descendant for that epoch. We then re-center the galaxy by recomputing the center of mass. Gas-phase metallicity profiles and gradients are measured as described in Section \ref{sect_method} This procedure ensures a consistent descendant choice and robust centering at every redshift before estimating the metallicity gradients.
}

\begin{figure}
	\includegraphics[width=0.5\textwidth]{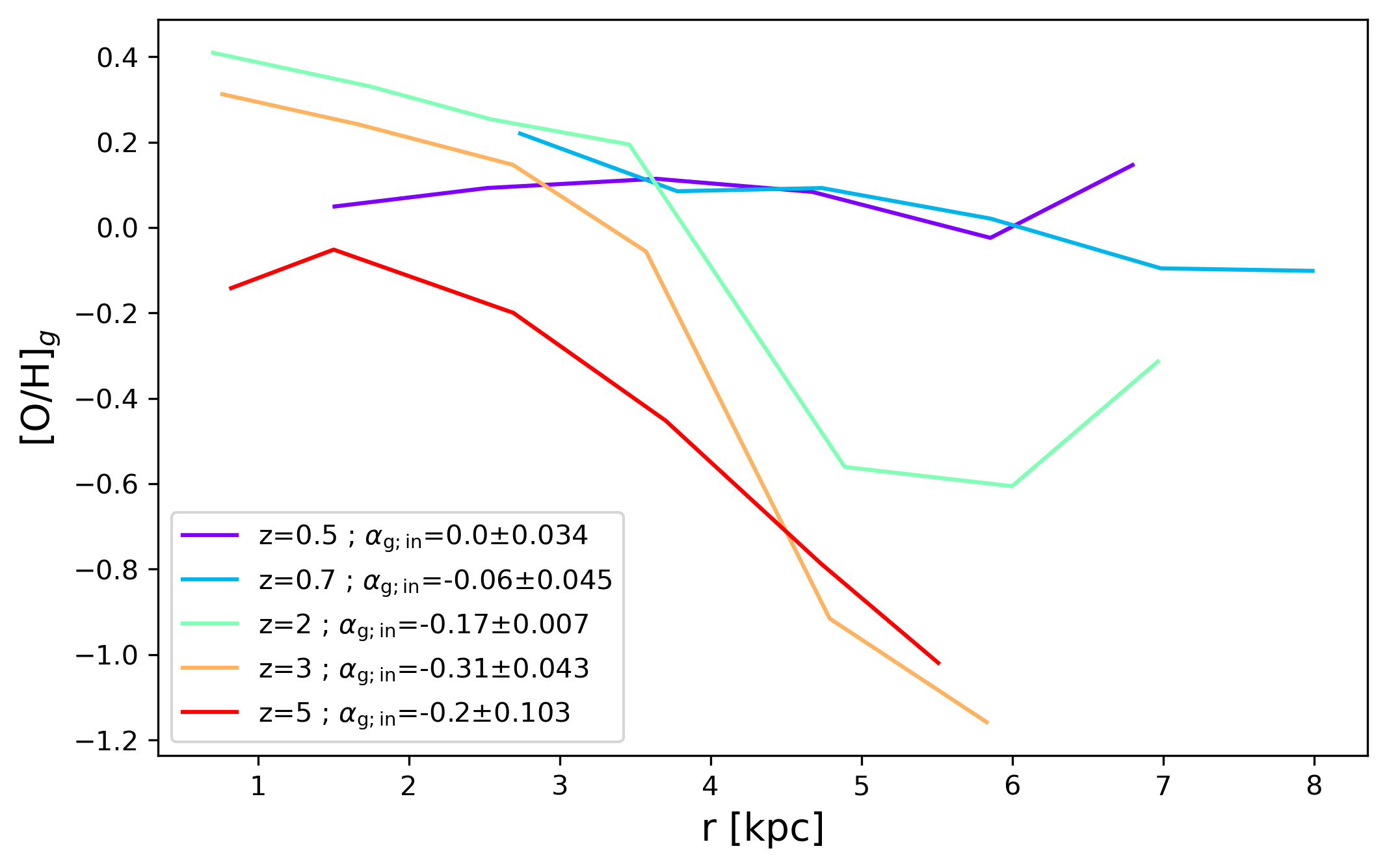}\caption{\di{Evolution of the SFR-weighted gas-phase metallicity profile of Galaxy A with the mechanical feedback, at redshifts $z = 0.7$, 2, 3, and 5. The slope of each profile, $\alpha_{\rm g,in}$ [dex/kpc], is obtained within 8 kpc and is listed in the legend. }}\label{Zprofil_galA}
\end{figure}

\begin{figure}
	\includegraphics[width=0.5\textwidth]{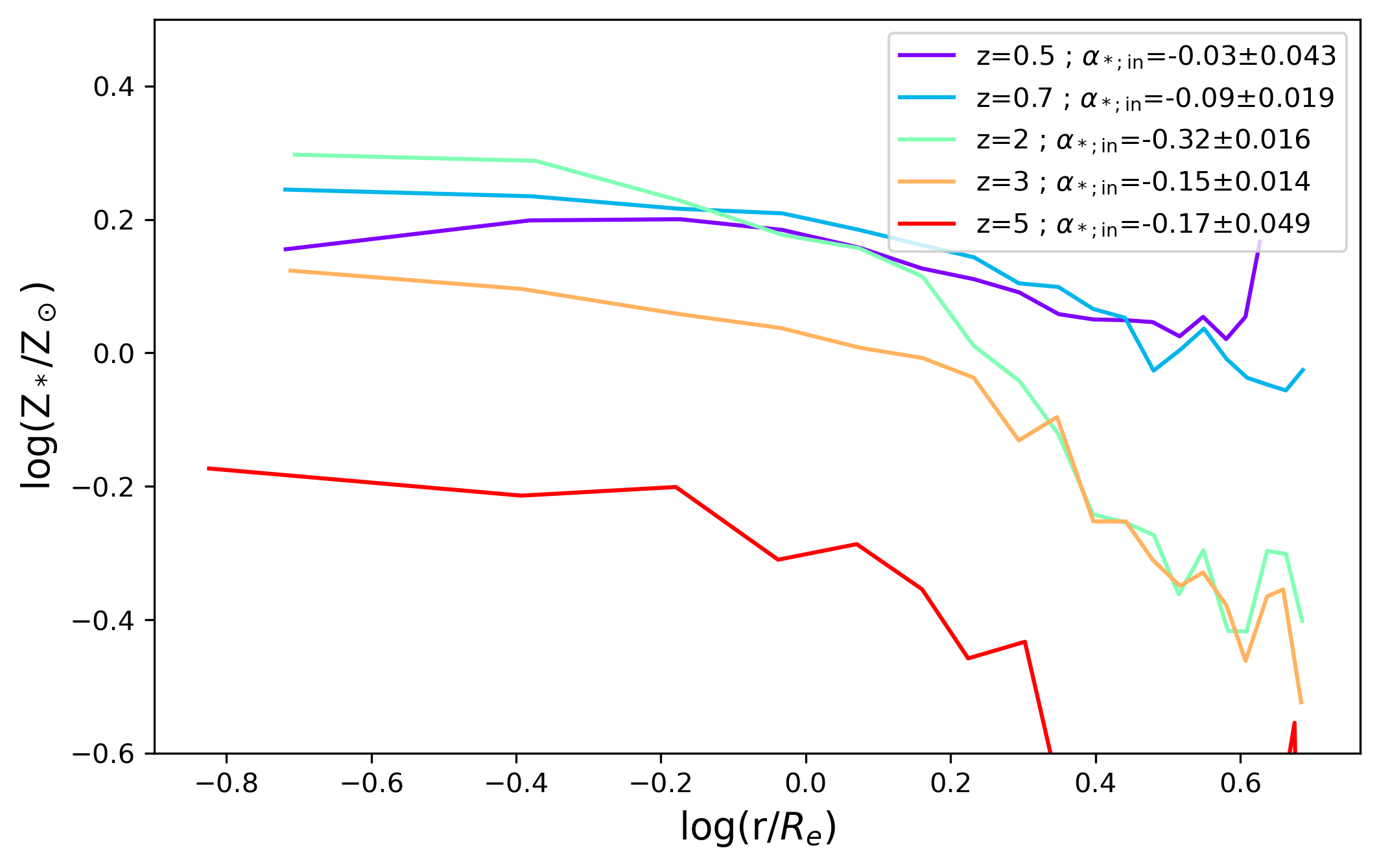}\caption{
    \di{{Same as Fig.\,\ref{Zprofil_galA} but for the luminosity-weighted stellar metallicity profile of Galaxy A.}
    The slope of each profile $\alpha_{\rm *,in}$ [dex/kpc], is obtained within 1.5 $R_{\rm e}$ (at each redshift) and is listed in the legend. }}\label{Zstrprofil_galA}
\end{figure}

\di{Figure \ref{Zprofil_galA} shows the evolution of SFR-weighted gas-phase metallicity profile of Galaxy A across five redshifts: $z = 5$ (red), 3 (orange), 2 (light green), 0.7 (cyan), and 0.5 (purple). The metallicity gradient $\alpha_{\rm g,in}$ is computed via a linear fit within 8 kpc at each redshift. The figure {clearly shows the flattening of the gradient as a function of time. In other words, the metallicity evolves more rapidly in the outskirts than in the centre.}
The slope is steep from $\alpha_{\rm g,in} = -0.20 \pm 0.103$ dex/kpc at $z = 5$ to a maximum of $-0.31 \pm 0.043$ at $z = 3$, and then gradually flattens, reaching $\alpha_{\rm g,in} = 0.00 \pm 0.034$ at $z = 0.5$. This progression is consistent with inside-out growth and the {increasing} impact of stellar feedback, which redistributes metals over time, as we as the flattenings due to mergers (see Figure \ref{galA_map_evo}).
{Figure \ref{Zstrprofil_galA} shows the evolution of stellar metallicity gradients, which remain negative at all times. Note that this is not the same massive galaxy that showed a positive gradient at $z=4$ in Figure \ref{grad_z_allM_str}. 
}}

\di{Between $z=0.5$ and $z=0.7$ {our measured $\alpha$-slopes indicate small flattening of the gradients both for gas and stars; however, the difference} is comparable to the temporal variation observed in individual systems (e.g., $\sim0.05$–0.2 $\mathrm{dex/kpc}$ in Fig. 12 of \citealt{Taylor_2017}) and therefore may not be significant. In other words, selecting a snapshot slightly earlier or later near $z\approx0.7$ could yield a slightly different $\alpha$ value. It is also possible that the galaxy is already in a steady state. 
By contrast, from $z\approx0.7$ to $z\approx2$ the gradient change is much larger but 
the intrinsic variability is also expected to be large around $z\sim2$, where mergers and bursty episodes drive ``zig–zag'' evolution in single-galaxy tracks \citep{Taylor_2017}. These temporal fluctuations are troublesome when following individual galaxies, {and are not included in the gradients values shown in Figures \ref{Z_R_galA} and \ref{Z_R_GalB}; however, they can be neglected in our main figures, where the results are averaged over many systems.} 
}

\section{Conclusions}\label{sect_conclusions}
We predict the evolution of stellar and gas-phase radial metallicity gradients across cosmic time, using our cosmological hydrodynamical simulations with the latest nucleosynthesis yields. Namely we show the dependence of the gradients on the stellar feedback models: thermal, stochastic and mechanical supernovae feedback models described in \citet{Ibrahim_Kobayashi_2024}. 

For the luminosity-weighted stellar metallicity, we find the radial profiles is best quantified with a logarithm fit (Eq.\ref{grad_str_Weighted}) of the inner gradient $\alpha_{*,\rm in}$ ($r<1R_{\rm e}$ and $r<1.5R_{\rm e}$) and the outer gradient $\alpha_{*,\rm out}$ ($r \in$ [$1R_{\rm e}$ and $2R_{\rm e}$] or $r \in$ [$1.5R_{\rm e}$ and $2R_{\rm e}$]), rather than the total gradient $\alpha_{*}$ (along the total projected radius). For the SFR-weighted gas-phase oxygen abundance, we use a linear fit (Eq.\ref{eq_grad_g_w}) to the inner gradient $\alpha_{\rm g,in}$ within $r<8$kpc {or $2.5R_{\rm e}$ to cover the same range at high-redshifts}.

\begin{itemize}
\item At $z=0$, our simulation with the mechanical feedback successfully reproduces the observed metallicity gradient vs mass relations of both stellar metallicity (Fig. \ref{Grad_M_z0}). Gradients are the steepest at intermediate-mass ($M_*\sim10^{10}M_\odot$) {due to on-going star formation at centre}. They flatten in massive galaxies, probably through major mergers \citep{Kobayashi2004,Taylor_2017}, \di{and in low-mass galaxies due to efficient feedback (Section \ref{sect_discussion}.1).} \dii{
We note that our low-mass galaxy sample is limited due to numerical resolution.
}

\item \di{Galaxies with steep negative gradients tend to have centrally concentrated gas and star formation (Figs. \ref{Zprofil_galA} and \ref{Zstrprofil_galA}), resulting in strong inner enrichment and limited metal transport to the outer regions. In contrast, galaxies with flatter or inverted gradients show more extended radial gas distributions and star formation, indicating efficient radial mixing likely driven by feedback.}

\item For massive galaxies ($M_* >10^{10} \rm M_\odot$) at $z=0.7$, inner stellar gradients ($\alpha_{*,\rm in}\sim-0.1$) are much flatter than outer ($\alpha_{*,\rm out}\sim-0.7$) and total gradients.
Gas-phase oxygen abundance gradients are also negative ($\alpha_{\rm g,in}\sim-0.1$ dex/kpc). The impact of stellar feedback on these gradients are subtle (Fig. \ref{Z_R_M>10}), which is expected due to their deep potential well.

\item For lower-mass galaxies ($M_* <10^{10} \rm M_\odot$) at $z=0.7$, the feedback models have a significant impact on the gradients. Stellar total gradients $\alpha_{*}$ are significantly flatter with the mechanical feedback by $\sim$0.2--0.3. Inner gradients $\alpha_{*,\rm in}$ are slightly flatter for the mechanical feedback by $\sim$0.02--0.03. 
Gas-phase gradient $\alpha_{\rm g,in}$ is also flatter with the mechanical feedback by $\sim$0.1 dex/kpc than the thermal and stochastic cases (Fig. \ref{Z_R_M<10}). These can be \dii{potentially} explained by the suppression of star formation at the centre and ejection of metals to the outskirts.

\item The galaxy stellar mass dependence of the stellar and gas-phase metallicity gradients with the three feedback models are shown up to $z=5$ (Fig. \ref{Grad_M}). 
From $z=0.7$ to $z=4$, we retrieve a relation between stellar mass and stellar metallicity gradient, where the gradient is steeper for lower-mass galaxies. This relation is stronger with the total gradients $\alpha_{*}$ than for the inner gradients $\alpha_{*,\rm in}$. 
Stellar outer gradients would be more informative, although these are very difficult to measure with high-redshifts observations.
For gas-phase, gradients are steepest at the intermediate mass, and become flatter toward lower and higher masses.
While the mechanical feedback results in gradients closest to observational data at $z=0.2$, the gradients are still steeper than observed at $z=2$ over $10^{9}$--$10^{10}$M$_\odot$. 

\item 
To explain recent and ongoing observational IFU data, the redshift evolution of the gas-phase oxygen abundance metallicity gradients up to $z=5$ is presented (Fig. \ref{grad_z_allM}). 
At all redshifts and with all three models, the gradient of the majority of galaxies in our simulations have negative gradients.
With the mechanical feedback, lower-mass galaxies have significantly flatter gradients up to $z=2$, whereas massive galaxies have flatter gradients at $z>2$.
These gradients are still steeper than recent IFU observations \citep[e.g.][]{Curti2020, Wang2022, Ju2024}, 
\dii{
although higher spatial resolution observations of gravitationally lensed galaxies with adaptive optics (e.g. \citealt{Yuan2011, Jones2013}) are also consistent with the $1\sigma$ scatter of our simulated galaxies.
}
\item The discrepancy of the gas-phase gradients at high-redshifts are unlikely due to the selection bias as shown with the the dependence on galaxy type (Fig. \ref{z_evo_type}).
Earlier type galaxies tend to have a gradient flatter by $\sim0.1$ dex/kpc compared to later type galaxies. 

\item
For future observations, the evolution of stellar metallicity gradients up to $z=5$ is also predicted (Fig. \ref{grad_z_allM_str}). The stellar gradients become flatter and flatter toward $z\sim4$. 
{Massive galaxies show inverse inner stellar gradients at $z\sim4$, }
\di{possibly because their centres formed most of their stars early from metal-poor gas, while later star formation in the outer regions used gas that was already enriched (Section \ref{sect_discussion}.2). }

\item
We find an evolutionary transition at $z\sim5$ with the mechanical feedback, where both stellar and gas-phase metallicity gradients become steep. \dii{This will be studied with larger volume simulations in our future work and } can be tested with future JWST observations.
\end{itemize}

Note that the metallicity profiles vary from galaxy to galaxy, so it is important to apply a proper fitting method to quantify the gradients. Above, we summarized the averaged results of our sample of galaxies. Nevertheless, to demonstrate this case-by-case scenario, we analyzed an example of massive and intermediate-mass galaxies (Galaxies A and B, respectively) in detail, including their kinematics (Appendix).
Galaxy A is a typical massive galaxy with negligible rotation. Even though different feedback models lead to a different gas distribution (Fig. \ref{map_highMass}), the metallicity gradients are largely unchanged (Fig.\ref{Z_R_galA}).
Galaxy B is a rotating intermediate-mass galaxy where inside-out quenching due to stellar feedback causes a {\it positive} inner gradient (Fig. \ref{Z_R_GalB}) inside the gas ring (Fig. \ref{map_lowMass}). 
This detailed feature might not be well resolved in lower-mass galaxies of our simulations, which might be the reason for the discrepancy of metallicity gradients at high redshifts. We will study this with a higher-resolution simulation in our future work.


\section*{Acknowledgements}
We thank E. C. Lake, J. Geach, and S. Bhattacharya, for fruitful discussions.
This work has made use of the University of Hertfordshire high-performance computing facility. 
This work used the DiRAC Memory Intensive service (Cosma8 / Cosma7 / Cosma6) at Durham University, managed by the Institute for Computational Cosmology on behalf of the STFC DiRAC HPC Facility (www.dirac.ac.uk). The DiRAC service at Durham was funded by BEIS, UKRI and STFC capital funding, Durham University and STFC operations grants. DiRAC is part of the UKRI Digital Research Infrastructure.
CK acknowledges funding from the UK Science and Technology Facilities Council through grants ST/V000632/1 and ST/Y001443/1. The work was also funded by a Leverhulme Trust Research Project Grant on ‘Birth of Elements’.

\section*{Data Availability}

The simulation data can be shared on request.



\bibliographystyle{mnras}
\bibliography{bibtex} 

\clearpage
\appendix

\section{Kinematics}\label{sect_kinematics}

\begin{figure*}
    \includegraphics[width=0.9\textwidth]{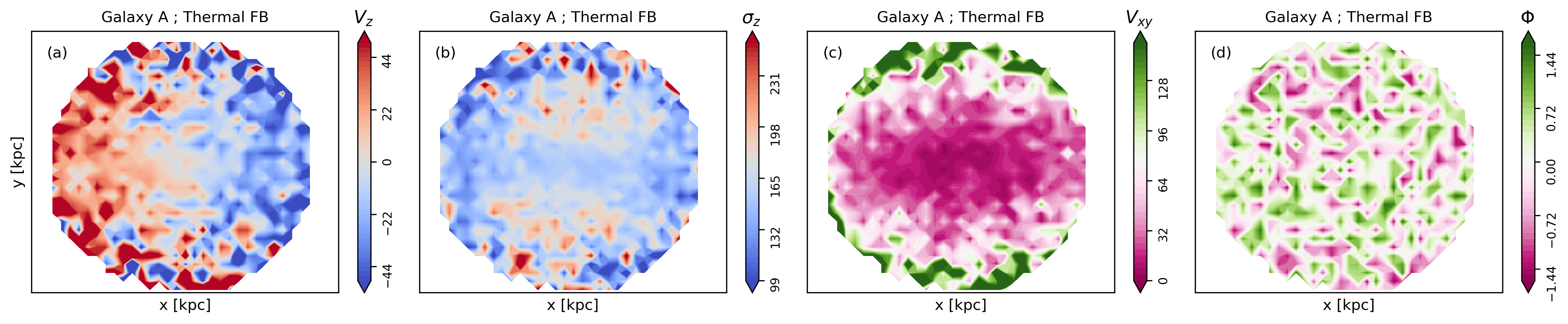}
    \includegraphics[width=0.9\textwidth]{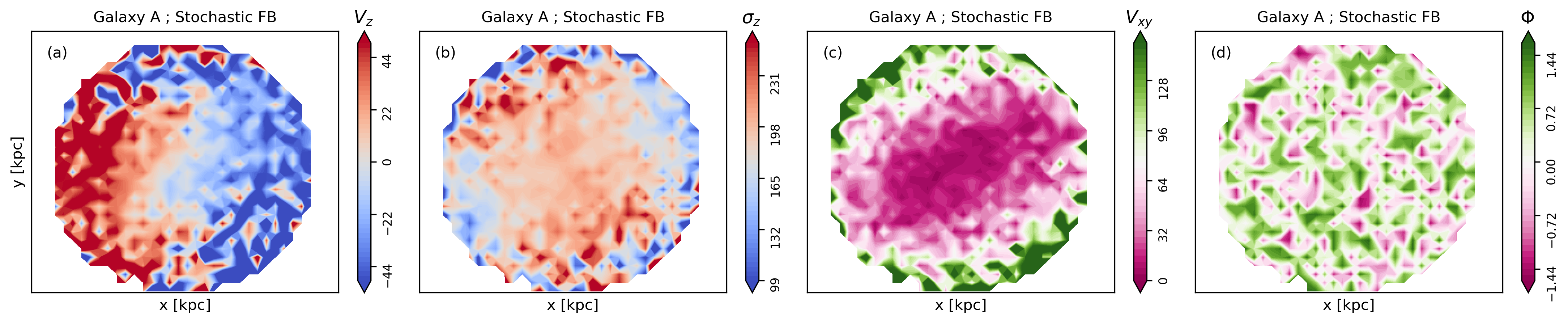}
    \includegraphics[width=0.9\textwidth]{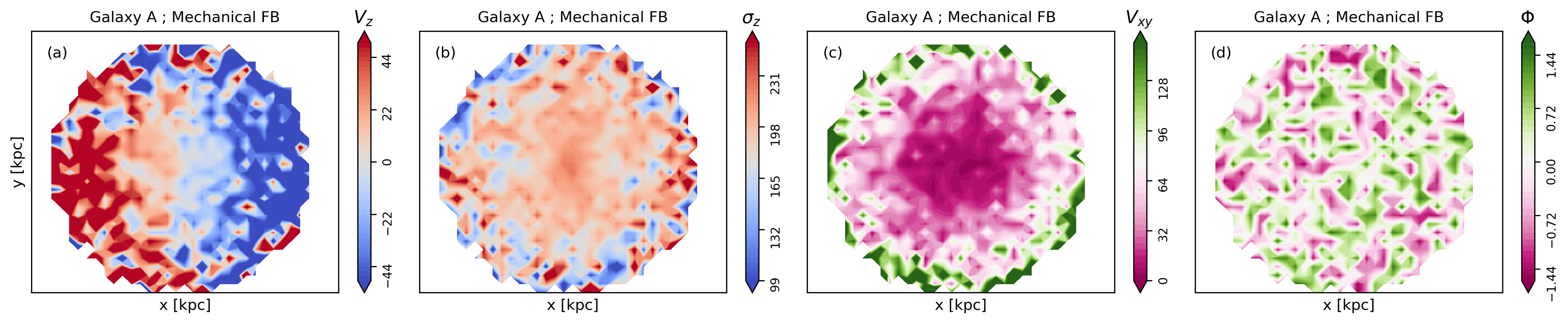}
\caption{Stellar kinematic maps of a massive galaxy A with the thermal, stochastic, and mechanical feedback simulations in the top, middle, and bottom rows, respectively. Each panel shows a projection along 20 kpc in the x and y axis for a map of the line of sight velocity <$V_z$> (first column), velocity dispersion $\sigma_z$ (second column), radial velocity $V_{xy}= \sqrt{<V_x>^2 + <V_y>^2}$ (third column), the angle $\phi = \tan^{-1}( <V_y> / <V_x>)$ showing the direction of motion in the plane (fourth column).
}
\label{Vmap_GalA_str}
\end{figure*}

Figure \ref{Vmap_GalA_str} shows the stellar projection map for the line of sight velocity $<V_z>$ (first column) of galaxy A, the radial velocity $V_{xy}= \sqrt{<V_x>^2 + <V_y>^2}$ (second column), the angle $\phi = \tan^{-1}( <V_y> / <V_x>)$ (third column) indicating the direction of motion such that $\phi=0^\circ$ means pure orbital velocity, while $\phi=90^\circ$ means pure radial velocity, and anything in between means the combination of both. And finally, the line-of-sight velocity dispersion $\sigma_z$ (fourth column). 
The line of sight velocity $<V_z>$ of stars for Galaxy A is close to zero near the centre, $V_{xy}$ is radially increasing from the centre and shows a higher velocity toward the north and south poles, and $\phi$ shows the direction of motion of the particles in the plane and shows no particular rotation for this galaxy.
These maps are obtained for galaxies observed through the $z$ axis of our simulation box, and a sphere of 20 kpc radius is projected on the $x-y$ plane. We tried a cylindrical selection (circle of $20$ kpc radius on the $xy$ plan with $\pm20$ kpc along the $z$ axis) to mimic observations, but we found no significant difference. 
\di{In this particular example, the feedback model does not appear to strongly affect the stellar kinematics, though this may reflect the specific mass and gas content of Galaxy A.}
Although, the stellar velocity dispersion is higher at the centre with the stochastic and mechanical feedback.

\begin{figure*}
    \includegraphics[width=0.9\textwidth]{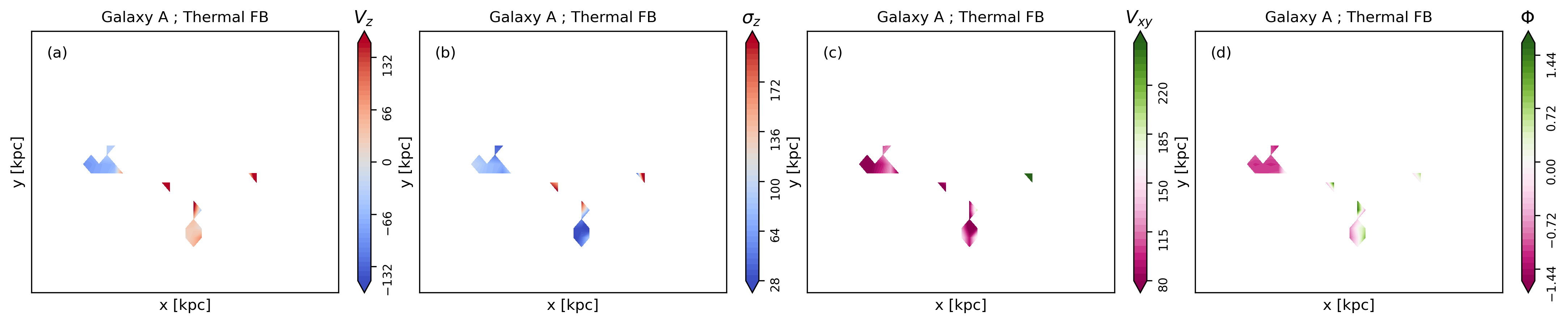}
    \includegraphics[width=0.9\textwidth]{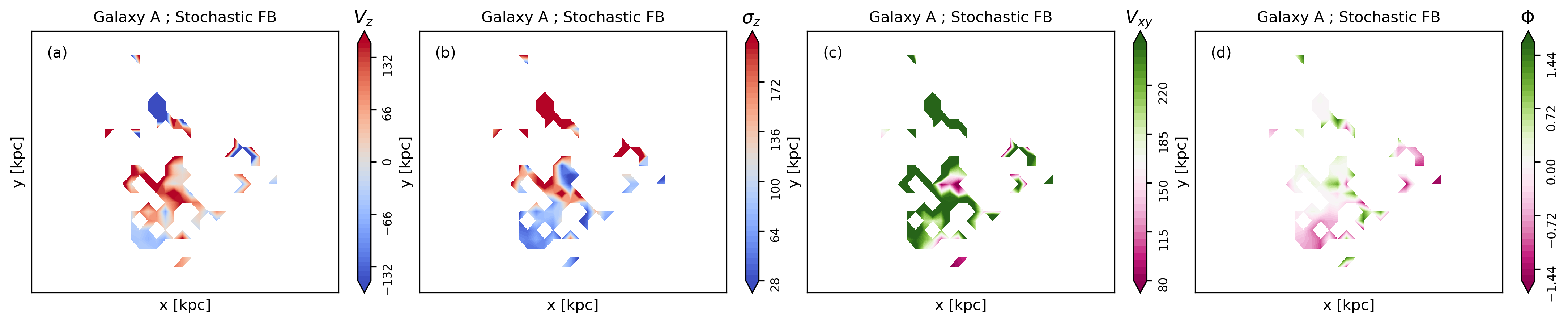}
    \includegraphics[width=0.9\textwidth]{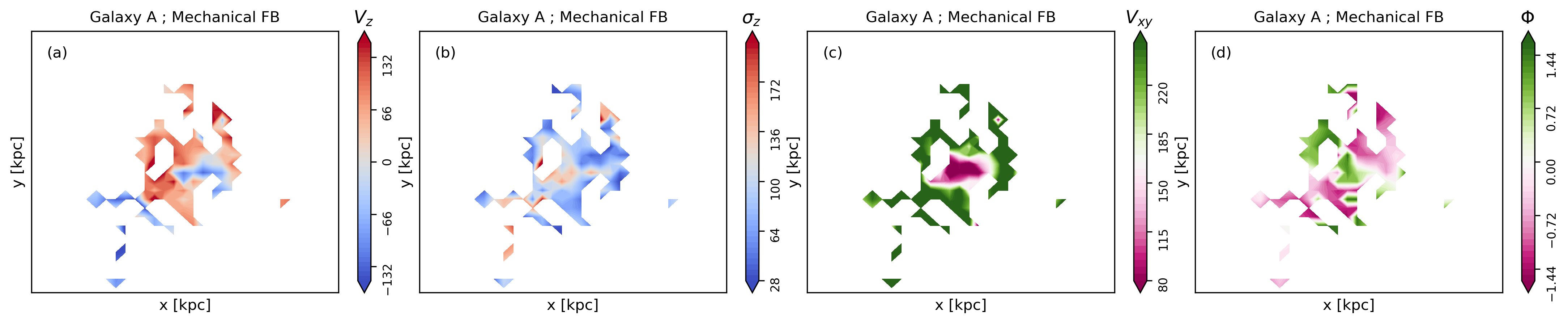}
\caption{
Same as Figure \ref{Vmap_GalA_str} but for gas kinematic maps of Galaxy A. 
}\label{Vmap_GalA_gas}
\end{figure*}

Figure \ref{Vmap_GalA_gas} is the same as Figure \ref{Vmap_GalA_str}, but for gas-phase. There is no significant rotation for stars and gas with all feedback models. The different distribution of gas among different feedback models does not affect the kinematics of stars either.


\begin{figure*}
    \includegraphics[width=0.9\textwidth]{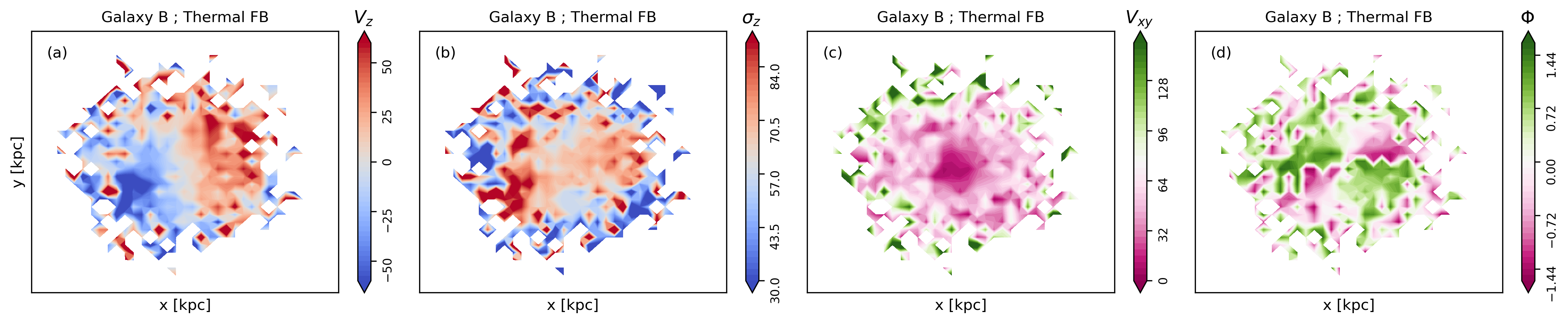}
    \includegraphics[width=0.9\textwidth]{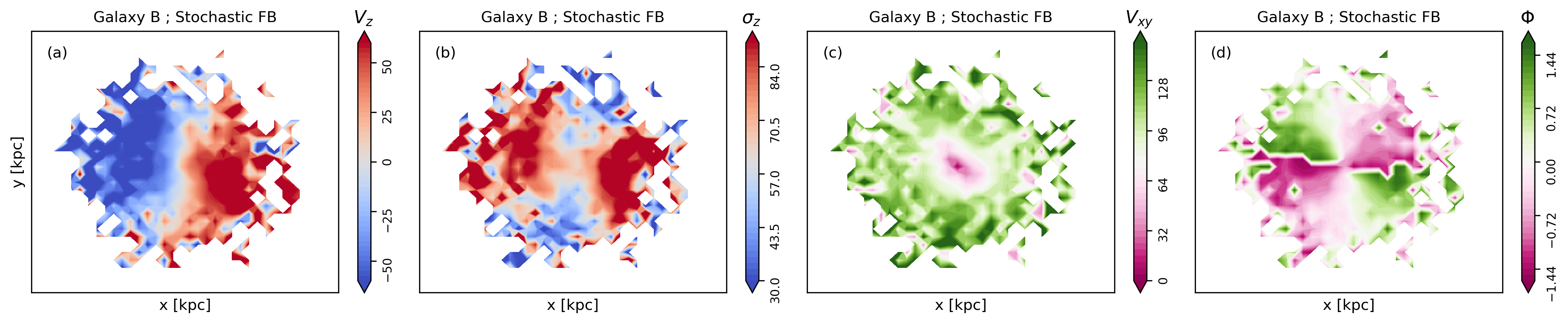}
    \includegraphics[width=0.9\textwidth]{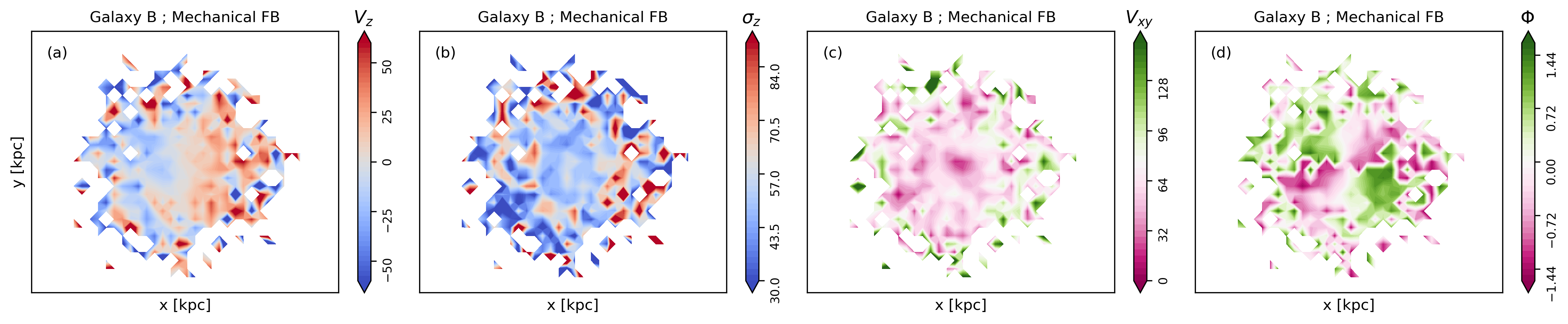}
    \caption{The same stellar kinematic maps as Figure \ref{Vmap_GalA_str} but for an intermediate-mass galaxy B. 
    }\label{Vmap_GalB_str}
\end{figure*}

\begin{figure*}
    \includegraphics[width=0.9\textwidth]{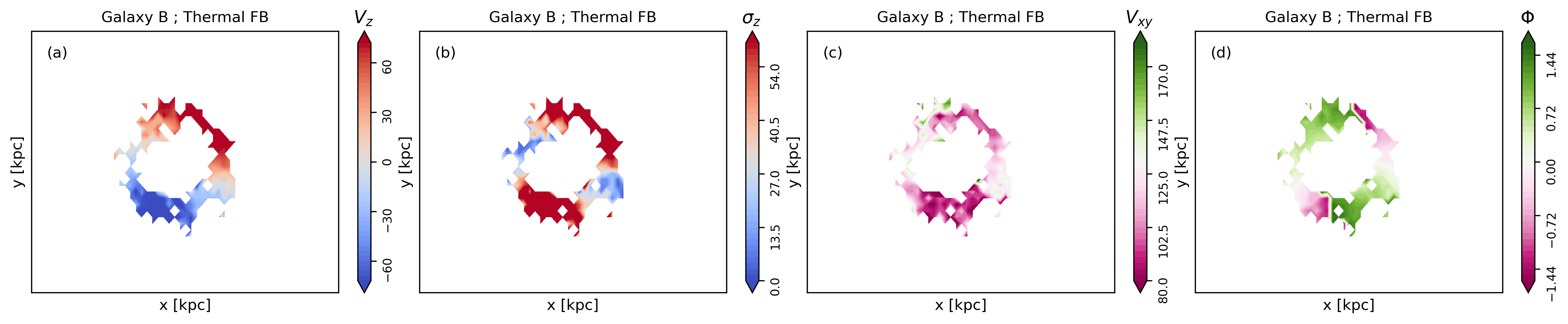}
    \includegraphics[width=0.9\textwidth]{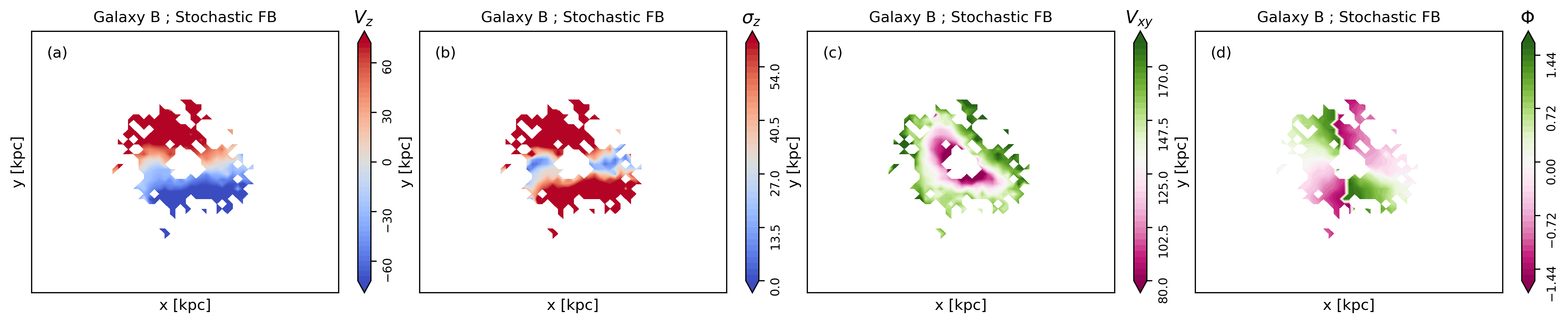}
    \includegraphics[width=0.9\textwidth]{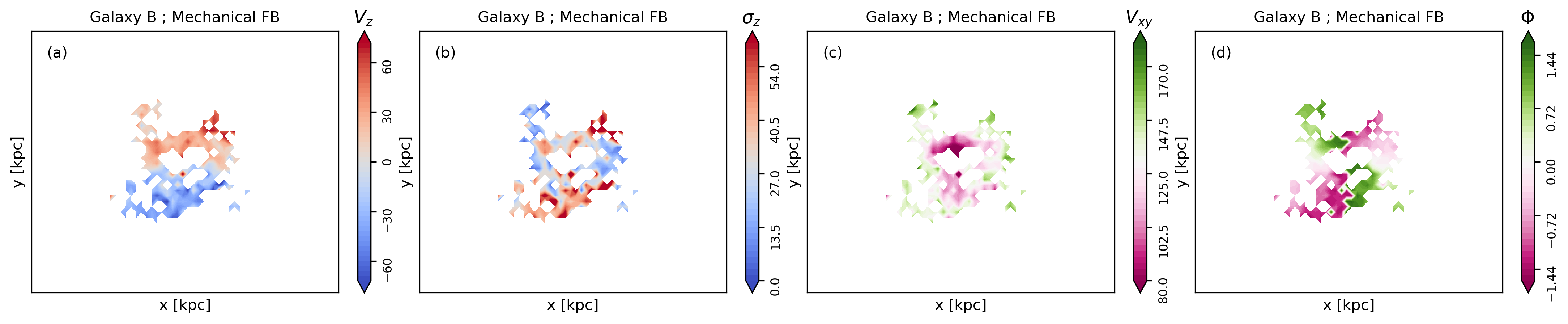}
    \caption{The same gas kinematic maps as Figure \ref{Vmap_GalA_str} but for Galaxy B.
    }\label{Vmap_GalB_gas}
\end{figure*}

Figures \ref{Vmap_GalB_str} and \ref{Vmap_GalB_gas} are the same as \ref{Vmap_GalA_str} and \ref{Vmap_GalA_gas}, respectively, but for an intermediate-mass galaxy B. This galaxy is viewed almost face-on, which makes $<V_z>$ close to zero along the disk, and $V_{xy}$ radially increasing from the centre. $\phi$ shows the direction of motion of stars and gas particles in the plane, suggesting that this galaxy is rotating.
Inside the ring structure of gas, there is a non-rotating stellar core with all feedback models.

\di{The velocity maps presented in this appendix are intended as illustrative examples of how stellar and gas kinematics vary across feedback models for two representative galaxies. While these cases provide useful insights into the diversity of kinematic structures in our simulations, they are not intended to draw general conclusions about the relationship between feedback, kinematics, and metallicity gradients based on these two cases alone. In particular, the limited gas content in some feedback models (e.g., mechanical feedback for Galaxy A) reduces the interpretability of the gas velocity fields. A more comprehensive, quantitative analysis of the role of kinematics in shaping metallicity gradients across our full galaxy sample will be explored in future work.}

\section{Star Formation Main Sequence}\label{sect_SFMS}
\begin{figure*}
    \begin{subfigure}{1.0\textwidth}
    \includegraphics[width=0.5\textwidth]{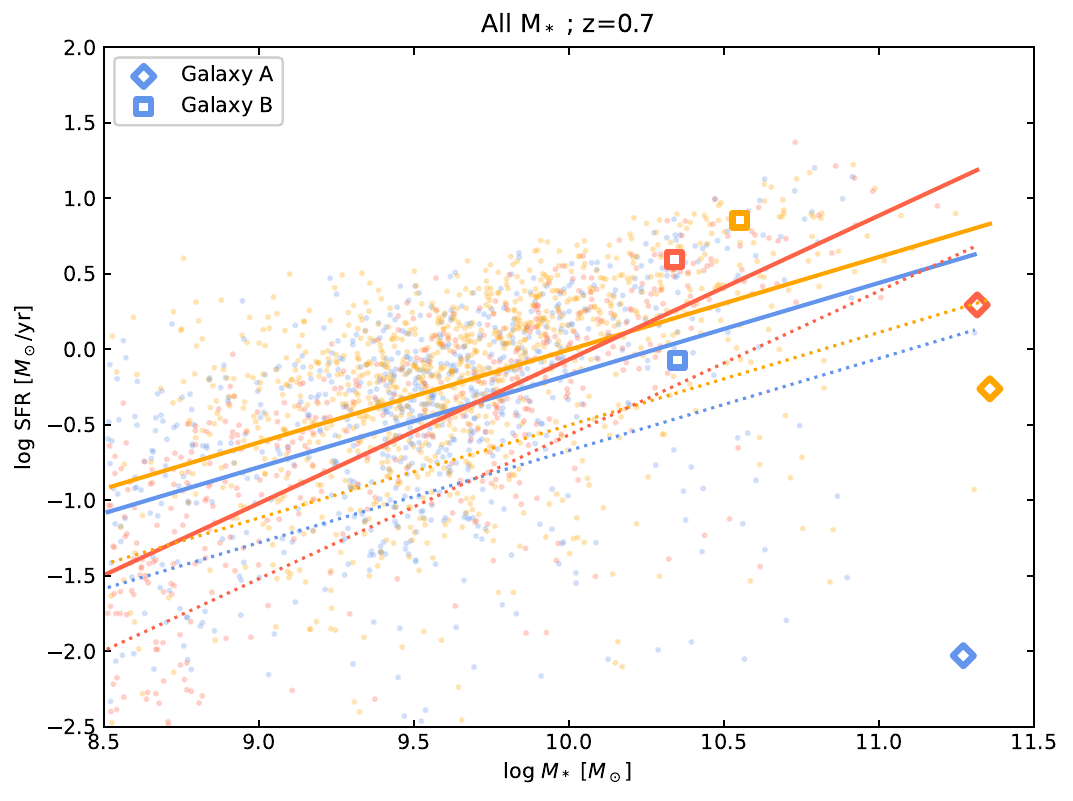}
    \includegraphics[width=0.46\textwidth]{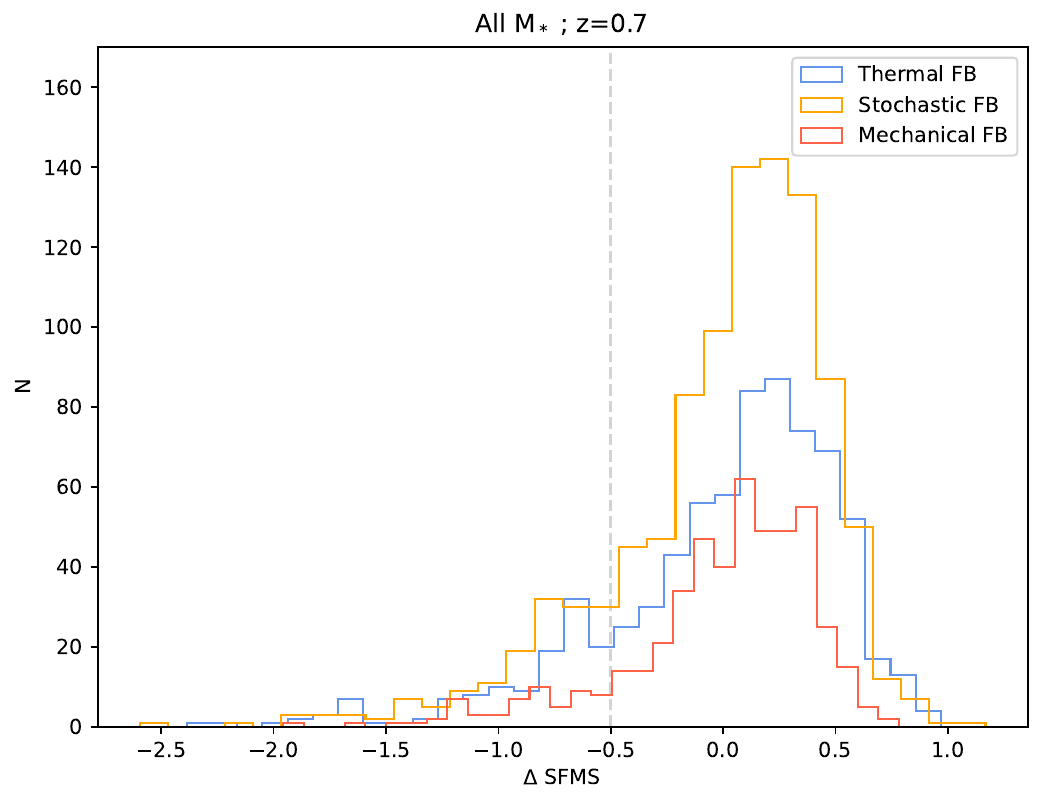}
    \end{subfigure}\caption{\textit{Left}: SFMS for all galaxies in our simulations with the thermal (blue), stochastic (orange), and mechanical (red) feedback models at $z=0.7$. The solid lines show the best fit to the data. The dotted lines are the perpendicular shift of the solid lines by $-0.5$ dex. The diamond and square symbols represent galaxy A and B, respectively in each model.
    \textit{Right}: Distribution of the perpendicular distances from the SFMS ($\Delta$SFMS) in our simulations with the thermal (blue), stochastic (orange), and mechanical (red) feedback models.}\label{SFMS}
\end{figure*}

All simulations with three different feedback models have a well-defined SFMS and a small number of ETGs below the SFMS. The left panel in Figure \ref{SFMS} shows the SFMS for the thermal (blue), stochastic (orange), and mechanical (red) feedback models. The points show individual galaxies, and the solid lines are the linear fits of these points. The dashed lines are parallel to the solid lines shifted down by $0.5$ dex/kpc, which defines our delimitation for ETGs and LTGs (Section \ref{gal_type}). \di{The diamond and square symbols represent galaxy A and B, respectively in each model.}
The right panel in Figure \ref{SFMS} is the distribution of the perpendicular distances $\Delta$SFMS. The grey dashed line shows the division between ETGs/LTGs at $\Delta$SFMS$=-0.5$. Although the simulation with the stochastic model has a larger number of galaxies, the fraction of ETGs is not so different and is {20, 14, and 12}\% for thermal, stochastic, and mechanical feedback, respectively.

\begin{figure*}
	\includegraphics[width=0.8\textwidth]{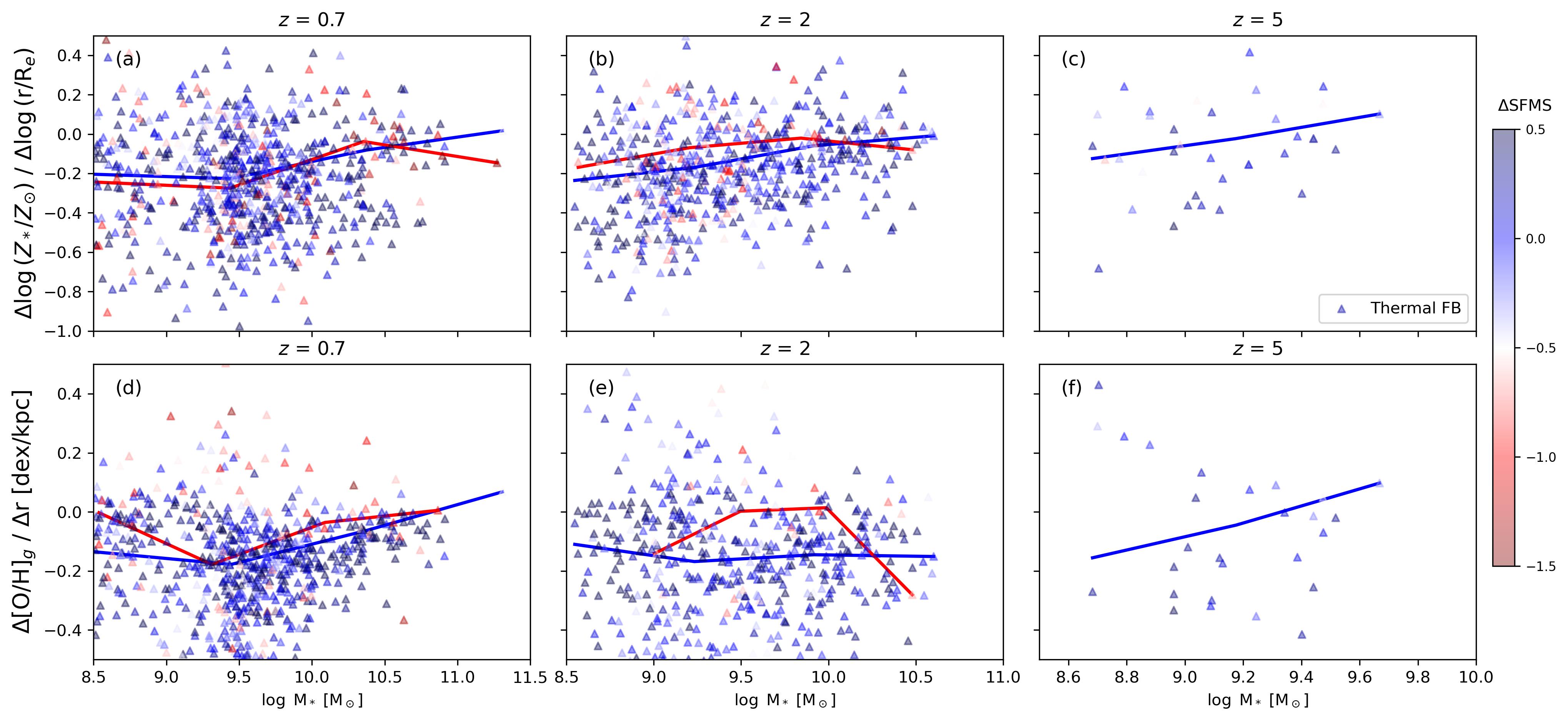}\caption{Same as Figure \ref{Grad_M}, but for the thermal feedback only, colour-mapped by $\Delta$SFMS for ETGs (red) and LTGs (blue).}\label{GradMassSFMS_Thermal}
    \includegraphics[width=0.8\textwidth]{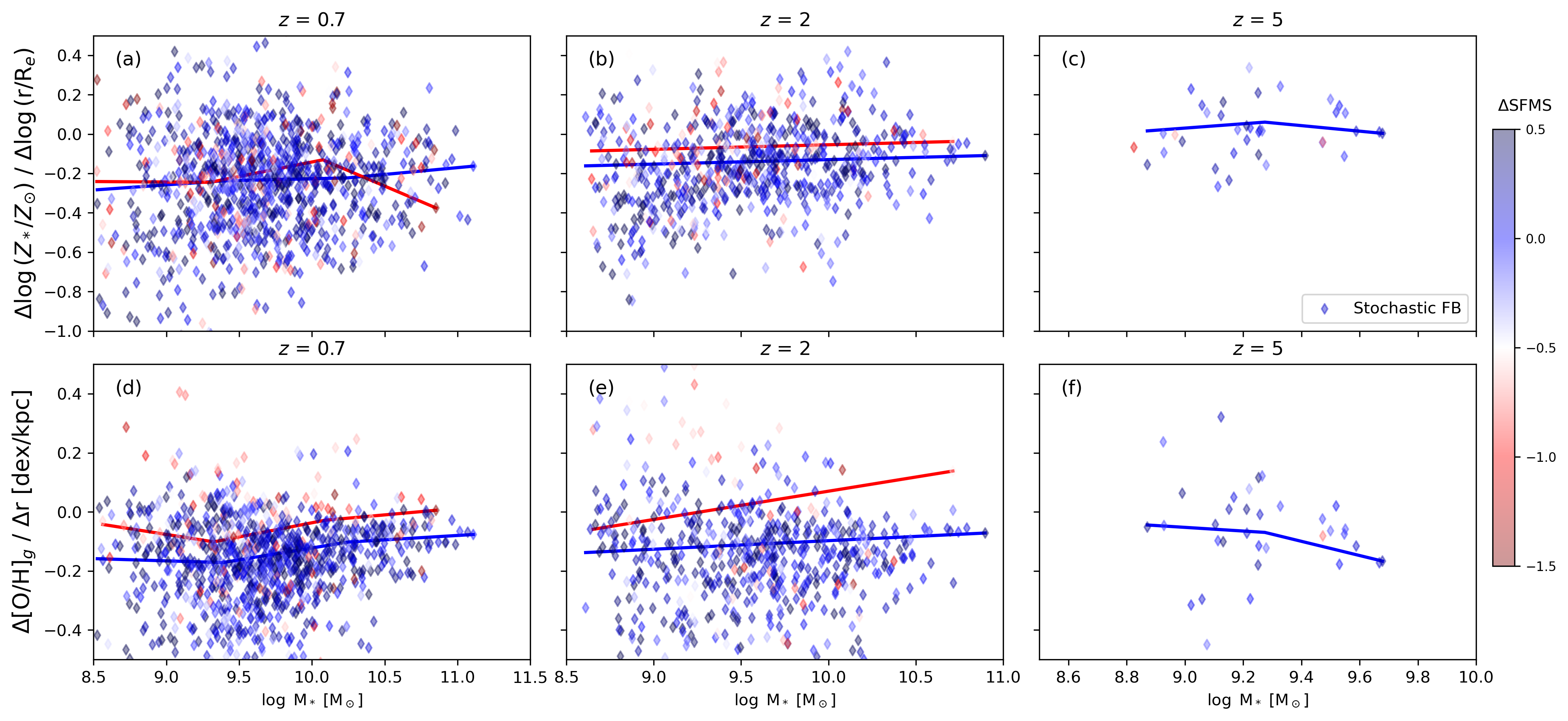}\caption{Same as Figure \ref{GradMassSFMS_Thermal}, but for the stochastic feedback.}\label{GradMassSFMS_Sto}
    \includegraphics[width=0.8\textwidth]{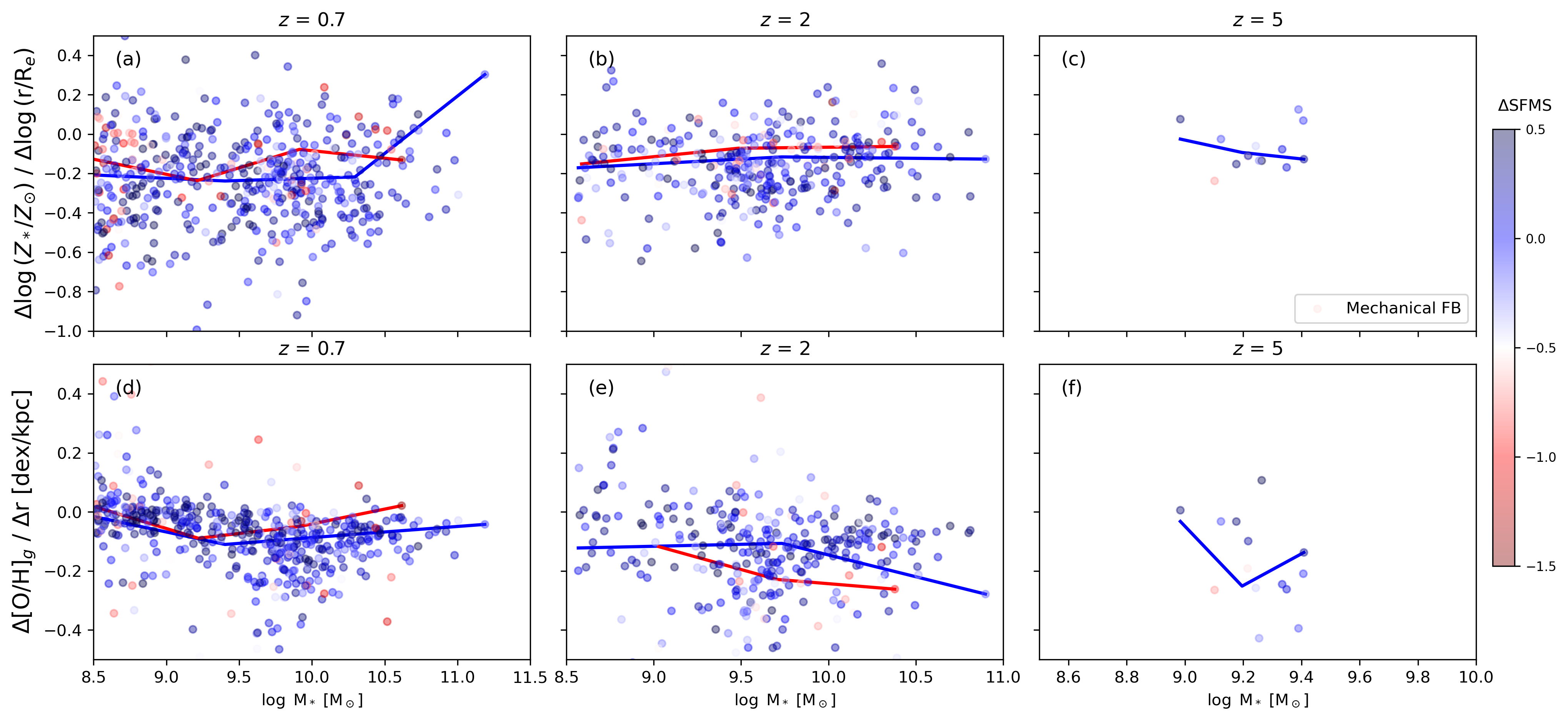}\caption{Same as Figure \ref{GradMassSFMS_Thermal}, but for the mechanical feedback.}\label{GradMassSFMS_Mec}
\end{figure*}

Figure \ref{GradMassSFMS_Thermal} is similar to Figure \ref{Grad_M}, where we show the stellar (top panels) and gas-phase (bottom panels) metallicity gradients as a function of galaxy stellar mass at $z=0.7$, $z=2$, and $z=4$ (first, second, and third columns, respectively) for our simulation with the thermal feedback. Each symbol is an individual galaxy, and the data is colour-coded by SFMS. The solid lines are the linear fits {median} of the galaxies with $\Delta$SFMS${\geq -0.5}$ (blue line) and $\Delta$SFMS$<0.5$ (red line).
As we have already seen, there is a weak correlation between stellar gradients and mass, where more massive galaxies tend to have flatter gradients. This is also the case if we split our sample into LTGs (blue triangles) and ETGs (red triangles). Overall, ETGs (the red lines) have flatter gradients than LTGs (blue lines) at all mass ranges and at all redshifts plotted here.
The gas-phase metallicity gradients are significantly flatter for ETGs (red), notably at the low mass end with a $\sim 0.1$ dex/kpc difference. The difference becomes small at $\log M_*/$M$_\odot > 10.5$. The range of gas-phase gradients for ETG is wider at $z=2$, with galaxies reaching $0.3$ dex/kpc. This continues at $z=4$, where ETGs seem to have flatter gas-phase gradients than LTGs.

Figure \ref{GradMassSFMS_Sto} and  \ref{GradMassSFMS_Mec} are the same as  \ref{GradMassSFMS_Thermal} with the stochastic and mechanical feedback models, respectively.  The stochastic feedback shows that massive ($\log M_*> 10$M$_\odot$) LTGs have gradients flatter than massive LTGs at all shown redshifts, which is not the case in thermal and mechanical models. The gas-phase metallicity gradient for ETGs at $z=2$ becomes much steeper toward low-mass galaxies, which may be due to the ejection of metals from the centre and could be a clear signature to test this feedback model.
The galaxy-type dependence of the metallicity gradients with mechanical feedback is, overall, similar to thermal feedback but with a slightly smaller difference between ETGs and LTGs. Overall, at all mass ranges, ETGs tend to show a flatter gradient than LTGs, which is expected and can be explained by the merging history of these galaxies. 

\section{\di{Tracking individual galaxies}}

\begin{figure*}
  \centering

  \begin{subfigure}[t]{0.32\textwidth}
    \includegraphics[width=\linewidth]{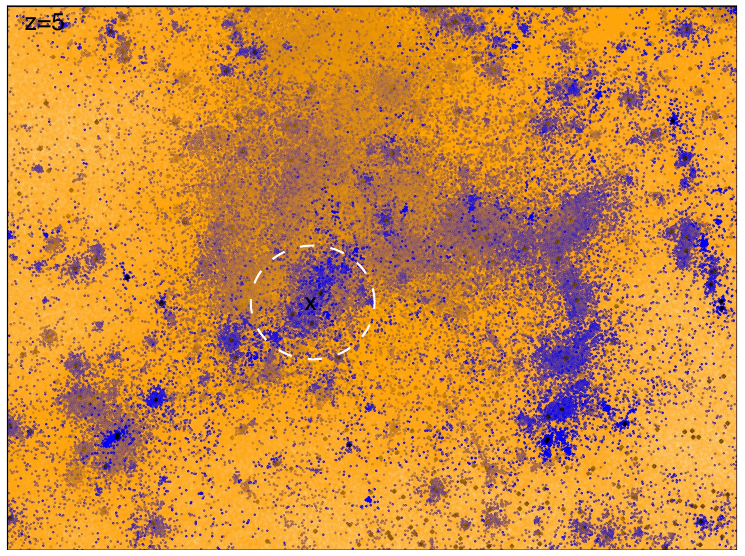}
  \end{subfigure}\hfill
  \begin{subfigure}[t]{0.32\textwidth}
    \includegraphics[width=\linewidth]{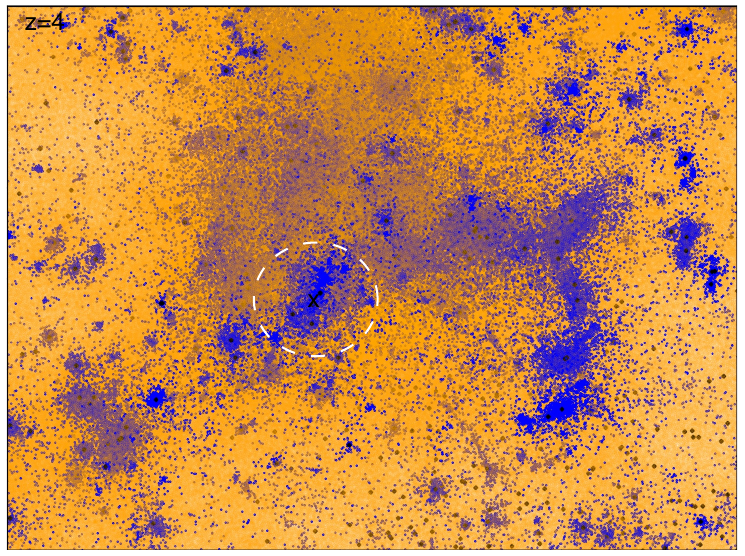}
  \end{subfigure}\hfill
  \begin{subfigure}[t]{0.32\textwidth}
    \includegraphics[width=\linewidth]{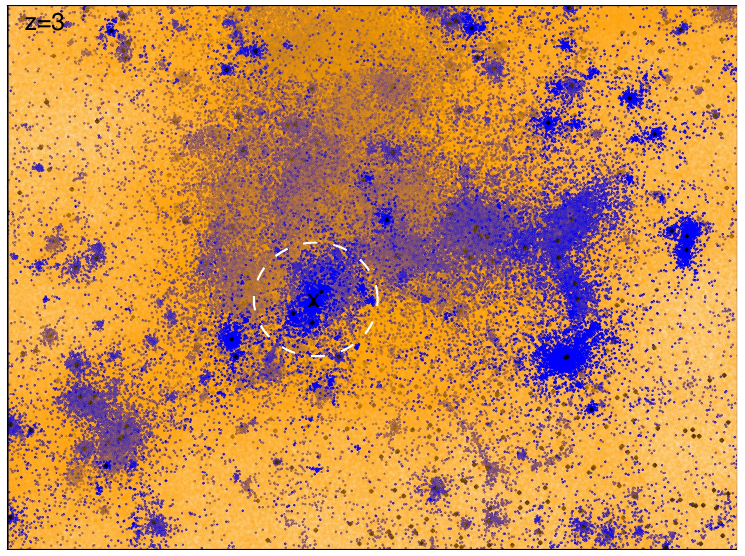}
  \end{subfigure}

  \vspace{0.5em}

  \begin{subfigure}[t]{0.32\textwidth}
    \includegraphics[width=\linewidth]{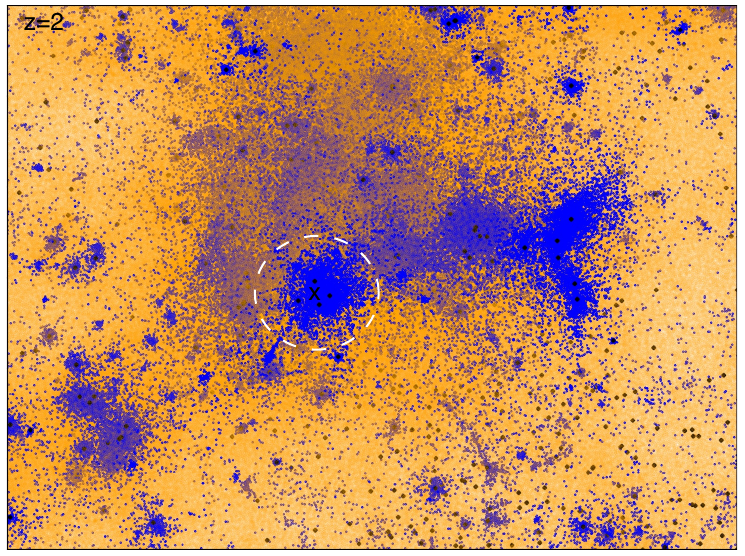}
  \end{subfigure}\hfill
  \begin{subfigure}[t]{0.32\textwidth}
    \includegraphics[width=\linewidth]{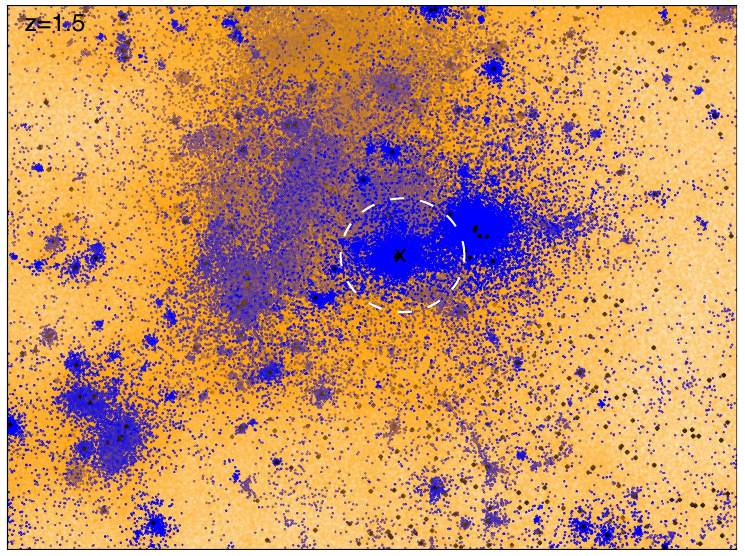}
  \end{subfigure}\hfill
  \begin{subfigure}[t]{0.32\textwidth}
    \includegraphics[width=\linewidth]{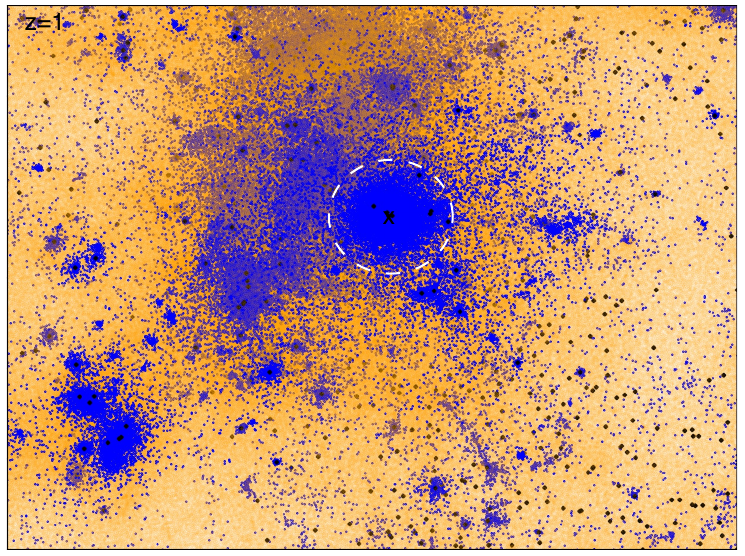}
  \end{subfigure}

  \vspace{0.5em}

  \begin{subfigure}[t]{0.32\textwidth}
    \includegraphics[width=\linewidth]{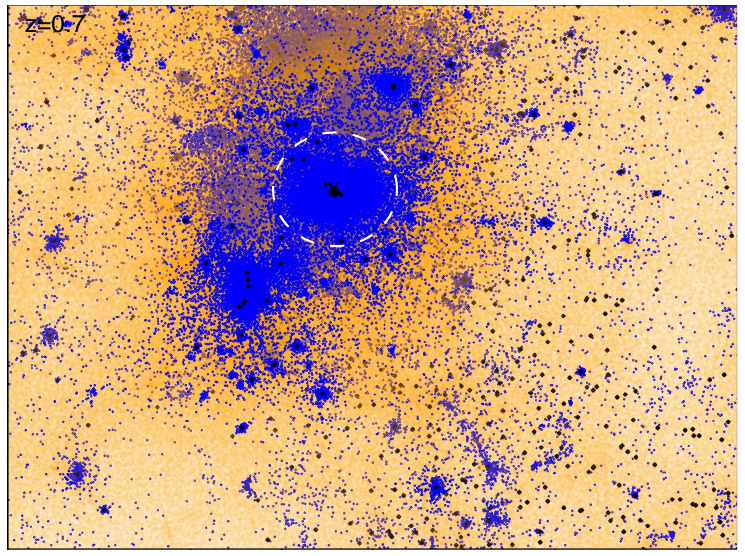}
  \end{subfigure}\hfill
  \begin{subfigure}[t]{0.32\textwidth}
    \includegraphics[width=\linewidth]{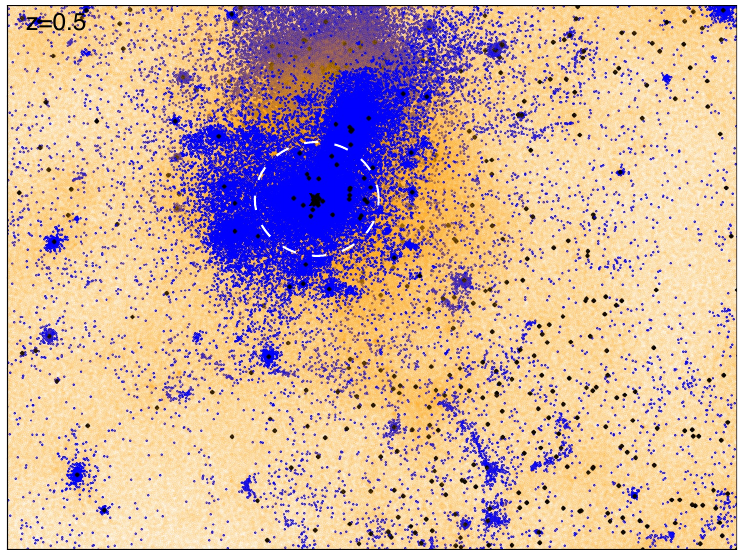}
  \end{subfigure}\hfill
  \begin{subfigure}[t]{0.32\textwidth}
    \includegraphics[width=\linewidth]{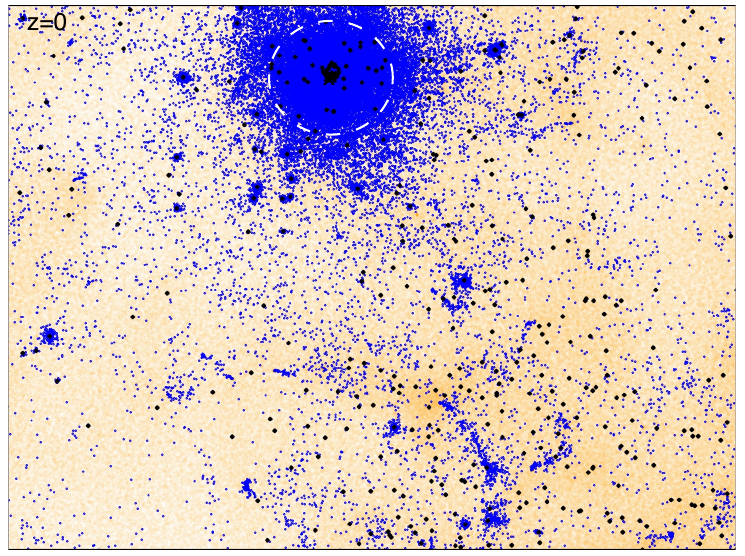}
  \end{subfigure}

  \caption{\di{Formation history of Galaxy A across cosmic time with the mechanical feedback. Each panel shows a projected view of the simulation volume {in a box of 4$\times$3 Mpc$^2$ in side}, with gas particles in orange, star particles in blue, and the central Friends-of-Friends group marked with a black cross. {The white circle shows the selected descendant of Galaxy A.}
  The panels are ordered chronologically from top left ($z = 5$) to bottom right ($z = 0$). }
    }
  \label{galA_map_evo}
\end{figure*}

\di{Figure \ref{galA_map_evo} shows the formation history of Galaxy A from $z=5$ (top left panel) to $z=0$ (bottom right panel) with the mechanical feedback. Each panel shows the cosmic map with gas particles (orange), star particles (blue), and the central Friends-of-Friends group (black cross).}

\di{The galaxy member star particles are selected at $z=0$, and traced using their ID numbers across redshift.{Each panel has a physical side length of 4$\times$3 Mpc$^2$. We adopt a fixed centre chosen to include all member particles of Galaxy A at $z=0.7$. All gas particles in the frame are also plotted in the background. The selected descendant is marked with a black cross surrounded by a white circle.}
This `merging tree' is used for making the metallicity gradient evolution in Figures \ref{Zprofil_galA} and \ref{Zstrprofil_galA}.}

\newpage


\bsp	
\label{lastpage}
\end{document}